\documentclass[preprint2]{emulateapj}

\newcommand{\chn}{{\it Chandra}}
\newcommand{\app}{Appendix A}

\slugcomment{version \today: fm}
\shorttitle{Large Scale Extragalactic Jets in the \chn~ era I}
\shortauthors{F. Massaro, D. E. Harris, C. C. Cheung}

\begin{document}
\title{Large Scale Extragalactic Jets in the \chn~era I: \\ data reduction and analysis}
\author{F. Massaro\altaffilmark{1}, D. E. Harris\altaffilmark{1}, C. C. Cheung\altaffilmark{2}}
\altaffiltext{1}{Harvard - Smithsonian Astrophysical Observatory, 60 Garden Street, Cambridge, MA 02138, USA}
\altaffiltext{2}{National Research Council Research Associate, National Academy of Sciences, Washington, DC 20001, resident at Naval Research Laboratory, Washington, DC 20375, USA}

\begin{abstract} 
In this paper we report the first stages of an investigation into the
X-ray properties of extragalactic jets.  Our approach is to subject
all sources for which X-ray emission has been detected by \chn\ to
uniform reduction procedures.  Using \chn\ archival data for 106 such
sources, we measure X-ray fluxes in three bands and compare these to
radio fluxes.  We discuss the sample, the reduction methods, and
present first results for the ratio of X-ray to radio flux for jet
knots and hotspots.
In particular, we apply statistical tests to various distributions of
key observational parameters to evaluate differences between the
different classes of sources.  Subsequent papers will deal with
various ramifications such as considerations of how well the
observational data fulfill expectations of the different radiation
processes proposed for the knots of FR\,I radio galaxies and quasars.
\end{abstract}


\section{Introduction}\label{sec:intro} 
Over the past decade, \chn~ observations of radio galaxies (RGs) and quasars (QSRs)
have provided crucial clues to investigate the nature of their extended components
(e.g., jets and hotspots). 
Since the \chn~discovery of X-ray emission in the 100 kpc radio jet of
PKS 0637-752 \citep{schwartz00}, 
the number of sources with X-ray detected extended components has increased 
from a handful to more than 100. For these radio sources with extended components detected in the X-ray band, 
we compiled their observed parameters, aiming at developing a {\it classification
criterion} for these extragalactic components, independent of their core properties,
and only based on the different radiative processes proposed to interpret
their X-ray emission.

In 1974, Fanaroff and Riley introduced a classification scheme for extragalactic radio sources. 
Investigating the complete sample of 3C
sources \citep{mackay71}, they chose two main classes noting that relative
positions of regions of high and low surface brightness in the
extended regions of extragalactic radio sources are correlated with
their radio luminosities.
In particular, they divided these radio sources into two classes using the
ratio $R_{FR}$ of the distance between the regions of highest surface
brightness on opposite sides of the central galaxy and/or quasar, to
the total extent of the source up to the lowest brightness contour in
the radio map.  Sources with $R_{FR}$ $\leq$ 0.5 were placed in Class
I (i.e., FR\,I) and sources with $R_{FR}$ $\geq$ 0.5 in Class II
(i.e., FR\,II).  In particular, at radio frequencies, FR\,Is show
surface brightness higher toward their cores while FR\,IIs toward
their edges \citep{fanaroff74}.
 
Fanaroff and Riley found that all sources with luminosity
$L_{\rm 178 MHz}$ $\leq$ 2 $\times$ 10$^{25}$ $h_{\rm 100}^{-2}$ W Hz$^{-1}$
str$^{-1}$ were classified as Class I while the brighter sources all were Class II. 
This luminosity boundary between them is not very sharp, and
there is some overlap in the luminosities of sources classified as
FR\,I or FR\,II on the basis of their structures.

Since the original work, the FR scheme remains successful in classifying RGs. 
Extending the FR morphological/radio luminosity classification to radio-loud sources in general, 
one needs to consider the source orientation with respect to the observer. 
Under such `unification schemes' \citep{urry95}, all radio-loud QSRs are posited to be FR\,II RG's 
observed at smaller angles to our line of sight  \citep[see also][]{barthel89}. 
Orr \& Browne (1982) introduced a sub-classification for QSRs distinguishing 
between core-dominated quasars (CDQ) and lobe-dominated quasars (LDQ), 
on the basis of the ratio between each source's radio core flux relative to the extended 
one \citep[see also][]{hine80,hough89}. The most core-dominated sources have the largest such ratios 
presumably due to Doppler boosting of emission from a relativistic jet aligned most closely to our line of sight.  
However, this classification based on core-dominance is not as useful as the basic FR\,I/FR\,II 
distinction since the determining parameters are more strongly dependent 
on the angular resolution and frequency of a particular observation, nor is there 
an obvious correlation with radio luminosity or other primary observational parameter \citep[e.g.,][]{landt06}.

The radio to optical emission arising from extended components of
RGs and QSRs is widely interpreted as synchrotron
radiation by relativistic particles \citep[e.g.,][]{meisenheimer89,scarpa02,sambruna04}, whereas the origin of X-ray
emission in their components (i.e., knots in jets and hotspots)
is still unclear, but certainly non-thermal \citep[see][]{harris02,harris06,worrall09}.
The main dichotomy lies in which emission mechanism, synchrotron or inverse
Compton (IC) scattering, dominates the X-ray emission of jets and
hotspots. The former describes emission from low power jets, typically
in FR\,Is, while the latter provides an explanation for high power
RGs (i.e., FR\,IIs) and QSRs \citep[see e.g.,][]{harris06}.

Over the past ten years, we have collected a sample of RGs
and QSRs for which the X-ray emission associated with their radio
jets or hotspots has been detected by \chn.  The main goal of this
endeavor (hereinafter {\it XJET
project}\footnote{\underline{http://hea-www.harvard.edu/XJET/}}) 
\citep{harris10,massaro10a,massaro11a} is the development of a new
classification criterion based not only on the radio morphology and
power but also on the X-ray properties of these extended components.
Our goal is to find criteria to distinguish between knots in jets and
hotspots, both in RGs and QSRs, linked with the
radiative process responsible for their X-ray emission.
Here, we present the X-ray data reduction and analysis procedures
together with some basic results.
The current work (Paper I) represents our initial effort, and although we have not yet developed
a new classification scheme as originally envisaged, we will explore additional lines of investigation 
in forthcoming works, such as including the information at 
optical frequencies to see if a more complete SED description 
will permit us a better differentiation of extragalactic jet components.
In particular, the XJET project guidelines can be summarized as follows:
in Paper II \citep{massaro11b}, we develop a statistical test for the IC/CMB model 
in QSR jets, while in Paper III \citep{massaro11c}, 
we plan to introduce the radio and X-ray emission from the cores as new parameters 
that could be used to classify RG and QSR jets 
and we will also provide measurements of the upper limits for
the undetected X-ray components in our XJET sample, 
taking into account the \chn\ sensitivity limits. 
In Paper IV \citep{harris11}, we will present a detailed
comparison between the distributions of the observed parameters 
and the expectations of both analytical and numerical calculations
for different radiative scenarios. Finally, the possible extension 
of the XJET project to the infrared and optical frequencies is already in progress.

Throughout, we assume a flat cosmology with $H_{\rm 0}=72$ km s$^{-1}$ Mpc$^{-1}$,
$\Omega_{\rm M}=0.26$ and $\Omega_{\Lambda}=0.74$ \citep{dunkley09}
and spectral indices, $\alpha$, are defined by flux density,
$S_{\nu}\propto\nu^{-\alpha}$. Unless otherwise stated, we use cgs units.

\begin{table*}
\caption{The basic parameters of the (00-08 h) sources considered in the XJET sample (see Section~\ref{sec:sample}).}
\begin{tabular}{|lccccccccccc|}
\hline
Name  & Class &   RA    &   DEC   & z  &  $D_{\rm L}$  &   scale  &    $N_{\rm H,Gal}$     & \chn   & CFS & CFM & Ref.\\
      &       & (J2000) & (J2000) &    &  [Mpc]  & [kpc/''] & [10$^{20}$~cm$^{-2}$] & Obs ID &     &     &     \\
\hline
\noalign{\smallskip}
3C~6.1          & FR~II & 00 16 31.147  & +79 16 49.88 & 0.8404 &  5297  & 7.58  & 15.5 & 3009 & 2.35 & 1.20 & H04\\
3C~9            & LDQ   & 00 20 25.220  & +15 40 54.73 & 2.0120 & 15777  & 8.43  & 3.57 & 1595 & 1.22 & 1.04 & 1  \\
3C~15           & FR~I  & 00 37 04.114  & -01 09 08.46 & 0.0730 &   321  & 1.35  & 2.32 & 2178 & 1.14 & 1.02 & 2  \\
3C~17           & FR~II & 00 38 20.528  & +02 07 40.49 & 0.2197 &  1066  & 3.47  & 1.99 & 9292 & 1.12 & 1.02 & M09\\
NGC~315         & FR~I  & 00 57 48.891  & +30 21 08.75 & 0.0165 &    70  & 0.33  & 5.83 & 4156 & 1.38 & 1.07 & 3  \\
3C~31           & FR~I  & 01 07 24.961  & +32 24 45.01 & 0.0167 &    71  & 0.33  & 5.36 & 2147 & 1.34 & 1.07 & 4  \\
0106+013        & CDQ   & 01 08 38.771  & +01 35 00.31 & 2.0990 & 16626  & 8.39  & 2.42 & 9281 & 1.14 & 1.02 & H11\\
3C~33           & FR~II & 01 08 52.878  & +13 20 14.38 & 0.0597 &   260  & 1.12  & 3.23 & 7200 & 1.20 & 1.05 & 5  \\
3C~47           & LDQ   & 01 36 24.423  & +20 57 27.40 & 0.4250 &  2297  & 5.48  & 4.95 & 2129 & 1.30 & 1.06 & H04\\
3C~52           & FR~II & 01 48 28.909  & +53 32 28.04 & 0.2854 &  1437  & 4.22  & 16.8 & 9296 & 2.49 & 1.23 & M10\\
4C~+35.03       & FR~I  & 02 09 38.553  & +35 47 51.04 & 0.0369 &   158  & 0.71  & 6.16 &  856 & 1.40 & 1.08 & W01\\
PKS~0208-512    & CDQ   & 02 10 46.283  & -51 01 02.95 & 0.9990 &  6575  & 7.98  & 1.84 & 4813 & 1.11 & 1.02 & M05\\
3C~61.1         & FR~II & 02 22 35.571  & +86 19 06.38 & 0.1878 &   893  & 3.07  & 7.87 & 9297 & 1.53 & 1.10 & M10\\
3C~66B          & FR~I  & 02 23 11.409  & +42 59 31.25 & 0.0215 &    91  & 0.42  & 7.67 &  828 & 1.52 & 1.09 & 6  \\
3C~83.1         & FR~I  & 03 18 15.669  & +41 51 27.91 & 0.0251 &   107  & 0.49  & 13.4 & 3237 & 2.11 & 1.19 & 7  \\
3C~105          & FR~II & 04 07 16.453  & +03 42 25.80 & 0.0890 &   396  & 1.62  & 11.5 & 9299 & 1.88 & 1.16 & M10\\
PKS~0405-123    & CDQ   & 04 07 48.432  & -12 11 36.69 & 0.5740 &  3305  & 6.47  & 3.48 & 2131 & 1.22 & 1.05 & S04\\
3C~109          & FR~II & 04 13 40.349  & +11 12 14.78 & 0.3056 &  1556  & 4.43  & 15.2 & 4005 & 2.30 & 1.21 & H04\\
PKS~0413-210    & CDQ   & 04 16 04.364  & -20 56 27.70 & 0.8080 &  5043  & 7.48  & 2.49 & 3110 & 1.15 & 1.03 & M05\\
3C~111          & FR~II & 04 18 21.309  & +38 01 36.28 & 0.0485 &   210  & 0.93  & 29.1 & 9279 & 4.30 & 1.40 & H11\\
3C~120          & FR~I  & 04 33 11.098  & +05 21 15.59 & 0.0330 &   141  & 0.64  & 10.6 & 3015 & 1.78 & 1.14 & 8  \\
3C~123          & FR~II & 04 37 04.367  & +29 40 13.67 & 0.2177 &  1055  & 3.45  & 18.0 &  829 & 2.62 & 1.24 & 9  \\
3C~129          & FR~I  & 04 49 09.072  & +45 00 39.35 & 0.0208 &    88  & 0.41  & 59.6 & 2218 &15.56 & 1.93 & 10 \\
Pictor A        & FR~II & 05 19 49.700  & -45 46 44.50 & 0.0350 &   150  & 0.68  & 3.12 &  346 & 1.18 & 1.03 & 11 \\ 
PKS~0521-365    & BL   & 05 22 57.992  & -36 27 30.62 & 0.0550 &   239  & 1.04  & 3.58 &  846 & 1.22 & 1.04 & 12 \\
0529+075        & CDQ   & 05 32 38.998  & +07 32 43.31 & 1.2540 &  8745  & 8.35  & 17.1 & 9289 & 2.54 & 1.22 & H11\\
PKS~0605-085    & CDQ   & 06 07 59.700  & -08 34 49.98 & 0.8700 &  5531  & 7.67  & 18.3 & 2132 & 2.64 & 1.25 & S04\\
3C~173.1        & FR~II & 07 09 18.187  & +74 49 31.91 & 0.2920 &  1476  & 4.29  & 4.60 & 3053 & 1.29 & 1.06 & H04\\
3C~179          & LDQ   & 07 28 10.901  & +67 48 47.70 & 0.8460 &  5341  & 7.60  & 4.38 & 2133 & 1.27 & 1.06 & S02\\
B2~0738+313     & CDQ   & 07 41 10.703  & +31 12 00.23 & 0.6310 &  3712  & 6.77  & 4.32 &  377 & 1.27 & 1.06 & 13 \\
3C~189          & FR~I  & 07 58 28.109  & +37 47 11.79 & 0.0428 &   184  & 0.82  & 4.68 &  858 & 1.29 & 1.05 & W01\\
\noalign{\smallskip}
\hline
\end{tabular}\\
~\\
Note: `CFS' and `CFM 'correspond to the correction factor for the Galactic absorption in the soft and in the medium bands, respectively
(see Section~\ref{sec:absorp} for more details).
~\\
References: H04 \citep{hardcastle04}, (1) \citep{fabian03a}, (2) \citep{kataoka03b}, M09 \citep{massaro09c}, (3) \citep{worrall03}, (4) \citep{hardcastle02a},
H11 \citep{hogan11}, (5) \citep{kraft07}, M10 \citep{massaro10b}, W01 \citep{worrall01}, M05 \citep{marshall05}, (6) \citep{hardcastle01b}, (7) \citep{sun05},
S04 \citep{sambruna04}, (8) \citep{harris04}, (9) \citep{hardcastle01a}, (10) \citep{harris02a}, (11) \citep{wilson01}, (12) \citep{birkinshaw02}, 
S02 \citep{sambruna02}, (13) \citep{siemiginowska03a}.
\label{tab:main1}
\end{table*}

\begin{table*}
\caption{The basic parameters of the (08-16 h) sources considered in the XJET sample (see Section~\ref{sec:sample}).}
\begin{tabular}{|lccccccccccc|}
\hline
Name  & Class &   RA    &   DEC   & z  &  $D_{\rm L}$  &   scale  &    $N_{\rm H,Gal}$     & \chn   & CFS & CFM & Ref.\\
      &       & (J2000) & (J2000) &    &  [Mpc]  & [kpc/''] & [10$^{20}$~cm$^{-2}$] & Obs ID &     &     &     \\
\hline
\noalign{\smallskip}
0827+243        & CDQ   & 08 30 52.086  & +24 10 59.83 & 0.9390 &  6085  & 7.85  & 2.92 & 3047 & 1.17 & 1.04 & 14 \\
4C~+29.30       & FR~I  & 08 40 02.356  & +29 49 02.49 & 0.0640 &   280  & 1.20  & 4.23 & 2135 & 1.27 & 1.05 & S04\\
3C~207          & LDQ   & 08 40 47.589  & +13 12 23.59 & 0.6800 &  4071  & 6.99  & 4.27 & 2130 & 1.27 & 1.05 & 15 \\
3C~212          & LDQ   & 08 58 41.460  & +14 09 44.79 & 1.0490 &  6991  & 8.07  & 3.79 &  434 & 1.23 & 1.04 & 16 \\
PKS~0903-573    & CDQ   & 09 04 53.184  & -57 35 05.85 & 0.6950 &  4183  & 7.06  & 27.0 & 3113 & 3.99 & 1.38 & M05\\
3C~219          & FR~II & 09 21 08.620  & +45 38 57.40 & 0.1744 &   823  & 2.89  & 1.35 &  827 & 1.08 & 1.02 & 17 \\
PKS~0920-397    & CDQ   & 09 22 46.413  & -39 59 34.96 & 0.5910 &  3425  & 6.56  & 17.7 & 5732 & 2.58 & 1.24 & M05\\
3C~227          & FR~II & 09 47 45.140  & +07 25 21.07 & 0.0861 &   383  & 1.57  & 2.11 & 6824 & 1.12 & 1.02 & H07\\
Q~0957+561      & LDQ   & 10 01 20.840  & +55 53 49.56 & 1.4100 & 10129  & 8.46  & 0.93 &  362 & 1.05 & 1.01 & 18 \\
4C~+13.41       & LDQ   & 10 07 26.099  & +12 48 56.19 & 0.2408 &  1183  & 3.72  & 3.56 & 5606 & 1.23 & 1.03 & 19 \\
PKS~1030-357    & CDQ   & 10 33 07.660  & -36 01 56.80 & 1.4550 & 10536  & 8.48  & 5.97 & 5730 & 1.39 & 1.08 & M05\\
1045-188        & CDQ   & 10 48 06.602  & -19 09 35.97 & 0.5950 &  3454  & 6.58  & 3.38 & 9280 & 1.19 & 1.03 & H11\\
PKS~1046-409    & CDQ   & 10 48 38.275  & -41 14 00.15 & 0.6200 &  3633  & 6.71  & 7.73 & 3116 & 1.52 & 1.09 & M05\\
4C~+20.24       & LDQ   & 10 58 17.870  & +19 51 50.86 & 1.1100 &  7504  & 8.17  & 1.69 & 7795 & 1.11 & 1.02 & 20 \\
3C~254          & LDQ   & 11 14 38.725  & +40 37 20.29 & 0.7340 &  4476  & 7.22  & 1.44 & 2209 & 1.09 & 1.02 & D03\\
PKS~1127-145    & CDQ   & 11 30 07.051  & -14 49 27.41 & 1.1800 &  8102  & 8.27  & 3.39 & 5708 & 1.21 & 1.04 & 21 \\
PKS~1136-135    & LDQ   & 11 39 10.701  & -13 50 43.41 & 0.5540 &  3165  & 6.35  & 3.31 & 3973 & 1.20 & 1.04 & S02\\
3C~263          & LDQ   & 11 39 57.023  & +65 47 49.50 & 0.6563 &  3897  & 6.89  & 0.90 & 2126 & 1.05 & 1.01 & H02\\
3C~264          & FR~I  & 11 45 05.008  & +19 36 22.76 & 0.0217 &    92  & 0.43  & 1.83 & 4916 & 1.10 & 1.01 & 22 \\
3C~265          & FR~II & 11 45 28.959  & +31 33 47.12 & 0.8110 &  5067  & 7.49  & 1.79 & 2984 & 1.11 & 1.02 & 23 \\
4C~+49.22       & CDQ   & 11 53 24.407  & +49 31 06.95 & 0.3340 &  1726  & 4.70  & 2.07 & 2139 & 1.13 & 1.02 & S02\\
PKS~1202-262    & CDQ   & 12 05 33.180  & -26 34 04.00 & 0.7890 &  4896  & 7.42  & 7.72 & 4812 & 1.52 & 1.09 & M05\\
3C~270          & FR~I  & 12 19 23.212  & +05 49 31.08 & 0.0074 &    31  & 0.15  & 1.75 &  834 & 1.11 & 1.02 & 24 \\
PG~1222+216     & CDQ   & 12 24 54.458  & +21 22 46.40 & 0.4320 &  2342  & 5.54  & 2.09 & 3049 & 1.13 & 1.02 & J06\\
M~84            & FR~I  & 12 25 03.729  & +12 53 13.20 & 0.0035 &    17  & 0.08  & 2.99 & 5908 & 1.18 & 1.04 & 25 \\
3C~273          & CDQ   & 12 29 06.704  & +02 03 08.63 & 0.1583 &   739  & 2.67  & 1.67 & 4876 & 1.11 & 1.02 & 26 \\
M~87            & FR~I  & 12 30 49.423  & +12 23 28.05 & 0.0043 &    16  & 0.08  & 1.94 & 2707 & 1.11 & 1.02 & 27 \\
PKS~1229-027    & CDQ   & 12 32 00.018  & -02 24 04.10 & 1.0450 &  6957  & 8.07  & 2.13 & 4841 & 1.13 & 1.02 & 28 \\
3C~275.1        & LDQ   & 12 43 57.676  & +16 22 53.40 & 0.5550 &  3172  & 6.36  & 1.77 & 2096 & 1.11 & 1.02 & C03\\
3C~280          & FR~II & 12 56 57.201  & +47 20 19.88 & 0.9960 &  6551  & 7.97  & 1.24 & 2210 & 1.08 & 1.02 & D03\\
1317+520        & CDQ   & 13 19 46.204  & +51 48 05.77 & 1.0600 &  7083  & 8.09  & 1.32 & 3050 & 1.08 & 1.02 & J06\\
3C~281          & LDQ   & 13 07 53.925  & +06 42 13.81 & 0.6020 &  3504  & 6.62  & 2.19 & 1593 & 1.02 & 1.12 & C03\\
Centaurus A     & FR~I  & 13 25 27.616  & -43 01 08.84 & 0.0018 &   3.8  & 0.02  & 8.09 & 8490 & 1.55 & 1.10 & 29 \\
4C~+65.15       & LDQ   & 13 25 29.714  & +65 15 13.16 & 1.6250 & 12096  & 8.51  & 1.91 & 7882 & 1.11 & 1.02 & 30 \\
3C~287.1        & FR~II & 13 32 53.257  & +02 00 45.60 & 0.2156 &  1043  & 3.42  & 1.63 & 9309 & 1.11 & 1.02 & M10\\
Centaurus B     & FR~I  & 13 46 49.036  & -60 24 29.41 & 0.0130 &    55  & 0.26  & 105.0& 3120 &68.10 & 2.89 & M05\\
4C~+19.44       & CDQ   & 13 57 04.438  & +19 19 07.35 & 0.7200 &  4370  & 7.16  & 2.49 & 7302 & 1.15 & 1.03 & S02\\
3C~294          & FR~II & 14 06 44.069  & +34 11 25.54 & 1.7860 & 13608  & 8.50  & 1.36 & 3207 & 1.09 & 1.02 & 31 \\
3C~295          & FR~II & 14 11 20.543  & +52 12 09.88 & 0.4500 &  2460  & 5.67  & 1.34 & 2254 & 1.08 & 1.02 & 32 \\
3C~296          & FR~I  & 14 16 52.953  & +10 48 26.76 & 0.0237 &   101  & 0.47  & 1.92 & 3968 & 1.12 & 1.02 & 33 \\
PKS~1421-490    & LDQ   & 14 24 31.925  & -49 13 54.77 & 0.6620 &  3939  & 6.91  & 15.7 & 5729 & 2.37 & 1.23 & 34 \\ 
3C~303          & FR~I  & 14 43 02.777  & +52 01 37.33 & 0.1410 &   651  & 2.42  & 1.71 & 1623 & 1.10 & 1.02 & 35 \\
1508+572        & CDQ   & 15 10 02.900  & +57 02 44.00 & 4.3000 & 39725  & 6.86  & 1.57 & 2241 & 1.10 & 1.02 & 36 \\
PKS~1510-089    & CDQ   & 15 12 50.536  & -09 05 59.73 & 0.3610 &  1892  & 4.95  & 6.89 & 2141 & 1.46 & 1.08 & S04\\
3C~321          & FR~II & 15 31 43.474  & +24 04 19.02 & 0.0960 &   430  & 1.73  & 3.82 & 3138 & 1.24 & 1.04 & H04\\
\noalign{\smallskip}
\hline
\end{tabular}\\
~\\
Note: `CFS' and `CFM 'correspond to the correction factor for the Galactic absorption in the soft and in the medium bands, respectively
(see Section~\ref{sec:absorp} for more details).
~\\
References: (14) \citep{jorstad04}, S04 \citep{sambruna04}, (15) \citep{brunetti02}, (16) \citep{aldcroft03}, M10 \citep{massaro10b}, 
M05 \citep{marshall05}, (17) \citep{comastri03}, H07 \citep{hardcastle07}, (18) \citep{chartas02}, (19) \citep{miller06}, H11 \citep{hogan11}, 
(20) \citep{schwartz06}, D03 \citep{donahue03}, (21) \citep{siemiginowska02}, S02 \citep{sambruna02}, H02 \citep{hardcastle02b}, (22) \citep{perlman10b},
(23) \citep{bondi04}, (24) \citep{chiaberge03}, J06 \citep{jorstad06}, (25) \citep{finoguenov08}, (26) \citep{marshall01}, (27) \citep{marshall02}, 
(28) \citep{tavecchio07}, C03 \citep{crawford03}, (29) \citep{kraft00}, (30) \citep{miller09}, (31) \citep{fabian03b}, (32) \citep{harris00},
(33) \citep{hardcastle05a}, (34) \citep{gelbord05}, (35) \citep{kataoka03a}, (36) \citep{siemiginowska03b}, H04 \citep{hardcastle04}, S04 \citep{sambruna04}.
\label{tab:main2}
\end{table*}

\begin{table*}
\caption{The basic parameters of the (16-24 h) sources considered in the XJET sample (see Section~\ref{sec:sample}).}
\begin{tabular}{|lccccccccccc|}
\hline
Name  & Class &   RA    &   DEC   & z  &  $D_{\rm L}$  &   scale  &    $N_{\rm H,Gal}$     & \chn   & CFS & CFM & Ref.\\
      &       & (J2000) & (J2000) &    &  [Mpc]  & [kpc/''] & [10$^{20}$~cm$^{-2}$] & Obs ID &     &     &     \\
\hline
\noalign{\smallskip}
3C~327          & FR~II & 16 02 27.370  & +01 57 56.24 & 0.1039 &   468  & 1.86  & 5.92 & 6841 & 1.38 & 1.06 & H07\\ 
4C~+00.58       & FR~I  & 16 06 12.687  & +00 00 27.22 & 0.0590 &   257  & 1.11  & 7.14 &10304 & 1.46 & 1.09 & 37 \\
3C~330          & FR~II & 16 09 34.930  & +65 56 37.78 & 0.5500 &  3137  & 6.33  & 2.66 & 2127 & 1.16 & 1.03 & H02\\
NGC~6251        & FR~I  & 16 32 31.981  & +82 32 16.34 & 0.0249 &   106  & 0.49  & 5.57 & 4130 & 1.35 & 1.07 & 38 \\
1642+690        & CDQ   & 16 42 07.866  & +68 56 39.75 & 0.7510 &  4605  & 7.28  & 5.15 & 2142 & 1.32 & 1.07 & S04\\
3C~345          & CDQ   & 16 42 58.810  & +39 48 37.00 & 0.5940 &  3447  & 6.58  & 1.14 & 2143 & 1.08 & 1.02 & S04\\
3C~346          & FR~I  & 16 43 48.613  & +17 15 49.54 & 0.1610 &   753  & 2.71  & 5.07 & 3129 & 1.32 & 1.06 & 39 \\
3C~349          & FR~II & 16 59 28.893  & +47 02 55.04 & 0.2050 &   986  & 3.29  & 1.88 & 9316 & 1.11 & 1.02 & M10\\
3C~351          & LDQ   & 17 04 41.417  & +60 44 30.65 & 0.3720 &  1960  & 5.05  & 2.02 & 2128 & 1.12 & 1.02 & H02\\
3C~353          & FR~II & 17 20 28.168  & -00 58 46.52 & 0.0304 &   130  & 0.59  & 9.37 & 7886 & 1.67 & 1.12 & 40 \\
4C~+62.29       & LDQ   & 17 46 14.033  & +62 26 54.77 & 3.8890 & 35233  & 7.15  & 3.09 & 4158 & 1.19 & 1.05 & 41 \\
1800+440        & CDQ   & 18 01 32.315  & +44 04 21.83 & 0.6630 &  3946  & 6.92  & 3.14 & 9286 & 1.18 & 1.03 & H11\\
3C~371          & BL   & 18 06 50.671  & +69 49 28.06 & 0.0510 &   221  & 0.97  & 4.16 & 2959 & 1.26 & 1.05 & 42 \\
3C~380          & CDQ   & 18 29 31.781  & +48 44 46.17 & 0.6920 &  4160  & 7.05  & 5.78 & 3124 & 1.37 & 1.08 & M05\\
3C~390.3        & FR~II & 18 42 08.981  & +79 46 17.21 & 0.0561 &   244  & 1.06  & 3.47 &  830 & 1.21 & 1.05 & H07\\
1849+670        & CDQ   & 18 49 16.072  & +67 05 41.68 & 0.6570 &  3902  & 6.89  & 5.00 & 9291 & 1.31 & 1.05 & H11\\
4C~+73.18       & CDQ   & 19 27 48.495  & +73 58 01.57 & 0.3020 &  1535  & 4.39  & 7.15 & 2145 & 1.47 & 1.09 & S04\\
3C~403          & FR~II & 19 52 15.796  & +02 30 24.39 & 0.0590 &   257  & 1.11  & 12.2 & 2968 & 1.96 & 1.17 & 43 \\
Cygnus A        & FR~II & 19 59 28.358  & +40 44 02.09 & 0.0560 &   244  & 1.06  & 27.2 &  360 & 3.99 & 1.37 & 44 \\
2007+777        & BL   & 20 05 30.965  & +77 52 43.21 & 0.3420 &  1775  & 4.78  & 8.35 & 5709 & 1.56 & 1.10 & 45 \\
PKS~2101-490    & CDQ   & 21 05 01.158  & -48 48 46.52 & 0.0630 &   275  & 1.18  & 2.91 & 5731 & 1.18 & 1.04 & M05\\
PKS~2153-69     & FR~II & 21 57 06.035  & -69 41 24.09 & 0.0283 &   121  & 0.55  & 2.64 & 1627 & 1.16 & 1.04 & 46 \\
2155-152        & CDQ   & 21 58 06.282  & -15 01 09.32 & 0.6720 &  4012  & 6.96  & 3.59 & 9284 & 1.21 & 1.03 & H11\\
PKS~2201+044    & CDQ   & 22 04 17.630  & +04 40 02.00 & 0.0270 &   115  & 0.53  & 4.35 & 2960 & 1.26 & 1.05 & 47 \\ 
2209+080        & CDQ   & 22 12 01.599  & +08 19 16.44 & 0.4850 &  2692  & 5.92  & 5.58 & 3051 & 1.35 & 1.07 & J06\\
2216-038        & CDQ   & 22 18 52.038  & -03 35 36.89 & 0.9010 &  5778  & 7.75  & 5.41 & 9285 & 1.34 & 1.06 & H11\\
3C~445          & FR~II & 22 23 49.548  & -02 06 13.22 & 0.0562 &   244  & 1.06  & 4.49 & 7869 & 1.26 & 1.04 & 48 \\ 
3C~452          & FR~II & 22 45 48.771  & +39 41 15.86 & 0.0811 &   359  & 1.49  & 9.64 & 2195 & 1.68 & 1.12 & H04\\
3C~454.3        & CDQ   & 22 53 57.748  & +16 08 53.56 & 0.8590 &  5444  & 7.64  & 6.63 & 4843 & 1.44 & 1.08 & M05\\ 
3C~465          & FR~I  & 23 38 29.383  & +27 01 53.27 & 0.0293 &   125  & 0.57  & 4.82 & 4816 & 1.30 & 1.05 & 49 \\
\noalign{\smallskip}
\hline
\end{tabular}\\
~\\
Note: `CFS' and `CFM 'correspond to the correction factor for the Galactic absorption in the soft and in the medium bands, respectively
(see Section~\ref{sec:absorp} for more details).
~\\
References: H07 \citep{hardcastle07}, (37) \citep{hodgeskluck04}, H02 \citep{hardcastle02b}, (38) \citep{evans05}, S04 \citep{sambruna04},
M10 \citep{massaro10b}, (39) \citep{worrall05}, (40) \citep{kataoka08}, (41) \citep{cheung06}, H11 \citep{hogan11}, (42) \citep{pesce01},
M05 \citep{marshall05}, H07 \citep{hardcastle07}, (43) \citep{kraft05}, (44) \citep{wilson00}, (45) \citep{sambruna08}, (46) \citep{ly05},
(47) \citep{sambruna07}, (48) \citep{perlman10a}, H04 \citep{hardcastle04}, (49) \citep{hardcastle05b}.
\label{tab:main3}
\end{table*}

\section{The XJET Sample}\label{sec:sample} 
The sample considered for our investigation consists
of 106 radio sources with a published \chn~
X-ray detection of a radio knot and/or hotspot, for a total of
236 components.  The selected sample contains: 22 FR\,I RGs, 29 FR\,II
RGs, 3 BL Lacs, and 52 QSRs (35 CDQs and 17 LDQs).  
In AGN unification schemes, QSRs and BL Lacs are broadly understood as 
FR\,IIs and FR\,Is, respectively, observed at smaller angles 
with respect to the line of sight \citep[e.g.,][]{urry95,landt04}.

The 236 components consist of 41 hotspots in FR\,II RGs and 21 in
QSRs; 58 knots in FR\,I RGs, 22 in FR\,II RGs, 68 in CDQs, 19 in LDQs,
and 7 in the BL Lacs.  All these knots and hotspots have been
classified on the basis of the radio morphology of their parent
source, adopting the definition suggested by Leahy et al. (1997) for
the hotspots, i.e., brightness peaks which are neither the core nor a part of the jet, 
usually lying where the jet terminates, and considering
all other discrete brightness enhancements as jet knots.  Here, we did not investigate
the extended radio/X-ray emission arising from lobes -- the typically
roughly symmetrical kpc-scale double radio structures
lying on both sides of the host RG and/or QSR.
The identification of BL Lacs have been adopted accordingly to the ROMA 
BZCAT classification scheme\footnote{\underline{http://www.asdc.asi.it/bzcat/}} \citep{massaro09a}.
Finally, we note that due to the small number of BL Lacs in our sample, and their limited number of components 
(i.e., only 7 out of 236), we did not consider the comparison between them class and the RGs or the QSRs.

\begin{figure}
\includegraphics[width=\columnwidth,angle=0]{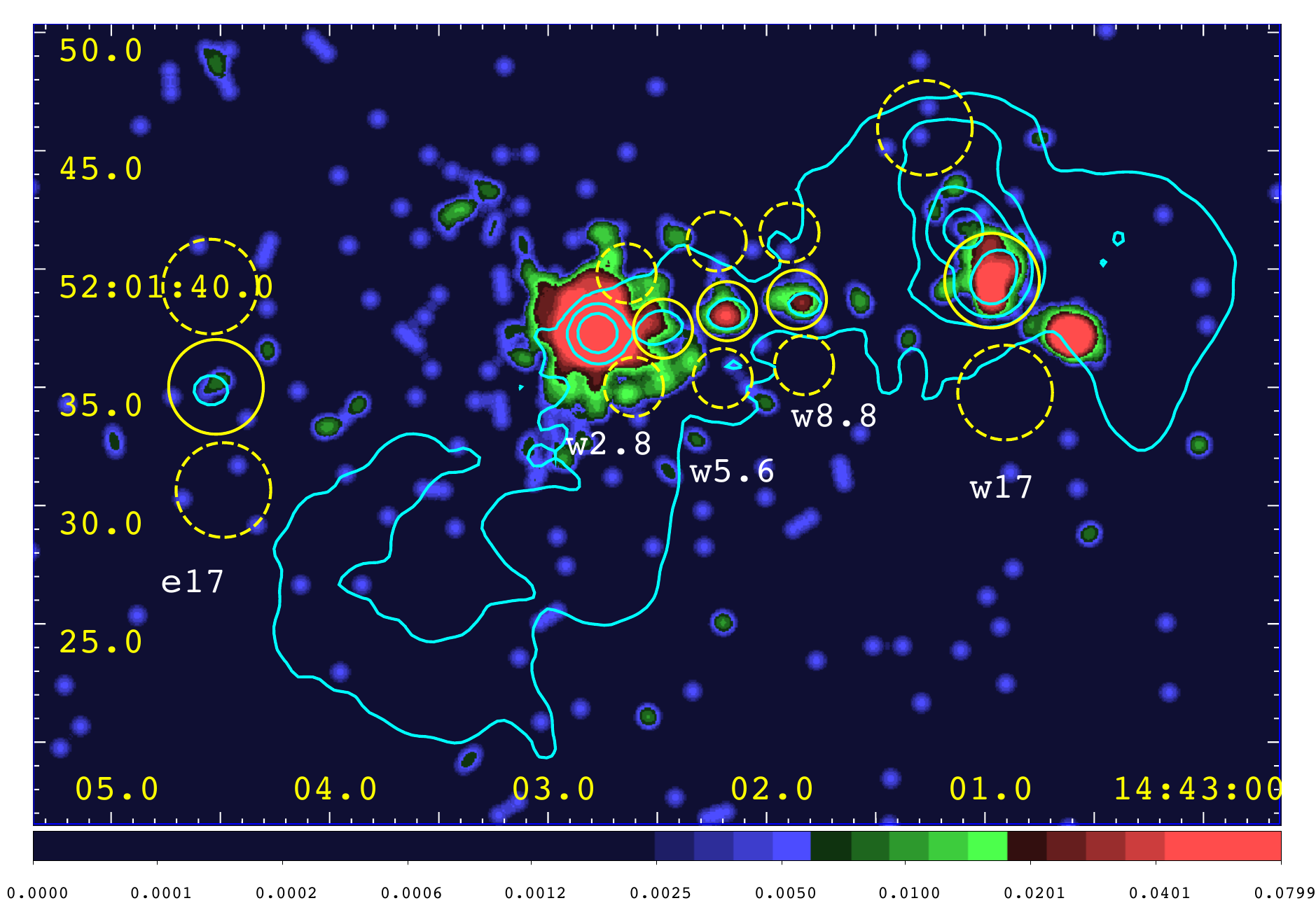}
\caption{An X-ray image of 3C303 with radio contours (1.4 GHz) overlayed.  The X-ray
image is for the band 0.5-7 keV and has been smoothed with a Gaussian
of FWHM=0.87$^{\prime\prime}$.  Radio contours start at 2 mJy/beam and
increase by factors of 4.  Photometric regions are shown in yellow,
with background regions dashed.  The labels are based on the cardinal
direction from the nucleus and the distance in arcsec from the
nucleus.  Note the northern background circle for w17 is moved further
north than usual so as to avoid the secondary hotspot.  When a region
is close to a bright nucleus, we deploy the background circles so as
to lie at the same distance from the nucleus as the 'on' region
(e.g. w2.8).  The logarithmic scale of the X-ray brightness runs from 0.002 (black) to
0.08 (red) but the peak intensity is 6.29 counts per pixel.  The pixel
size in this image is 0.0615$^{\prime\prime}$.
Coordinates are in J2000.0 equinox.}
\label{fig:303regions_rgb}
\end{figure}

The following conventions for the knots 
has been adopted for all the figures:
\begin{itemize}
\item{FR\,I RGs = `k1' (filled black squares)}
\item{FR\,II RGs = `k2' (filled red squares)}
\item{CDQs = `kqc' (filled green squares)}
\item{LDQs = `kql' (filled blue squares)}
\end{itemize}
while we label hotspots as follows:
\begin{itemize}
\item{FR\,II RGs = `hs2' (open red circles)}
\item{QSRs = `hsq' (open blue circles)}
\end{itemize}
We also introduce a nomenclature for indicating each component in every source. 
The name of each knot or hotspot is a combination of one letter
indicating the orientation of the radio structure and one number indicating distance from the core in arcsec.
The case of 3C\,303 is shown as an example in Figure~\ref{fig:303regions_rgb}.

\begin{figure}
\includegraphics[width=\columnwidth,angle=0]{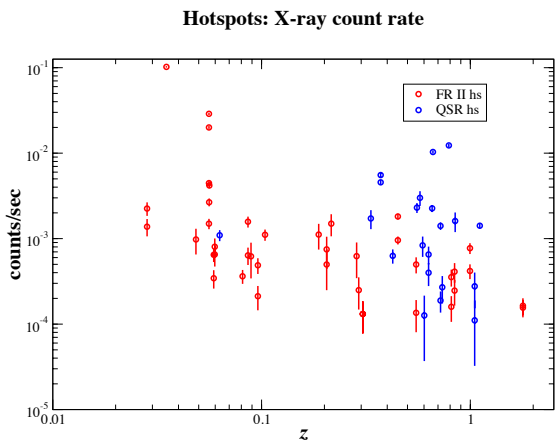}
\caption{The X-ray count rate of hotspots in both RGs and QSRs with respect to the redshift.}
\label{fig:hs_cts}
\end{figure}

\begin{figure}
\includegraphics[width=\columnwidth,angle=0]{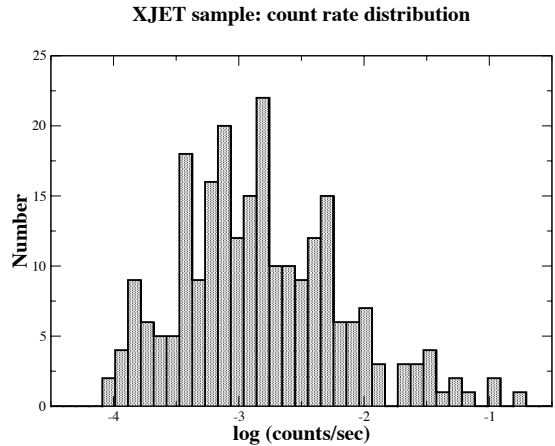}
\caption{The count rate distribution for whole sample of XJET components detected in the X-rays.}
\label{fig:cts}
\end{figure}

All the components identified in our analysis have been previously discovered as reported in the referenced
papers (see Table \ref{tab:main1}, \ref{tab:main2} and \ref{tab:main3}) with
four exceptions, namely: n46.4 in 3C\,109, n3.5, n5.6 in M\,84 and w6.0 in 3C\,280, that we detected 
because in this work we used \chn~observation with longer exposure than the previous ones.
\begin{figure}
\includegraphics[width=\columnwidth,angle=0]{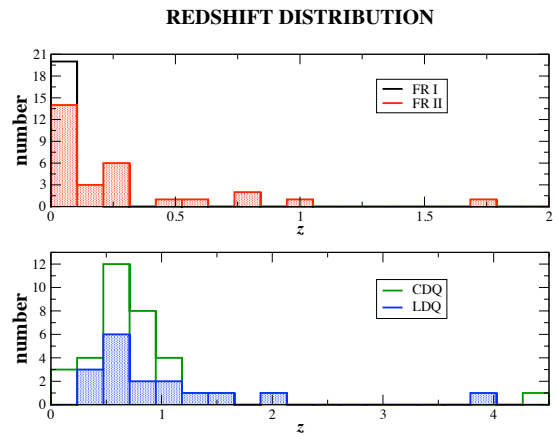}
\caption{The redshift distributions for the RGs (upper panel) and QSRs (lower panel)
in the XJET selected sample.}
\label{fig:redshift}
\end{figure}
In Table \ref{tab:main1}, \ref{tab:main2} and \ref{tab:main3} we summarize the whole XJET sample and the main
properties of the sources. We report the name, the source class (e.g., FR\,I, LDQ
etc.), the radio coordinates adopted for the registration of the X-ray
image (see Section~\ref{sec:xray} for details), the redshift ($z$), and the
corresponding scale (kpc/$^{\prime\prime}$) and luminosity distance, $D_{\rm L}$ \citep{wright06}
\footnote{For the 3 nearby objects (Centaurus A, M\,87, M\,84), distances from the literature 
are adopted in favor of those computed based on their redshifts.}, the
Galactic absorption \citep{kalberla05}, the \chn~observation
ID in which the radio component is detected in the X-rays, and the
correction factors computed to correct the X-ray fluxes for the
Galactic absorption in the soft and the medium bands (see
Section~\ref{sec:absorp}).
In Figure~\ref{fig:redshift}, we show the redshift distributions of RGs and QSRs 
in our selected sample.
\begin{figure}
\includegraphics[width=\columnwidth,angle=0]{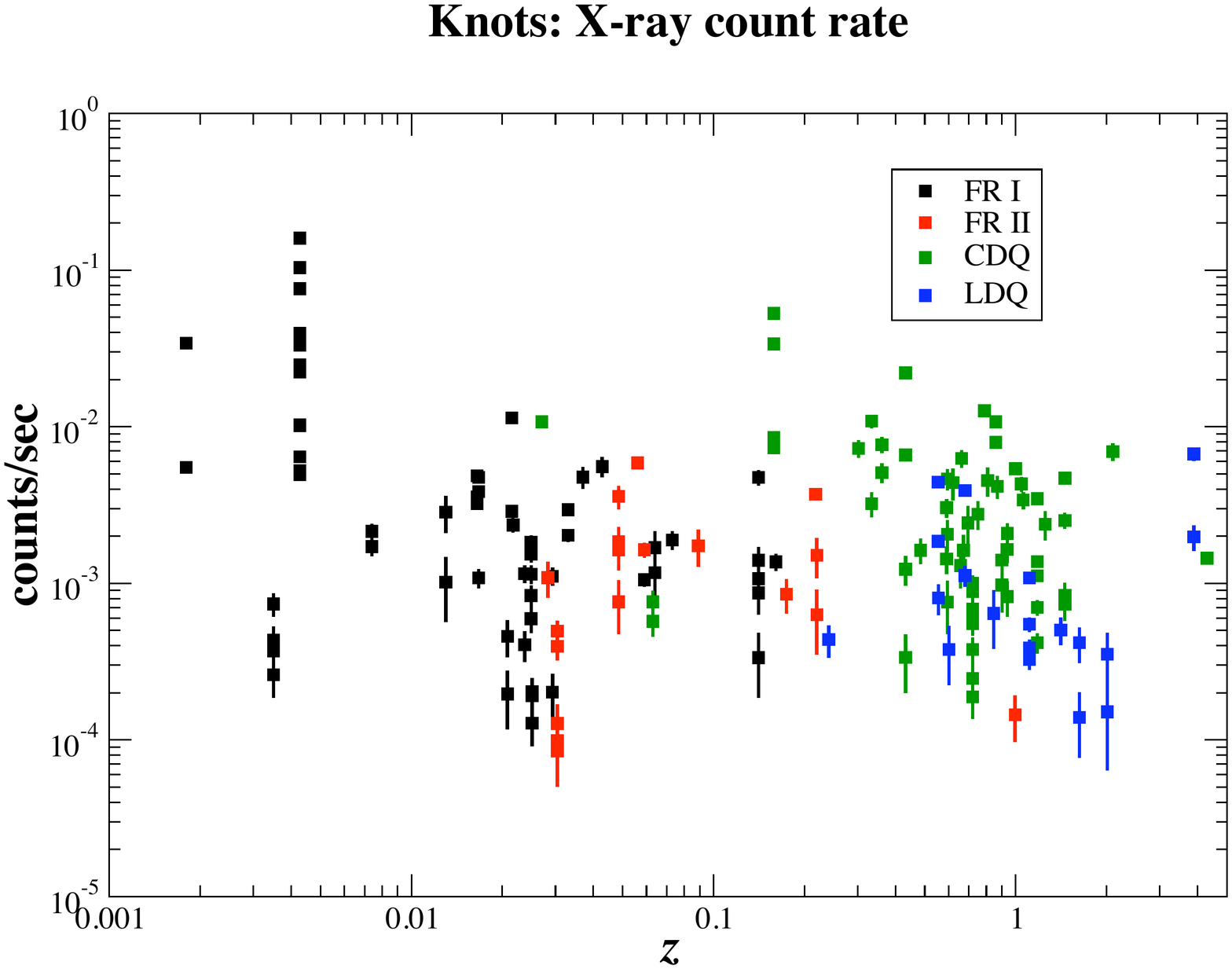}
\caption{The X-ray count rate of knots in RGs and QSRs with respect to the redshift.}
\label{fig:ks_cts}
\end{figure}
We show the plot of the X-ray count rate of each component vs. the redshift
for both knots and hotspots in Figure~\ref{fig:ks_cts} and Figure~\ref{fig:hs_cts}, respectively
and the distribution of the X-ray count rate for the whole XJET sample in Figure~\ref{fig:cts}.
In detail, the distribution of the count rate for the whole XJET sample has an average count rate 
of 5.5$\times$10$^{-4}$ $\pm$ 7.4$\times$10$^{-6}$ s$^{-1}$ with a variance 3.9$\times$10$^{-6}$ s$^{-1}$,
corresponding to about 0.5 counts in 1 ksec similar to the peak of the X-ray count rate distribution, 
that is $\sim$ 1 count per 1 ks of exposure (see Figure~\ref{fig:cts}).
Note that the individual X-ray count rate distributions for the different components are reported in
Figure~\ref{fig:counts} of \app. 
The majority ($\sim$3/4) of the exposures are $\sim$5-50 ksec in length with longer exposures up to 100-150 ksec for a handful of objects.
The apparent lack of dependence of count rate on redshift is most
likely caused by the limited sensitivity of most Chandra observations.
For example, there are presumably large numbers of FRI jet knots at
z>0.1 with count rates less than 10${-4}$ counts/sec.

\section{X-ray Data Reduction and Analysis}\label{sec:xray} 
The \chn~data reduction has been performed following the standard
reduction procedure described in the \chn~ Interactive Analysis of
Observations (CIAO) threads
\footnote{\underline{http://cxc.harvard.edu/ciao/guides/index.html}}, 
using CIAO v4.2 and the Chandra Calibration
Database (CALDB) version 4.2.2. 
All the data reduction and the data analysis procedures 
described in the following sections were adopted
previously in our studies of 3C~305 \citep{massaro09b} and 3C~17 \citep{massaro09c},
and for the ongoing \chn~ 3C snapshot survey \citep{massaro10b}.

Specifically, level 2 event files were generated using the ${\tt acis\_process\_events}$
task, after removing the hot pixels with ${\tt acis\_run\_hotpix}$.  Events
were filtered for grades 0, 2, 3, 4, 6 and we removed pixel randomization.

Light curves for every dataset were extracted and checked for high
background intervals that have been excluded if the background count
rate was found high over the whole back illuminated 
chip in the 7 -- 10 keV energy range accordingly to the CIAO 
threads\footnote{\underline{http://cxc.harvard.edu/ciao/threads/filter\_ltcrv/}}.

Astrometric registration was achieved by changing the appropriate
keywords (RA\_NOM, DEC\_NOM, TCRVL11, TCRVL12) 
in the fits header so as to align the nuclear X-ray position with that of the radio 
(i.e., the world coordinate system (WCS) of the X-ray image 
was shifted so it would be the same as the radio image).

\subsection{Fluxmaps}\label{sec:fluxmaps} 
Following the standard reduction, we created fluxmaps in 3 defined bands 
(soft, medium, and hard, in the ranges 0.5 -- 1, 1 -- 2, and 2 -- 7 keV, respectively) 
by dividing the data with
monochromatic exposure maps (with nominal energies of soft=0.8 keV,
medium=1.4 keV, and hard=4 keV).  The exposure maps and the flux maps
were regridded to a common pixel size which was usually 1/4 the size
of a native ACIS pixel (native=0.492$^{\prime\prime}$).  For sources
of large angular extent we used 1/2 or no regridding.  To obtain maps
with brightness units of ergs~cm$^{-2}$~s$^{-1}$~pixel$^{-1}$, we
multiplied each event by the nominal energy of its respective band.

\subsection{Photometry}\label{sec:photom} 
X-ray detected components were identified via visual inspection of the fluxmaps, 
referring to the detections published in the original references.
To measure observed X-ray fluxes, we construct appropriate regions (usually
circular) as well as two adjacent background regions of the same size 
(see e.g., Figure~\ref{fig:303regions_rgb}). 
The shape and the sizes of regions selected for our flux measurements
are reported in \app~together with the radio and the X-ray fluxes
for each component.

The net X-ray flux in each region for each band were measured using 
{\tt funtools}\footnote{\underline{http://www.cfa.harvard.edu/$\sim$john/funtools}}.
After applying a small correction which is the ratio of the mean
energy of the events within the 'on' aperture to the nominal energy
applied earlier to all events, a 1$\sigma$ error is assigned based
on the usual Poisson statistic $\sqrt{\rm number-of-counts}$ in the on and
background regions.

\subsection{Absorption corrections}\label{sec:absorp} 
To estimate the factors required to correct the observed X-ray fluxes for
Galactic absorption we adopted the following method.  We chose an
arbitrary value of the intrinsic flux ($F_{\rm int}$) in each band
(i.e., soft, medium, hard) and we computed the ratio between 
$F_{\rm int}$ and the absorbed flux $F_{\rm abs}$ for different values of
$N_{\rm H,Gal}$ using the {\it WEBPIMMS} tool\footnote{\underline{http://heasarc.nasa.gov/Tools/w3pimms.html}}.  
We considered $N_{\rm H,Gal}$ in the range between 10$^{20}$ cm$^{-2}$ and 10$^{22}$
cm$^{-2}$, as for the sources in our sample (see Table \ref{tab:main1},
Table \ref{tab:main2} and Tabel \ref{tab:main3}).  
We repeated this procedure for three representative values of the X-ray spectral
index, $\alpha_{\rm X}$~=~0.5,~1.0 and 1.5.
\begin{figure}
\includegraphics[width=\columnwidth,angle=0]{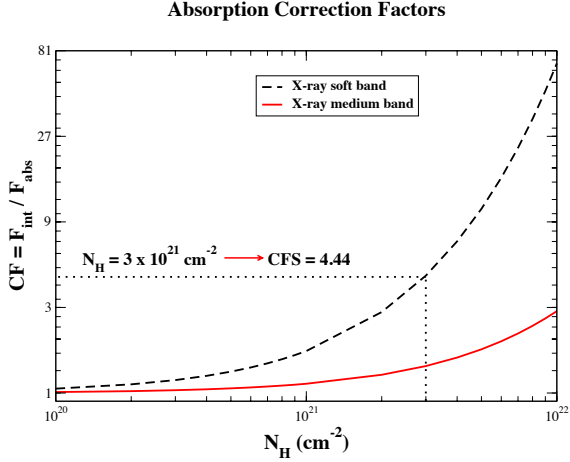}
\caption{The correction factor, $CF$, estimated for the soft and the medium band, assuming an
X-ray spectral index $\alpha_{\rm X} = 1$, as a function of different values of the Galactic column density $N_{\rm H}$.
Our method used to derive the correction factor for the soft band is illustrated (see Section~
\ref{sec:absorp} for details).}
\label{fig:absorption}
\end{figure}

In Figure~\ref{fig:absorption}, we show the plot of the correction factor, $F_{\rm int}/F_{\rm abs}$ vs. $N_{\rm H,Gal}$ for the
soft (0.5 -- 1 keV) and the medium (1.0 -- 2.0 keV) bands evaluated assuming $\alpha_{\rm X} = 1$.  
For each source in Table \ref{tab:main1},
Table \ref{tab:main2} and in Table \ref{tab:main3}, this ratio (again, for $\alpha_{\rm X} = 1$) 
appears in the column `CFS' and `CFM' for the X-ray soft and the medium band, respectively.
For the range of $N_{\rm H,Gal}$ considered, we found $<$ 7\% difference in the corrections assuming $\alpha_{\rm X}$ = 0.5 and 1.5.
In addition, for our choice of the hard band (i.e., 2 -- 7 keV), we found that the
correction factor for $N_{\rm H,Gal}\leq\,$10$^{22}$ cm$^{-2}$ is always less
than 1\% different from unity for all three choices of the
spectral index ($\alpha_{\rm X}$ = 0.5, 1.0, and 1.5).
Thus the only significant corrections occur for the soft and the medium bands.

The X-ray fluxes reported (i.e., absorbed) in \app~are those measured for each
component while those corrected for Galactic absorption were used in calculating the 
X-ray to radio flux ratios (see Section~\ref{sec:ratios}) and the 
X-ray hardness ratios (see section~\ref{sec:hardness}).

\section{Flux ratios}\label{sec:ratios} 
We developed one initial {\it parameter} to begin the investigation
of the selected knots and hotspots.  Using radio maps available in the
public archives
(e.g., NVAS\footnote{\underline{http://www.aoc.nrao.edu/$\sim$vlbacald/}},
NED\footnote{\underline{http://nedwww.ipac.caltech.edu/}}, and 
MERLIN\footnote{\underline{http://www.jb.man.ac.uk/cgi-bin/merlin\_retrieve.pl}}) 
or kindly provided by our colleagues, we measured the radio fluxes and computed 
$\nu_{\rm R}~\times~S_{\nu}(\nu_{\rm R})$
where the radio frequency of the maps used were, $\nu_{\rm R}$ = 1.4, 5, or 8 GHz.  We
then calculated the X-ray-to-radio flux ratio ($\rho$)
\begin{equation}
\rho~=~\frac{F_{0.5-7 keV}}{\nu_R\,S(\nu_R)}~~,
\label{eq:ratio}
\end{equation}
where the X-ray fluxes (0.5 -- 7 keV) are the totals from the three band fluxmaps 
corrected for Galactic absorption (see Section \ref{sec:absorp} for details).  
Observationally, this flux ratio is, to first order, independent of the redshift and
it also is the same as the luminosity ratio, that hereinafter will be simply referred to as the $ratio$.
All the main results derived from our analysis are reported in the Section~\ref{sec:results}.

We also note that for the X-ray to radio flux ratios, we utilized the integrated 0.5--7 keV 
X-ray fluxes (i.e., $F_{0.5-7 keV}$) whereas in other works, the monochromatic ($\nu\,S_{\nu}$) 
fluxes at both X-rays and radio are used \citep[e.g.][]{cheung04,kataoka05,marshall11}.
For the range of considered X-ray spectral indices, $\alpha_{\rm X} = 0.5 - 1.5$, 
these monochromatic X-ray fluxes scale as: $\nu\,S_{\nu}$ (1 keV) = (0.26 -- 0.48) $\times$ $F_{0.5-7 keV}$.
Consequently the intrinsic dispersion in $\alpha_{\rm X}$ cannot be responsible for the observed dispersion of the ratios $\rho$
(see Section \ref{sec:results}).

\begin{figure}
\includegraphics[width=\columnwidth,angle=0]{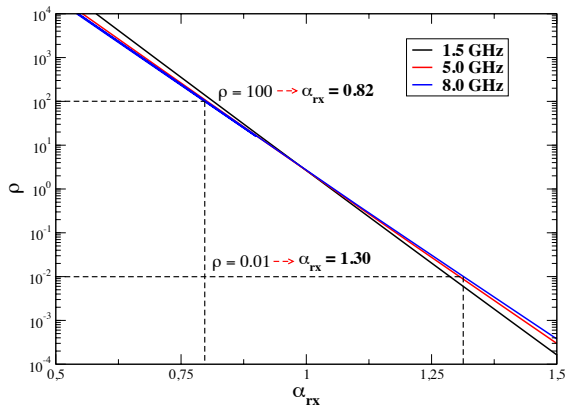}
\caption{The relation between the luminosity ratios, $\rho$ (see Equation~\ref{eq:ratio})
and the conventional radio-to-X-ray spectral index $\alpha_{rx}$
estimated for three different radio frequencies (see Section~\ref{sec:ratios} for
more details).}
\label{fig:slope}
\end{figure}

Finally, we show the relation between our ratios $\rho$ and the conventional 
radio-to-X-ray spectral index $\alpha_{rx}$. 
As is well known, a slope of -1 in a power-law radiation spectrum (i.e.
$S_{\nu}=k~\nu^{-1}$, with k the normalization) leads to equal energy
per decade.  However, since we have defined the ratio of X-ray flux to radio
flux to be the measured flux in the 0.5 to 7 keV band divided by $\nu_R\,S(\nu_R)$
(i.e. an approximation to a radio flux), a value of $\rho$ = 1 does not
correspond to $\alpha_{rx}$=1.  The actual relation is:
\begin{equation}
\rho~=~\nu_R^{(\alpha_{rx}-1)}\cdot\int_{\nu_1}^{\nu_2}\nu^{-\alpha_{rx}}\,d\nu~~,
\label{eq:alfa}
\end{equation}
where $\nu_r$ is the observed radio frequency (Hz) and the integral is taken
over the band from $\nu_1=1.21\times 10^{17}$ to $\nu_2=1.69\times 10^{18}$ Hz.
In Figure~\ref{fig:slope} we show the above relationship (i.e., Equation~\ref{eq:alfa}) 
for the three radio bands we have used in this work, namely: 1.5, 5, and 8 GHz.

Because most of our measured values of $\rho$ fall between 0.01 and 100,
it is fairly obvious that our total range can be described by a range
of $\alpha_{rx}$ from 0.7 to 1.3.  Since we have not included radio
knots which are not detected at X-rays, there will be larger values of
$\alpha_{rx}$ than 1.3.  However, jet components with $\alpha_{rx}$
significantly less than 0.7 have yet to be found. 

Finally, we note that we did not apply any K-correction to our fluxes in order to estimate the 
luminosities in both the radio and the X-ray bands.
Given the redshift distribution of our XJET sample, with only one exception, the highest value of
$z$ is $\sim$ 2; consequently, for values of the spectral slope $\alpha$ between 0.5 and 1.5, the 
K-correction ranges between 0.58 and 1.73 and it does not affect our results, because it cannot be responsible 
for the large scatter of the observed luminosities and/or ratios (see Section~\ref{sec:results}). 
In particular, for the value $\alpha$ = 1, used for the X-ray band to estimate the 
absorption correction factors (see Section~\ref{sec:absorp}), no K-correction is needed
\citep[e.g.,][]{hogg00,hogg02}.

\section{X-ray Hardness ratios}\label{sec:hardness} 
We also evaluated the X-ray hardness ratios (HRs) using the
hard ($H$), medium ($M$), and soft ($S$) X-ray fluxes corrected for the Galactic absorption,
with the following relations:
\begin{equation}
HR_1=\frac{H-M}{H+M},\,
HR_2=\frac{H-S}{H+S},\,
HR_3=\frac{M-S}{M+S}.
\label{eq:hardness}
\end{equation}

\begin{figure}
\includegraphics[width=\columnwidth,angle=0]{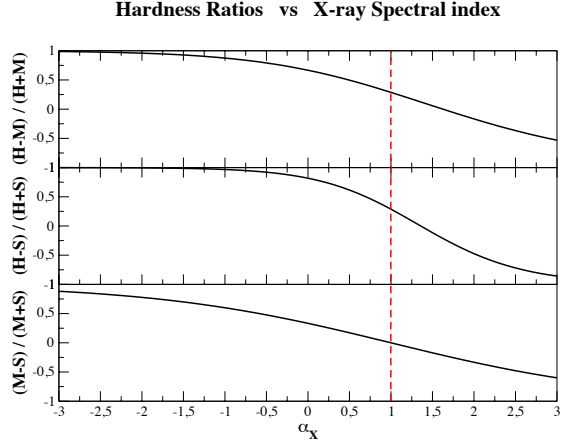}
\caption{The relation between the hardness ratios, $HR_1$ (upper panel),
$HR_2$ (middle panel) and $HR_3$ (lower panel),
defined by Equation (\ref{eq:hardness}) and the X-ray spectral slope $\alpha_{\rm X}$.}
\label{fig:hardness}
\end{figure}

\begin{figure}
\includegraphics[width=\columnwidth,angle=0]{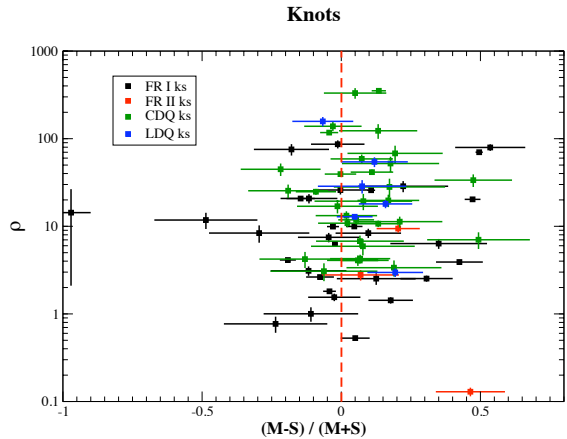}
\caption{The ratios $vs$ the hardness ratios $HR_3$ (see Equation~\ref{eq:hardness}) 
for knots. We only considered those components for which the error on the $HR_3$
is lower than 0.2. The red dashed line corresponds to $\alpha_{\rm X}$ = 1 
(see Figure~\ref{fig:hardness} for more details).}
\label{fig:ks_hr}
\end{figure}

\begin{figure}
\includegraphics[width=\columnwidth,angle=0]{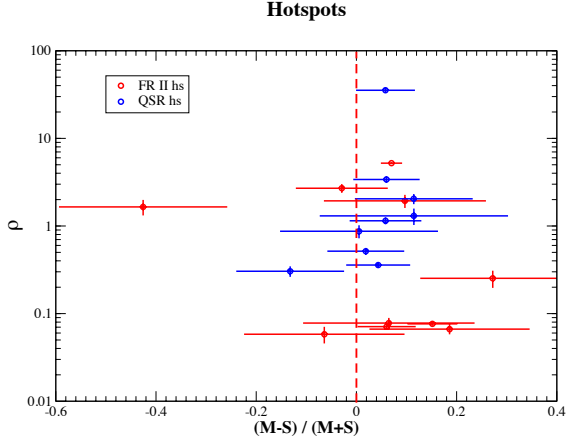}
\caption{The ratios $vs$ the hardness ratios $HR_3$ (see Equation~\ref{eq:hardness}) 
for hotspots. We only considered those components for which the error on the $HR_3$
is lower than 0.2. The red dashed line corresponds to $\alpha_{\rm X}$ = 1 
(see Figure~\ref{fig:hardness} for more details).}
\label{fig:hs_hr}
\end{figure}

\noindent We note that because HRs are defined using X-ray fluxes, there is a
relationship between the X-ray spectral index ($\alpha_{\rm X}$) and HRs for any given selection of
energy bands (e.g., Figure \ref{fig:hardness}). In this way, the 
X-ray HRs serve as a reasonable surrogate for
the $\alpha_{\rm X}$ of these extended components
because the short exposures of the available X-ray observations
combined with their low intrinsic flux (i.e., $\sim$10$^{-15}$
erg~s$^{-1}$~cm$^{-2}$) do not allow us to estimate values
of $\alpha_{\rm X}$ with the usual spectral tools (e.g. Sherpa, XSPEC etc.) for the whole sample.  
However, as the current HR values do not provide robust constraints on $\alpha_{\rm X}$, 
this conclusion should be revisited over a wide sample of source when deep X-ray 
observations will be available that will further limit the error on the HRs,
also involving the Bayesian analysis \citep{park06}.
In Figure~\ref{fig:ks_hr} and Figure~\ref{fig:hs_hr} we report the ratios with respect to the hardness ratios $HR_3$ only 
for those sources for which the error on $HR_3$ is lower than 0.2, no significant trend is evident.

We compare all the values of $\rho$ and HRs dividing the knots and the
hotspots of our sources in 6 categories as defined in Section \ref{sec:sample}. 
We did not find any significant difference in the distributions of the HRs between RGs and QSRs.

Finally, we note that given the small number of BL Lacs in our sample, and the limited number of their components 
(i.e., only 7 out of 236), we did not consider the comparison between this class of object and the RGs or the QSRs
not only regarding the HRs but also with respect to the other parameters.

\section{The KS test and the Monte Carlo Simulations}\label{sec:stat} 
To search for possible differences or similarities between knots and
hotspots both in RGs and QSRs, we compared the
distributions of the observed parameters such as the luminosities
(both $L_{\rm R}$ and $L_{\rm X}$), their ratios ($\rho$), and their HRs as defined in
Equation~\ref{eq:hardness}.

To perform our analysis, we adopted a Kolmogorov-Smirnov (KS) test,
measuring the distance $D_{\rm KS}$ between the normalized cumulative
distributions of parameters for two different samples of components 
and estimating the associated probability to test our hypothesis.  
However, because our selected sample is not 
statistically complete, our analysis could be affected by some biases.
Consequently, it is possible to measure a large value of $D_{\rm KS}$ between two
selected cumulative distributions, suggesting that they are different,
simply because of the lack of sources in a particular bin of our
histograms. This could strongly affect our analysis and our results.
To check the significance of results provided by the KS test,
we developed a Monte Carlo method to take into account this
effect and to estimate its relevance.

\begin{figure}
\includegraphics[width=\columnwidth,angle=0]{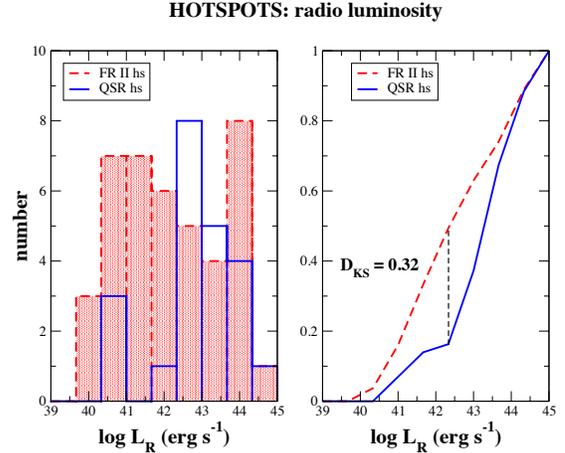}
\caption{Left panel) The $L_{\rm R}$ distributions for the hotspots in the FR\,II RGs (dashed red lines)
and LDQs (blue solid line). Right panel) The normalized cumulative distributions of the radio luminosities
for the hotspots in FR\,IIs and in LDQs.}
\label{fig:Lrad_hs}
\end{figure}

We illustrate this method for the simple case to test if the
distributions of radio luminosities $L_{\rm R}$ in RGs and
QSRs are similar or different (see Figure~\ref{fig:Lrad_hs}).

First, we performed the KS test and we measured the $D_{\rm KS}$ in the
$L_{\rm R}$ normalized cumulative distributions.  Second, we randomly
simulate the two distributions of $L_{\rm R}$ for both cases of hotspots in
RGs and QSRs, with the same number of components (i.e., 41 hs2 and 21 hsq).
We adopted two different shapes for the simulated distributions,  
the log-uniform and the log-normal. The former having simply the same 
maximum and minimum value of the observed distribution
while the latter with the same variance, the same median of
the observed distribution and spanning the same range of
luminosities (see Table~\ref{tab:properties}).  
Then, we measured the $D_{\rm KS,simul}$ variable between the two
simulated distributions.

\begin{figure}
\includegraphics[width=\columnwidth,angle=0]{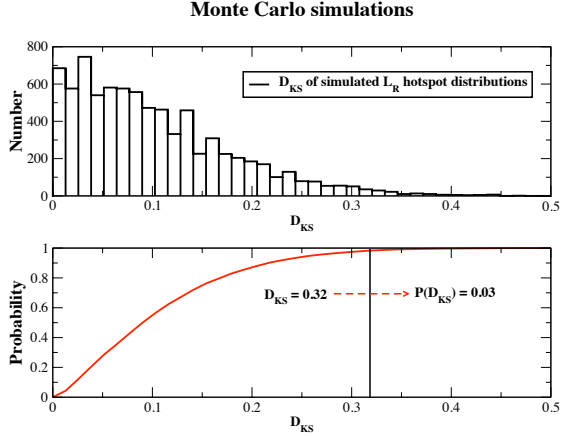}
\caption{Upper panel) The $D_{\rm KS}$ between the two cumulative normalized $L_{\rm R}$
distributions of hotspots in FR\,IIs and LDQs 
simulated via the Monte Carlo method as described in Section \ref{sec:stat}.
Lower panel) The probability to obtain randomly the observed $D_{\rm KS}$
assuming a log-uniform shape for the simulated $L_{\rm R}$ hotspots distributions
(see Section~\ref{sec:stat} for more details).}
\label{fig:Monte Carlo}
\end{figure}

We repeated the simulation 8000 times and we built the distribution of
the $D_{\rm KS,simul}$ distance (see Figure \ref{fig:Monte Carlo}, for the case of the log-uniform distribution).  
Finally, we estimated the probability to obtain, randomly, the observed $D_{\rm KS}$, and this
provides us the level of confidence of our KS test (see Figure \ref{fig:Monte Carlo}, lower panel).

The levels of confidence (i.e., probabilities) derived from Monte Carlo simulations 
performed to generate the two hotspot distributions of $L_{\rm R}$, 
run adopting both the log-uniform and log-normal
function are reported in Table~\ref{tab:prob1} and Table~\ref{tab:prob2} for the two different cases.
Adopting the log-uniform function for the simulated distributions in the case of the ratio comparison,
because these depend only by the maximum and the minimum value of $\rho$,
that is roughly the same for all components does not provide a meaningful and significative check of the KS test.
Thus, we do not report the $P(D_{\rm KS})^u$ for the ratio comparison.
This problem does not occur in the case of the log-normal distribution because it is described by 
more parameters than the log-uniform.

In both Tables~\ref{tab:prob1} and ~\ref{tab:prob2},
Col. (1) and Col. (2) show the two components compared. Col. (3) reports the observed parameter and Col. (4)
the KS variable $D_{\rm KS}$ measured of the observed distributions. 
Col.(5) show the probability of the Monte Carlo simulations
assuming a log-uniform distribution (i.e., $P(D_{\rm KS})^u$) 
while that of a log-normal is reported in Col. (6) (i.e., $P(D_{\rm KS})^g$).

Finally, we note that, unfortunately, due to the short exposures of X-ray observations for the majority of the selected sources,
a statistical analysis of their HRs cannot be performed. 
Consequently, no firm conclusions can be obtained 
from our investigation on the average spectral behavior of different source classes
and no results are reported in the following sections. 
The errors on the single source HRs are too large
to allow a statistically significant comparison between their distributions.

\section{Results}\label{sec:results} 
The average properties (i.e., max, min values, median and variance of $L_{\rm R}$, $L_{\rm X}$ and $\rho$) 
for all the class components are reported in Table~\ref{tab:properties},
while the level of confidences derived from our Monte Carlo simulations are shown in Table~\ref{tab:prob1}
and \ref{tab:prob2} as described in Section~\ref{sec:stat}.
Here, we discuss the main observational results derived from our investigation.

When comparing the radio and the X-ray luminosities possible biases could arise from the 
the different redshift distributions of RGs and QSRs. For example, QSRs could appear more luminous than FR\,Is 
simply because they lie at higher redshift, where the low luminosity RGs would be too faint for detection.
However, this will not affect the comparison of the ratios.
The ratios could be affected by biases of the incompleteness of our source sample,
but this problem have been addressed by using the MonteCarlo simulations (see Section~\ref{sec:stat}).
Finally, we note that to make the redshift distribution more uniform
future Chandra observations of both QSRs and RGs have been proposed.

\begin{table*}
\caption{The average properties of the extended components.}
\begin{tabular}{|lcccccccc|}
\hline
Class          & Num. & log$\,L_{\rm R,min}$    & log$\,L_{\rm R,max}$   & $<log\,L_{\rm R}>$ & log$\,L_{\rm X,min}$   & log$\,L_{\rm X,max}$   & $<log\,L_{\rm X}>$ & $<log\,\rho>$ \\
               &      & erg\,s$^{-1}$       & erg\,s$^{-1}$      & erg\,s$^{-1}$  & erg\,s$^{-1}$      & erg\,s$^{-1}$      & erg\,s$^{-1}$  &               \\
\hline
\noalign{\smallskip}
FR I  ks  (k1) & 58   & 36.9 & 41.7 & 38.82(1.05) & 37.0 & 42.1 & 40.04(0.24) & +1.13(0.26) \\ 
FR II ks  (k2) & 22   & 38.9 & 43.3 & 40.05(1.80) & 39.1 & 42.5 & 41.18(0.81) & +0.58(0.79) \\ 
\noalign{\smallskip}
CDQ   ks  (kqc) & 68  & 39.3 & 44.3 & 41.11(0.97) & 40.4 & 45.2 & 43.20(0.75) & +1.73(0.48) \\ 
LDQ   ks  (kql) & 19  & 40.3 & 44.4 & 42.67(1.28) & 41.4 & 45.2 & 43.64(0.65) & +1.15(0.33) \\ 
\hline
\noalign{\smallskip}
FR II hs  (hs2)& 41   & 39.9 & 44.4 & 42.17(1.69) & 40.0 & 43.2 & 41.99(0.19) & -0.11(0.79) \\ 
QSR   hs  (hsq)& 21   & 40.5 & 43.9 & 42.86(0.99) & 40.6 & 44.3 & 43.70(0.53) & +0.24(0.61) \\ 
\noalign{\smallskip}
\hline
\end{tabular}\\
~\\
Note: for each average parameter we also report the variance of its distribution in parenthesis. 
\label{tab:properties}
\end{table*}

\subsection{Hotspots}\label{sec:hs} 
The components that can be morphologically classified as hotspots are only present in FR IIs 
(hs2) and in QSRs (hsq).
We compared the distributions of their radio and X-ray luminosity (i.e., $L_{\rm R}$, $L_{\rm X}$) 
and also of their ratios (i.e., $\rho$) (see Figure~\ref{fig:rg_hs} in \app).
\begin{figure}
\includegraphics[width=\columnwidth,angle=0]{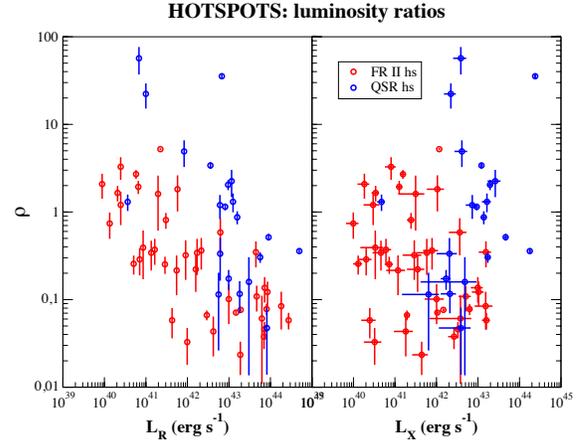}
\caption{The behavior of ratios measured for the hotspots in FR\,IIs (hs2) and in QSRs (hsq) 
with respect to the radio luminosity $L_{\rm R}$ (left panel) and the X-ray luminosity $L_{\rm X}$ (right panel).}
\label{fig:hs_ratios}
\end{figure}

We found that hotspots in FR IIs and in LDQs do not show significant differences 
in the distributions of $L_{\rm R}$ and $L_{\rm X}$,
and they appear to have also similar $\rho$ distribution.
We also note that the hs2 are the only components for which the average 
value of $\rho$ is lower than one (see Table~\ref{tab:properties}).
\begin{figure}
\includegraphics[width=\columnwidth,angle=0]{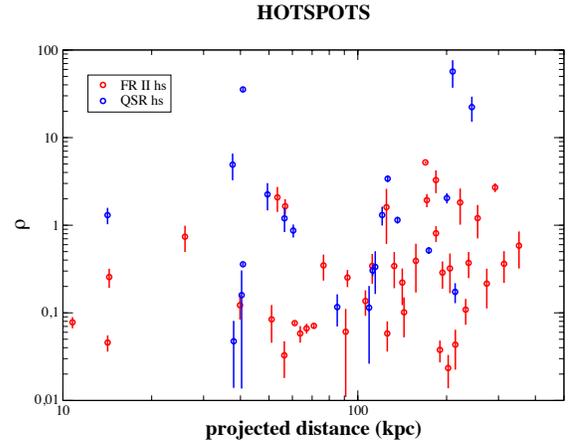}
\caption{The ratios measured for the hotspots in FR\,IIs (hs2) and in QSRs (hsq) as
function of the projected distance from the core in kpc.}
\label{fig:ratios_d_hs}
\end{figure}
In Figure~\ref{fig:hs_ratios} we show the ratios $vs$ both $L_{\rm R}$ and $L_{\rm X}$ for the hotspots.
There is a marginal trend between $\rho$ and $L_{\rm R}$ where hotspots in both RGs and QSRs,
with high values of $\rho$ have typically low values of $L_{\rm R}$, in agreement with the results
found in Hardcastle et al. (2004). However, because the two variables $\rho$ and $L_{\rm R}$
are not independent the estimate of their correlation coefficient will not be statistically meaningful.
In Figure~\ref{fig:ratios_d_hs} we report the $\rho$ values as a function of the projected distance,
where no clear trend has been found.

\subsection{Knots}\label{sec:ks} 
First, we compared knots between RGs (i.e., FR\,Is vs FR\,IIs) 
and between QSR classes (CDQs vs LDQs), then we considered 
also the comparison between RGs and QSRs.
\begin{figure}
\includegraphics[width=\columnwidth,angle=0]{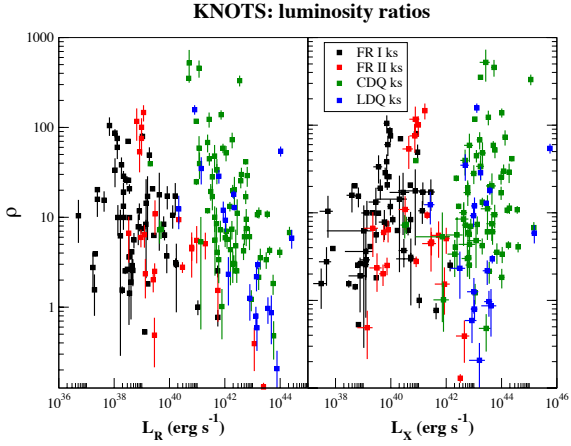}
\caption{The behavior of ratios measured for the knots in RGs and in QSRs 
with respect to the radio luminosity $L_{\rm R}$ (left panel) and the X-ray luminosity $L_{\rm X}$ (right panel).}
\label{fig:ks_ratios}
\end{figure}

1.) {\it k1 vs k2.} Knots in FR\,Is are less powerful in $L_{\rm R}$ than knots in FR\,IIs (see Figure~\ref{fig:rg_ks} in \app).
There is no significant difference in $L_{\rm X}$ and $\rho$ between k1 and k2 components 
even if the probability given from our Monte Carlo method is 
not very high, as reported in Table~\ref{tab:prob1}.
However, the FR\,II $L_{\rm R}$ distribution is not well sampled as is the case for FR\,Is, that has about 3 times the number 
of components with respect to the former, requiring a deeper investigation to confirm these results.

2.) {\it kqc vs kql.} The situation of QSR knots is very different with respect to that of RGs
(see Figure~\ref{fig:qsr_ks} in \app).
We did not find any difference between knots in CDQs and LDQs. All their distributions of
$L_{\rm R}$, $L_{\rm X}$ and $\rho$ are identical, within the level of confidence provided by the 
Monte Carlo simulations (see Table~\ref{tab:prob1}).

3.) {\it k1 vs kq.} Comparing the FR\,I with the CDQ knots we found that: k1 components are 
systematically different in $L_{\rm R}$ and $L_{\rm X}$ distributions than kqc components
with a high level of confidence (see Table~\ref{tab:prob1}). CDQ knots appear to be systematically brighter than FR\,I knots
in both the radio and the X-ray band.
On the other hand, their ratio distributions are similar
(see Figure~\ref{fig:k1}, Figure~\ref{fig:k3} in \app).
\begin{figure}
\includegraphics[width=\columnwidth,angle=0]{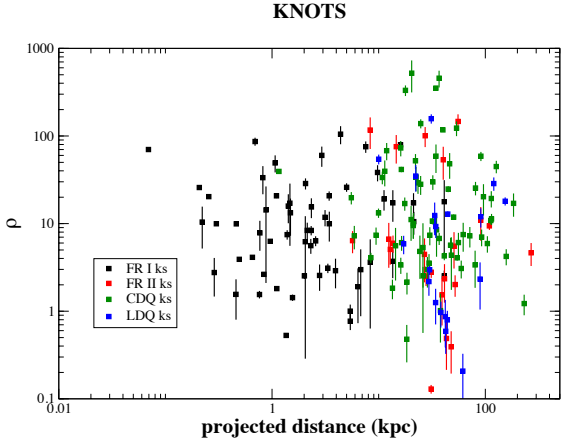}
\caption{The ratios measured for the hotspots in RGs and in QSRs as
function of the projected distance from the core in kpc.}
\label{fig:ratios_d_ks}
\end{figure}
The same behavior has been found comparing knots in FR\,Is and in LDQs, where they appear to be 
significantly different and brighter in both the $L_{\rm R}$ and $L_{\rm X}$ distributions but having similar ratios.

4.) {\it k2 vs kq.} The comparison between knots in FR\,IIs and QSR is the same for both classes (i.e., CDQ and LDQ)
(see Figure~\ref{fig:k2}, Figure~\ref{fig:k4} in \app).
The k2 components show similar $\rho$ distributions, while they appear to be significantly fainter than kq components in 
both $L_{\rm X}$ and $L_{\rm R}$ distributions even if with a small difference in the probability provided by the Monte Carlo method
(see Table~\ref{tab:prob1}).
\begin{table}
\caption{The KS test and the results of the Monte Carlo method (see Section~\ref{sec:stat})
performed to compare the RG and QSR parameter distributions of knots and hotspots, separately.}
\begin{tabular}{|cccccc|}
\hline
Type 1 & Type 2 & Param. & $D_{\rm KS}$ & $P(D_{\rm KS})^u$ & $P(D_{\rm KS})^g$ \\
\hline
\noalign{\smallskip}
hs2 & hsq    & log$\,L_{\rm R}$  & 0.32 & 0.03 & 0.35 \\ 
hs2 & hsq    & log$\,L_{\rm X}$  & 0.57 & 0.00 & 0.33 \\ 
hs2 & hsq    & log$\,\rho$       & 0.17 & ---  & 0.55 \\ 
\hline
\noalign{\smallskip}
k1  & k2     & log$\,L_{\rm R}$  & 0.25 & 0.04 & 0.05 \\ 
k1  & kqc    & log$\,L_{\rm R}$  & 0.82 & 0.00 & 0.00 \\ 
k1  & kql    & log$\,L_{\rm R}$  & 0.83 & 0.00 & 0.00 \\ 
k2  & kqc    & log$\,L_{\rm R}$  & 0.60 & 0.00 & 0.00 \\ 
k2  & kql    & log$\,L_{\rm R}$  & 0.70 & 0.00 & 0.00 \\ 
kqc & kql    & log$\,L_{\rm R}$  & 0.22 & 0.07 & 0.68 \\ 
\noalign{\smallskip}
k1  & k2     & log$\,L_{\rm X}$  & 0.14 & 0.24 & 0.68 \\ 
k1  & kqc    & log$\,L_{\rm X}$  & 0.88 & 0.00 & 0.00 \\ 
k1  & kql    & log$\,L_{\rm X}$  & 0.91 & 0.00 & 0.00 \\ 
k2  & kqc    & log$\,L_{\rm X}$  & 0.86 & 0.00 & 0.00 \\ 
k2  & kql    & log$\,L_{\rm X}$  & 0.95 & 0.00 & 0.00 \\ 
kqc & kql    & log$\,L_{\rm X}$  & 0.03 & 0.82 & 0.94 \\ 
\noalign{\smallskip}
k1  & k2     & log$\,\rho$  & 0.08 & --- & 0.73 \\ 
k1  & kqc    & log$\,\rho$  & 0.04 & --- & 0.88 \\ 
k1  & kql    & log$\,\rho$  & 0.11 & --- & 0.71 \\ 
k2  & kqc    & log$\,\rho$  & 0.11 & --- & 0.66 \\ 
k2  & kql    & log$\,\rho$  & 0.05 & --- & 0.75 \\ 
kqc & kql    & log$\,\rho$  & 0.14 & --- & 0.43 \\ 
\noalign{\smallskip}
\hline
\end{tabular}\\
\label{tab:prob1}
\end{table}
In Figure~\ref{fig:ks_ratios} we show the ratios $vs$ both the $L_{\rm R}$ and $L_{\rm X}$ luminosities
for the knots, while in Figure~\ref{fig:ratios_d_ks} we report the $\rho$ values as a function of the projected distance,
in both cases no clear trend or correlation has been found between the observed parameters.

Finally, we note the presence of a few red filled squares with $\rho < $ 1, shown in both Figure~\ref{fig:ks_ratios}
and Figure~\ref{fig:ratios_d_ks}, these k2 components belong to three different FR\,II RGs, as the case of 3C\,353, 
that with $z =$ 0.0304 is also the second closest FR II RG in our sample
(the closest is PKS2153-69 at $z =$ 0.0283).

These k2 components have been detected because the exposures of these source observations
are the longest among the whole FR\,II RGs,
a crucial test to verify if other k2 knots show similar values of ratios could be provided by deeper observations
of radio sources in the same class. 
These `low ratio' knots in these FR\,II RGs represent a clear example of a bias that could appear in the KS test.
In fact, they can make the k2 ratio distribution different from the real one, because the number 
of components in FR\,IIs is lower with respect to the other classes (i.e., 20), resulting in a large value of the $D_{\rm KS}$
variable (see Table~\ref{tab:prob1}).

However, our Monte Carlo method developed to take into account this problem, shows that
the differences in $\rho$ between k2 and other knots are marginal.
A better investigation of the k2 ratio distribution might be possible in the future if/when deeper \chn~observations will be 
performed.

\subsection{Hotspots vs Knots}\label{sec:hk} 
We compared hotspots and knots in RGs and QSRs and between them.

1.) {\it k1 vs hs2.} We found that hs2 components are different in $L_{\rm R}$ and $\rho$ with respect to k1 ones, being less bright at radio frequencies,
while their $L_{\rm X}$ distribution is roughly similar (see Figure~\ref{fig:hk1}).

2.) {\it k2 vs hs2.} No signifcant differences have been found in both the $L_{\rm X}$ 
distributions comparing hotspots and knots of  
FR\,II RGs (see Figure~\ref{fig:hk2}), while their $L_{\rm R}$ and $\rho$ distributions appear different as for k1 components. 

3.) {\it kq vs hs2.}  Comparing hotspots in RGs (hs2) with knots in QSR (i.e., both kqc and kql), 
we found no significant differences in their $L_{\rm R}$ distribution
while they appear to be extremely different in the $L_{\rm X}$ distributions, where kq components are brighter than hs2 ones.
Consequently also their $\rho$ distributions are significantly different (see Figure~\ref{fig:hk3} and Figure~\ref{fig:hk4}).

4.) {\it k1 vs hsq.} Knots in FR \,Is appear very different in $L_{\rm R}$, $L_{\rm X}$  and $\rho$ distributions
with respect to hotspots in QSRs (see Figure~\ref{fig:hk5}). FR\,I knots are systematically fainter than hsq components
at radio and X-ray frequencies.

5.) {\it k2 vs hsq.} In the case of FR\,II knots (i.e., k2), the situation is similar to that of k1 components
in comparison with hsq. Even if k2 are different in $L_{\rm R}$ and in $L_{\rm X}$ with respect to hsq,
their $\rho$ distribution appear to be similar (see Figure~\ref{fig:hk6}) within our level of confidence (see Table~\ref{tab:prob2}).
However, we note that this comparison regards the two classes with the smallest number of components,
that could be more affected by statistical biases. Thus to confirm this result an investigation on a larger sample 
than the one considered is necessary. 

6.) {\it kqc vs hsq.}  The behavior of hotspots vs knots in QSRs appear to be different from that in RGs.
We did not find any significant difference between hsq and kqc components in their $L_{\rm R}$ and $L_{\rm X}$ distributions,
there is only a significant difference in their $\rho$ distribution (see Figure~\ref{fig:hk7}).
In addition, kqc components have $\rho$ values systematically higher than 1,
while the average $\rho$ of hsq is closer to zero (see Table~\ref{tab:properties}).

7.) {\it kql vs hsq.} Also the $L_{\rm R}$ and $L_{\rm X}$ distributions of hsq and kql components are similar within the probabilities 
as indicated by our statistical analysis, while a marginal difference has been found in the $\rho$ distribution
(see Figure~\ref{fig:hk8}), less significant than that found between kqc and hsq components, 
because of the smaller number of knots considered.

Finally, we note that the most statistically significant differences we found regard the comparison between 
all the parameter distributions of knots in RGs and hotspots in QSRs.
Both the kqc and kql components are significantly different in all the $L_{\rm R}$, $L_{\rm X}$, $\rho$ distributions,
with respect to the hs2 ones.
\begin{table}
\caption{The KS test and the results of the Monte Carlo method (see Section~\ref{sec:stat})
performed to compare the parameter distributions between knots and hotspots.}
\begin{tabular}{|cccccc|}
\hline
Type 1 & Type 2 & Param. & $D_{\rm KS}$ & $P(D_{\rm KS})^u$ & $P(D_{\rm KS})^g$ \\
\hline
\noalign{\smallskip}
k1   & hs2   & log$\,L_{\rm R}$  & 0.78 & 0.00 & 0.00 \\ 
k2   & hs2   & log$\,L_{\rm R}$  & 0.55 & 0.00 & 0.00 \\ 
\noalign{\smallskip}
k1   & hs2   & log$\,L_{\rm X}$  & 0.60 & 0.00 & 0.06 \\ 
k2   & hs2   & log$\,L_{\rm X}$  & 0.21 & 0.10 & 0.64 \\ 
\noalign{\smallskip}
k1   & hs2   & log$\,\rho$  & 0.74 & --- & 0.00 \\ 
k2   & hs2   & log$\,\rho$  & 0.66 & --- & 0.00 \\ 
\hline
\noalign{\smallskip}
kqc  & hs2   & log$\,L_{\rm R}$  & 0.16 & 0.16 & 0.54 \\ 
kql  & hs2   & log$\,L_{\rm R}$  & 0.20 & 0.14 & 0.51 \\ 
\noalign{\smallskip}
kqc  & hs2   & log$\,L_{\rm X}$  & 0.59 & 0.00 & 0.06 \\ 
kql  & hs2   & log$\,L_{\rm X}$  & 0.68 & 0.00 & 0.00 \\ 
\noalign{\smallskip}
kqc  & hs2   & log$\,\rho$  & 0.77 & --- & 0.00 \\ 
kql  & hs2   & log$\,\rho$  & 0.62 & --- & 0.00 \\ 
\hline
\noalign{\smallskip}
k1   & hsq   & log$\,L_{\rm R}$  & 0.84 & 0.00 & 0.00 \\ 
k2   & hsq   & log$\,L_{\rm R}$  & 0.73 & 0.00 & 0.00 \\ 
\noalign{\smallskip}
k1   & hsq   & log$\,L_{\rm X}$  & 0.83 & 0.00 & 0.00 \\ 
k2   & hsq   & log$\,L_{\rm X}$  & 0.64 & 0.00 & 0.00 \\ 
\noalign{\smallskip}
k1   & hsq   & log$\,\rho$  & 0.42 & --- & 0.04 \\ 
k2   & hsq   & log$\,\rho$  & 0.38 & --- & 0.36 \\ 
\hline
\noalign{\smallskip}
kqc  & hsq   & log$\,L_{\rm R}$  & 0.38 & 0.00 & 0.10 \\ 
kql  & hsq   & log$\,L_{\rm R}$  & 0.19 & 0.23 & 0.35 \\ 
\noalign{\smallskip}
kqc  & hsq   & log$\,L_{\rm X}$  & 0.12 & 0.31 & 0.51 \\ 
kql  & hsq   & log$\,L_{\rm X}$  & 0.17 & 0.30 & 0.41 \\ 
\noalign{\smallskip}
kqc  & hsq   & log$\,\rho$  & 0.50 & --- & 0.00 \\ 
kql  & hsq   & log$\,\rho$  & 0.33 & --- & 0.49 \\ 
\noalign{\smallskip}
\hline
\end{tabular}\\
\label{tab:prob2}
\end{table}

\section{Discussion and Summary}\label{sec:discuss} 
The chief reason we investigated the ratios of X-ray to radio
fluxes was the expectation that there would be substantial differences between
components commonly thought to be synchrotron X-ray sources (e.g., FR\,I
knots) and those thought to come from inverse Compton scattering off 
the Cosmic Microwave Background (IC/CMB) 
\citep{bergamini67,tavecchio00,celotti01} dominated emission
(e.g., QSR knots).  In this section we compare the observed results with
various ``expectations" associated with the two emission processes.

In a sense, the most surprising result is that the $\rho$
distributions for FR\,I knots is essentially indistinguishable from that
for QSR knots.  If the X-rays from both classes of objects are dominated
by synchrotron emission there would be no reason for surprise: the
observed spread in $\rho$ would reflect the spread in the ratio of
amplitudes of the electron spectra between the energies responsible
for the X-rays and those responsible for the radio.  However, if the
X-ray emission from QSR knots is actually IC/CMB, we must deduce that
the factors contributing to $<\rho>$ and the width of the $\rho$
distribution conspire to produce the same results for FR\,I and QSR
knots.  In the synchrotron case these factors are those described
above.  

In the IC/CMB model, $\rho$ is mainly dependent on three factors: 
the ratio of amplitudes of the electron energy spectra around $\gamma\approx$ 10$^2$ emitting in the X-rays 
to those producing the observed radio emission at GHz frequencies (i.e., $\gamma\approx\,$ 10$^4$), 
the ratio of the bulk Lorentz factor $\Gamma$ over the magnetic field $B$, as $(\Gamma/B)^2$, 
and an additional broadening of the $\rho$ distribution arising from the dispersion in the distribution of the angle
between the velocity vector of the jet knot and the line-of-sight in different sources \citep{harris10}.  
Our expectation was that these completely different factors would produce substantially different
$\rho$ distributions, however, we found no significant differences between the QSRs and FR\,Is in this respect.

Another difference between the two emission processes is the additional
bulk Lorentz factor in the IC/CMB model stemming from the fact that head-on IC
scattering is more probable than over-taking scattering (i.e., ``extra-beaming factor'')
\citep{dermer95,harris02,massaro09c}.  Statistically, this feature should manifest most prominently when
comparing sources with small angles between our line-of-sight and the jet to
those with larger angles.  In our sample, we have both LDQs and CDQs, 
and it is commonly believed that
the determining factor for this division is that CDQ have jets closer to the line-of-sight.
Thus we might expect larger $\rho$ values for CDQs than
LDQs.  This prediction was not found in our current data.
A possible interpretation of the lack of differences between knots in CDQs and in LDQs is that 
the sampling of Chandra observations (and subsequent x-ray jet detections compiled in XJET) 
have tended toward sampling of the most `aligned' sources (i.e., the most core-dominated) 
in general. Future observations of LDQs characterized with systematically 
lower radio core-dominance values than currently probed may make any 
differences apparent. Alternative explanations include 
that most jets of CDQs bend significantly between the inner (pc-scale) and the outer (kpc-scale).

One of the few real differences in $\rho$ distributions occurs when
one compares FR\,II hotspots to knots (both in RGs and QSRs).  
Most FR\,II hotspots have $\rho<$ 1 whereas most knots have $\rho >$ 1.  
This effect finds a reasonable cause in our current understanding of X-ray emission from
hotspots \citep{hardcastle04}.  The more powerful radio hotspots are
consistent with a synchrotron self-Compton (SSC) model for the
dominant X-ray emission process.    
As shown in Figure~\ref{fig:hs_ratios}, the FR\,II hotspots
with $\rho >$ 1 all have $L_{\rm r} < 10^{42}$ erg s$^{-1}$, and these lower
luminosity hotspots are those for which SSC is thought not to be the
dominant X-ray emission process \citep{hardcastle04}
but rather, an additional emission mechanism (synchrotron or additional IC) may be responsible for the observed X-rays.

More details about the interpretations of the above results
and the comparison between the theoretical expectations and
the observational evidences will be discussed in forthcoming papers \citep{harris11}
where these results will be also compared with those on RGs found by Kataoka \& Stawarz (2005).
Finally, a possible statistical test for the IC/CMB process in QSRs 
will be described in \citep{massaro11b}.

\acknowledgments 
We thank the anonymous referee for useful comments that led to improvements in the paper.
We are extremely grateful to our friends and colleagues:
T. Aldcroft, 
M. Birkinshaw, 
A. Bliss, 
K. Blundell, 
D. Evans, 
J. Gelbord,
G. Giovannini, 
M. Hardcastle, 
P. Kharb,
R. Kraft, 
S. Jorstad, 
R. Laing, 
J. Leahy, 
M. Lister, 
C. Ly, 
A. Marscher, 
H. Marshall, 
B. Miller,
R. Morganti, 
D. Schwartz,
A. Siemiginowska, 
and
D. Worrall, 
for providing radio maps; their contribution has been crucial to carry on our investigation. 
F. Massaro thanks A. Cavaliere, G. Brunetti, G. Giovannini and G. Migliori for fruitful discussions,
and R. D'Abrusco for his suggestions on compiling the tables.\\
This research has made use of SAOImage DS9, developed by the Smithsonian
Astrophysical Observatory (SAO) and the NASA/IPAC Extragalactic Database
(NED) which is operated by the Jet Propulsion Laboratory, California
Institute of Technology, under contract with the National Aeronautics and Space Administration.  
TOPCAT\footnote{\underline{http://www.star.bris.ac.uk/$\sim$mbt/topcat/}} 
\citep{taylor2005} was used extensively in this work for the preparation and manipulation of the tabular data.\\
Several radio maps were downloaded from the
NVAS (NRAO VLA Archive Survey) and from the MERLIN archive.
The National Radio Astronomy Observatory is operated by Associated Universities, Inc., under contract with the National Science Foundation.  
MERLIN is a National Facility operated by the University of Manchester at Jodrell Bank Observatory on behalf of STFC.
The Australia Telescope is funded by the Commonwealth of Australia for operation as a National Facility managed by CSIRO. 
This research has made use of data obtained through the High Energy Astrophysics 
Science Archive Research Center Online Service, provided by the NASA/Goddard Space Flight Center.\\
F. Massaro acknowledges the Fondazione Angelo Della Riccia for the grant awarded him to support 
his research at SAO during 2011.
The XJET website is partially supported by NASA grant AR6-7013X and NASA contract NAS8-39073.
The work at SAO was supported by NASA-GRANTS GO8-9114A and NNX10AD50G. 
F. Massaro acknowledges the Foundation BLANCEFLOR Boncompagni-Ludovisi, n'ee Bildt 
for the grants awarded him in 2009 and in 2010 to support his research at SAO.

{\bf Facilities:}  \facility{CXO (ACIS)}, \facility{VLA}, \facility{MERLIN}, \facility{ATCA}

~

\appendix
In Tables~\ref{tab:app1}, \ref{tab:app2}, \ref{tab:app3}, \ref{tab:app4}, \ref{tab:app5}, \ref{tab:app6} 
of this appendix we report: the source name together with the component type 
and the region size used for the flux measurements,
the value of the radio flux densities at given radio frequency, 
the observed X-ray fluxes in each band (i.e., soft, medium, hard and total, see Section~\ref{sec:xray} for more details) 
and the values of the ratios corrected for the Galactic absorption (see Section~\ref{sec:absorp}), for each components in every source
of our XJET selected sample.
When a dashed line is shown, it implies that the flux is consistent with zero within 1$\sigma$ error.

\begin{figure*}
\begin{tabular}{cc}
\includegraphics[height=6.cm,width=6.3cm,angle=0]{./Lrad_hs.pdf}
\includegraphics[height=6.cm,width=6.3cm,angle=0]{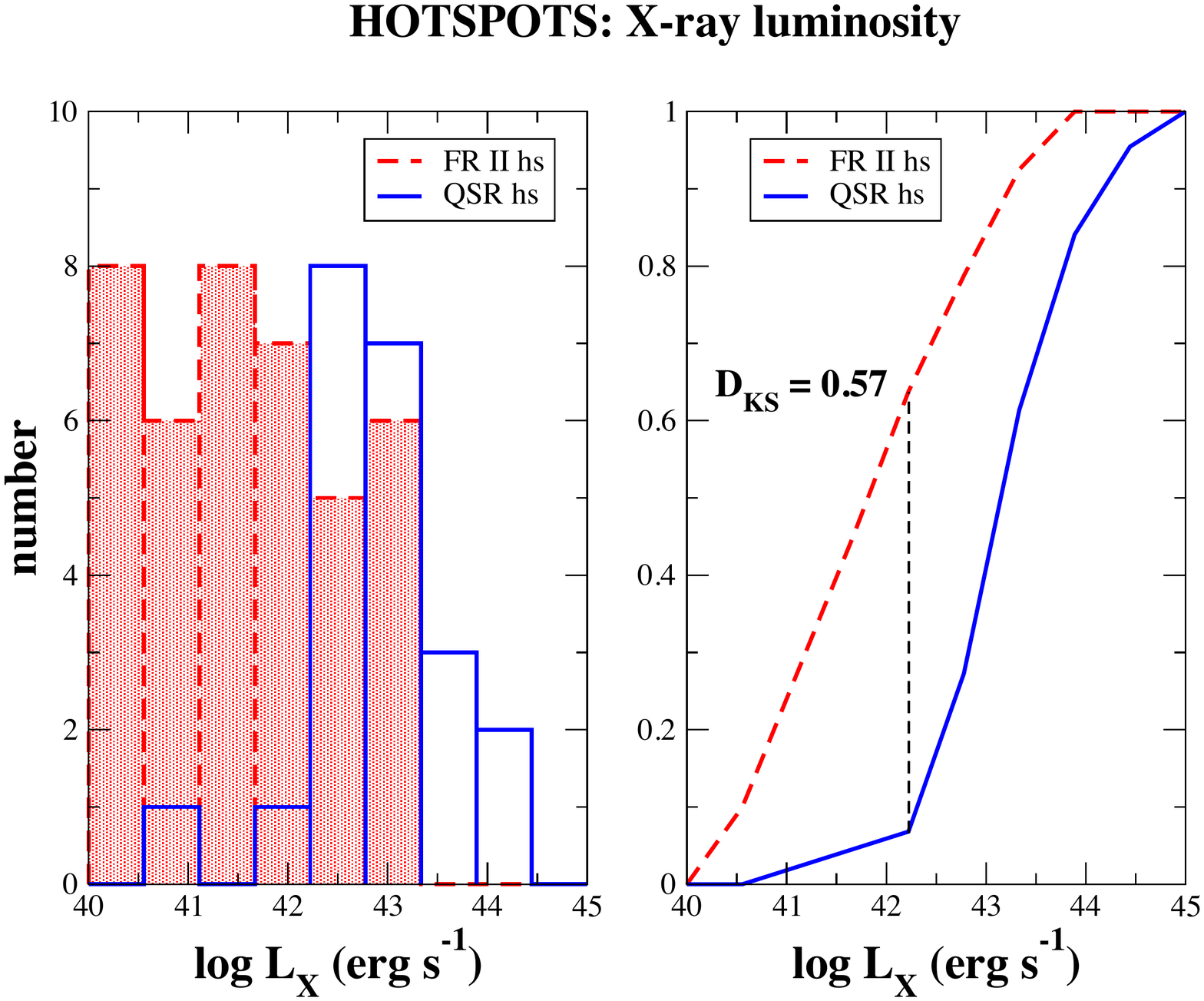}
\includegraphics[height=6.cm,width=6.3cm,angle=0]{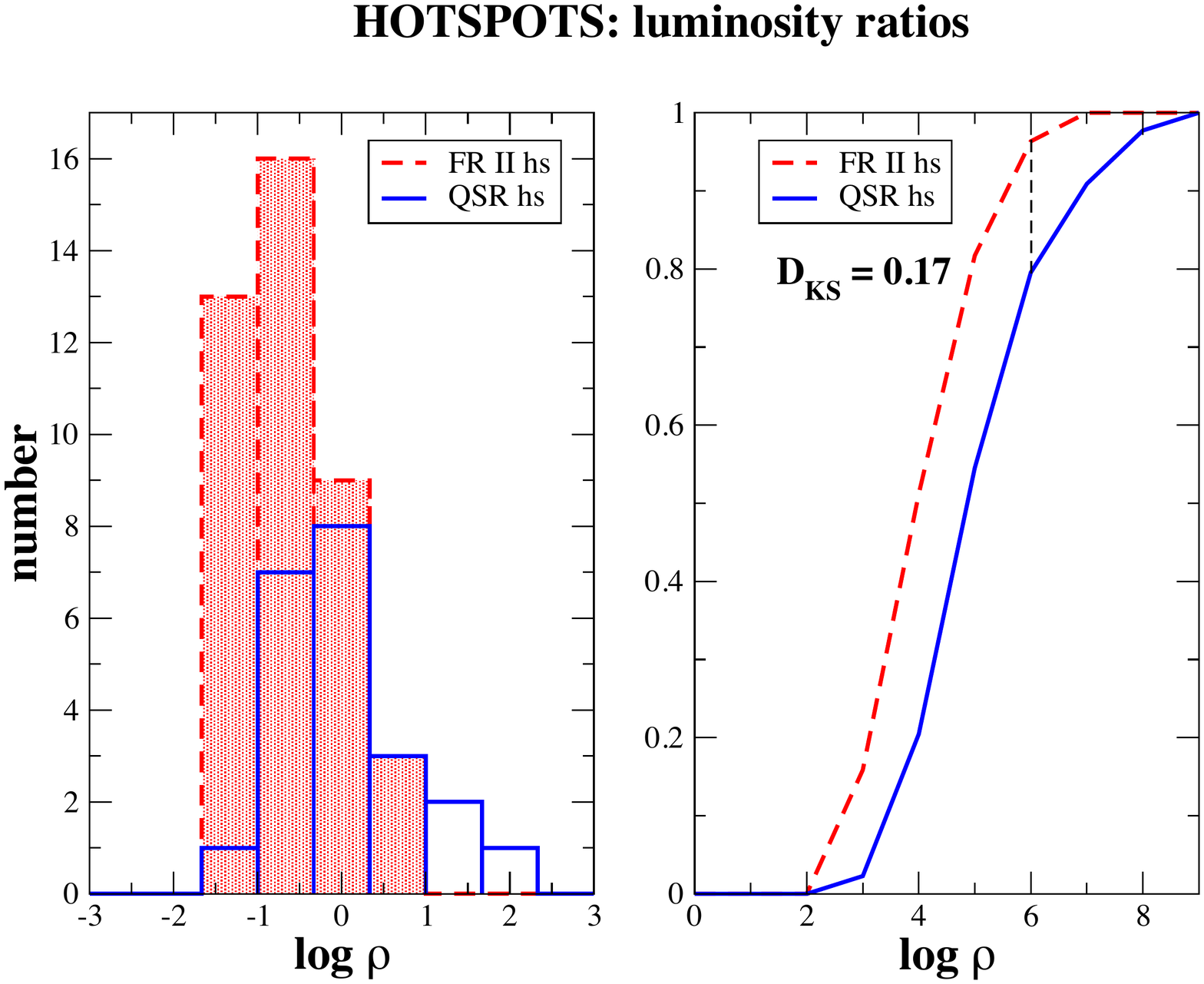}
\end{tabular}
\caption{
a) The distributions of radio luminosities $L_{\rm R}$ of hotspots in RGs and QSRs.
b) The normalized cumulative distributions of radio luminosities for hotspots in RGs and QSRs.
c) The distributions of X-ray luminosities $L_{\rm X}$ of hotspots in RGs and QSRs.
d) The normalized cumulative distributions of X-ray luminosities for hotspots in RGs and QSRs.
e) The distributions of $\rho$ of hotspots in RGs and QSRs.
f) The normalized cumulative distributions of luminosity ratios for hotspots in RGs and QSRs.
}
\label{fig:rg_hs}
\end{figure*}

\begin{figure*}
\begin{tabular}{cc}
\includegraphics[height=6.cm,width=6.3cm,angle=0]{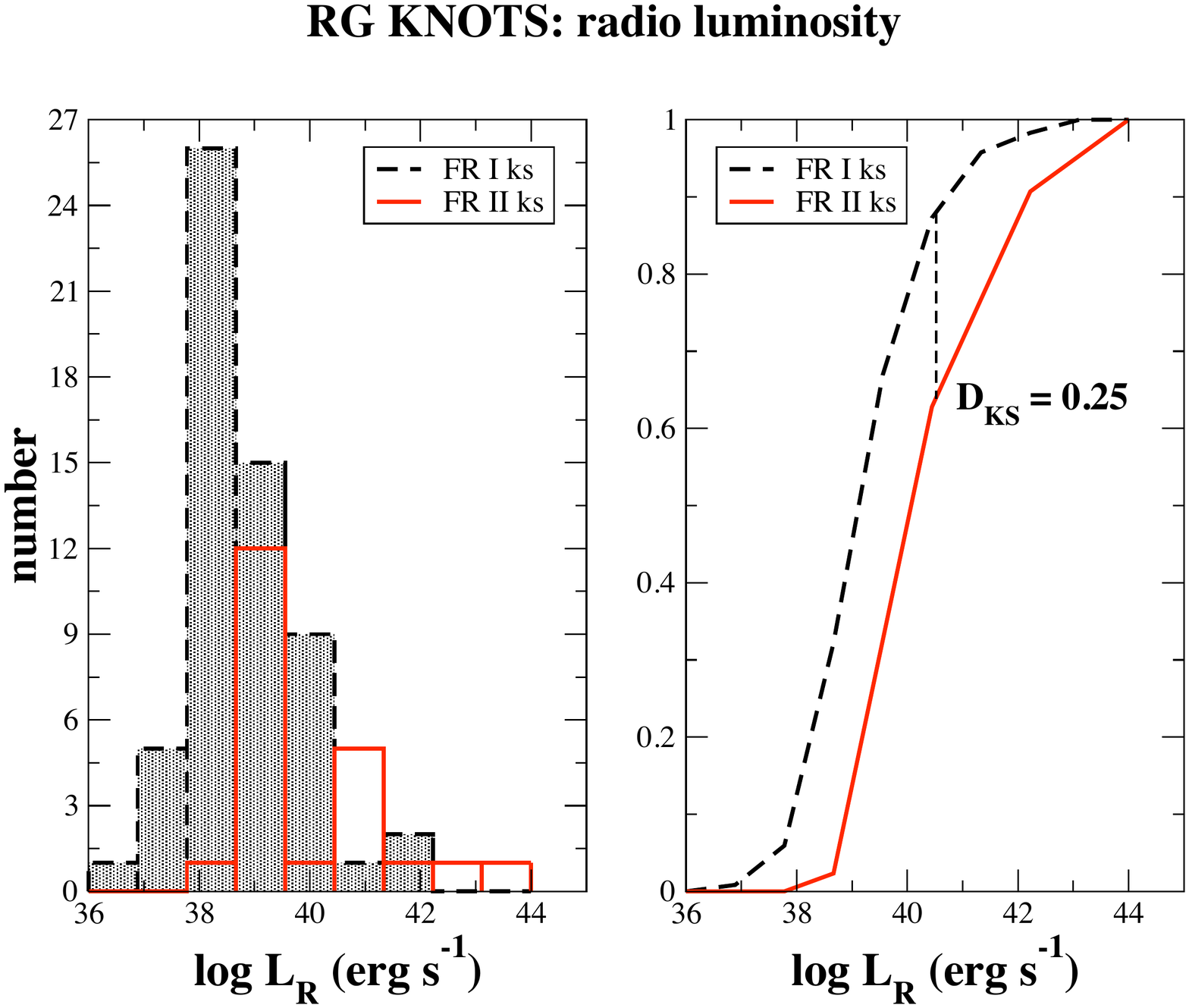}
\includegraphics[height=6.cm,width=6.3cm,angle=0]{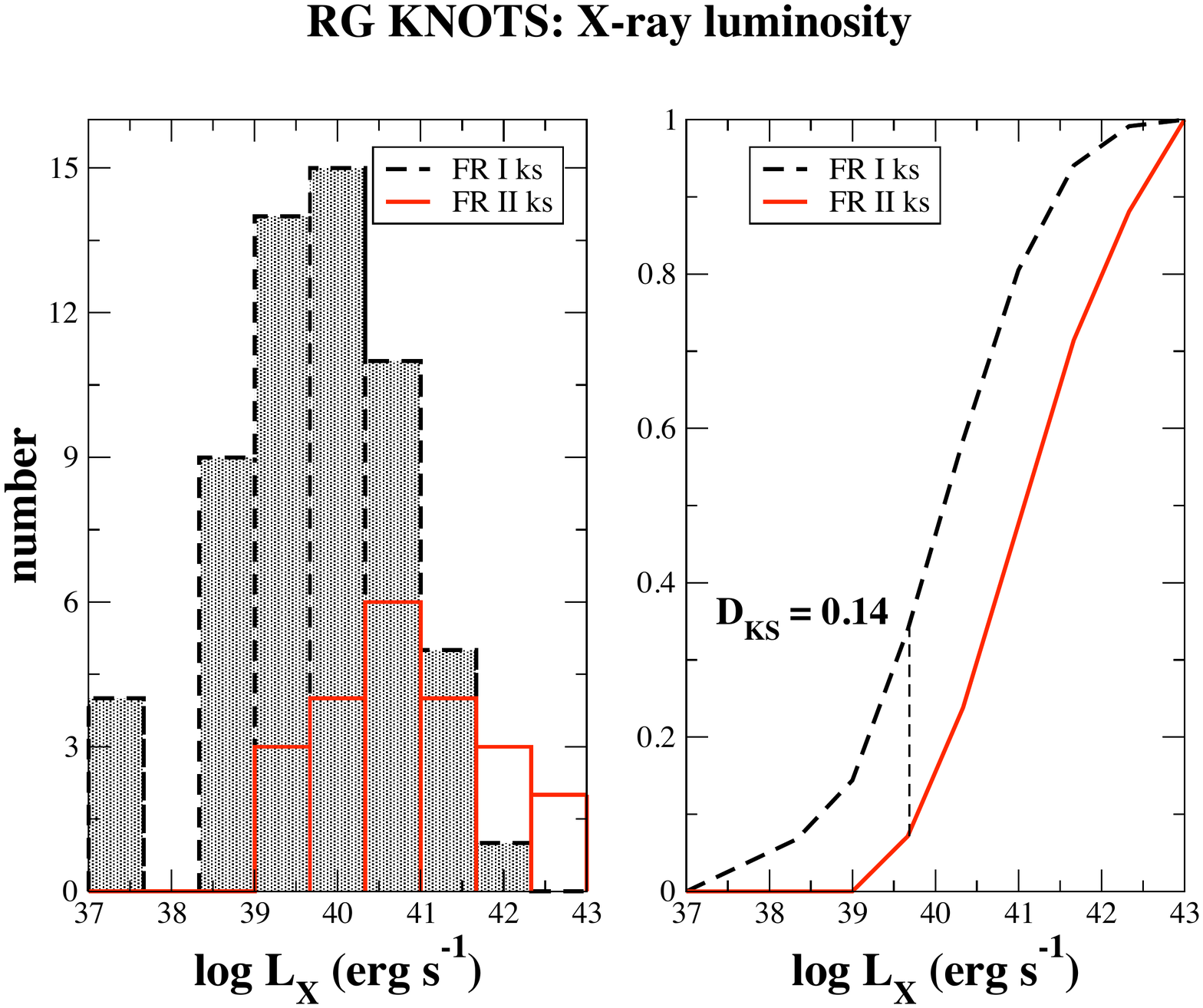}
\includegraphics[height=6.cm,width=6.3cm,angle=0]{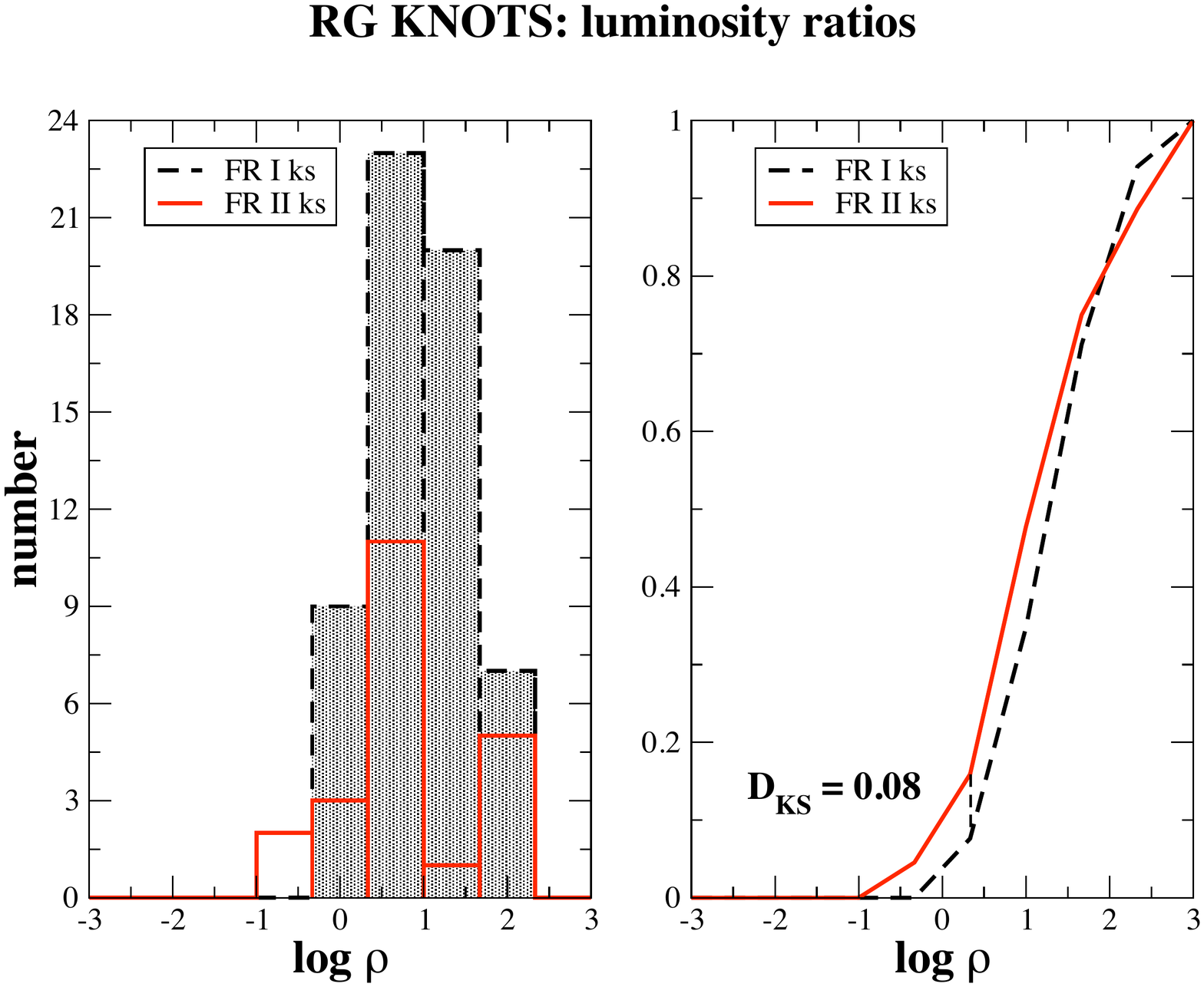}
\end{tabular}
\caption{
a) The distributions of radio luminosities $L_{\rm R}$ of knots in RGs.
b) The normalized cumulative distributions of radio luminosities for knots in RGs.
c) The distributions of X-ray luminosities $L_{\rm X}$ of knots in RGs.
d) The normalized cumulative distributions of X-ray luminosities for knots in RGs.
e) The distributions of $\rho$ of knots in RGs.
f) The normalized cumulative distributions of luminosity ratios for knots in RGs.
}
\label{fig:rg_ks}
\end{figure*}

\begin{figure*}
\begin{tabular}{cc}
\includegraphics[height=6.cm,width=6.3cm,angle=0]{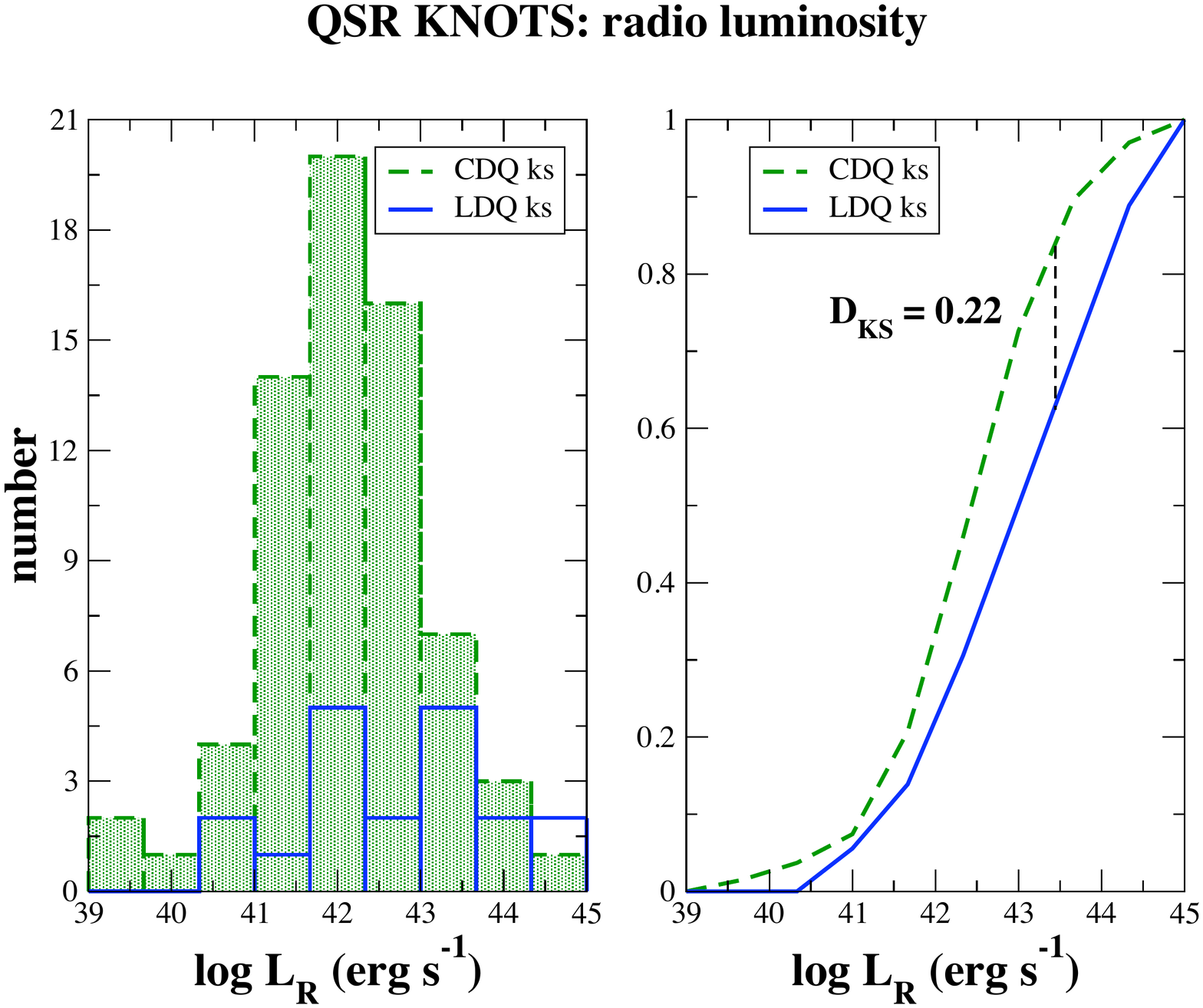}
\includegraphics[height=6.cm,width=6.3cm,angle=0]{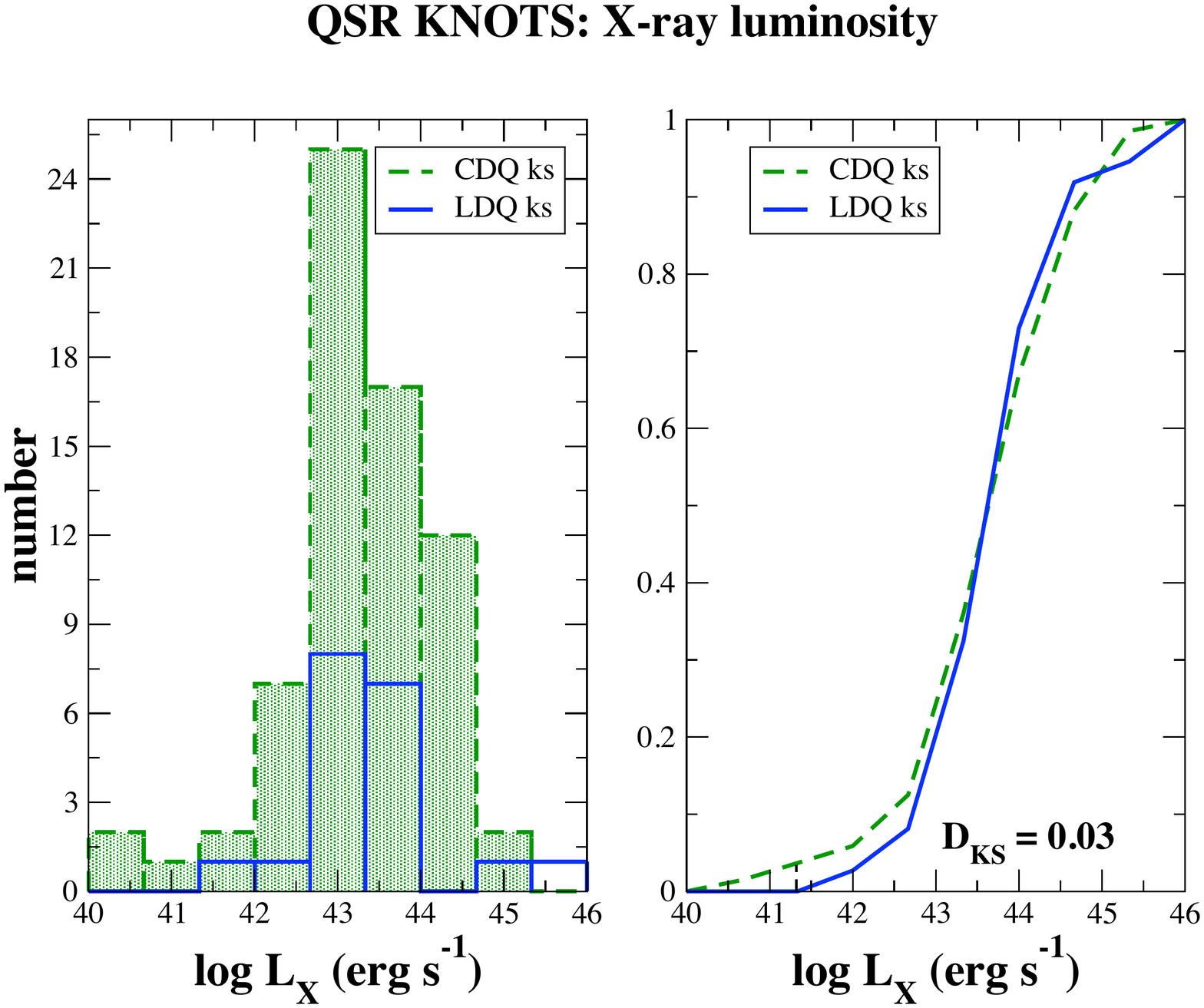}
\includegraphics[height=6.cm,width=6.3cm,angle=0]{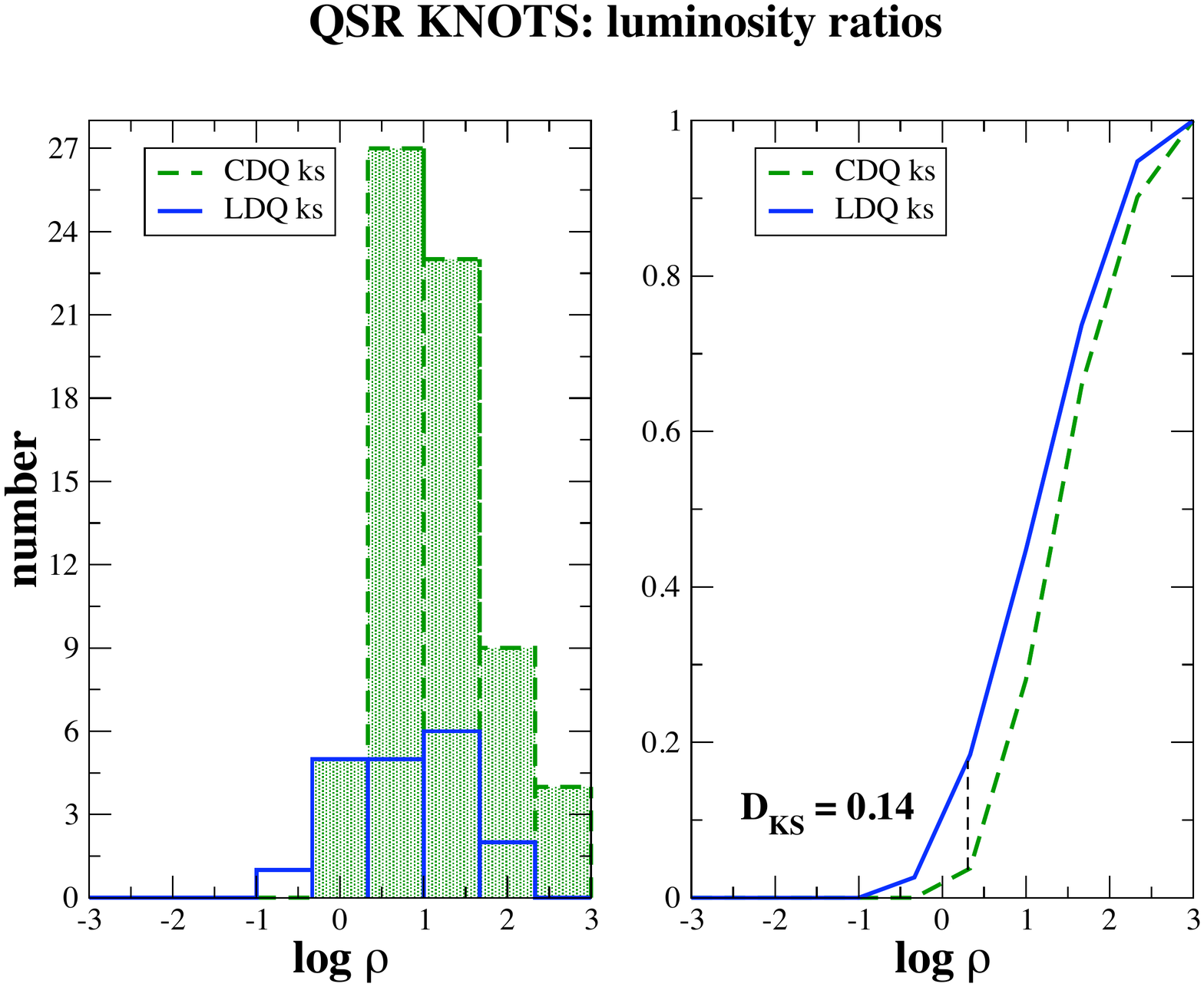}
\end{tabular}
\caption{
a) The distributions of radio luminosities $L_{\rm R}$ of knots in QSRs.
b) The normalized cumulative distributions of radio luminosities for knots in QSRs.
c) The distributions of X-ray luminosities $L_{\rm X}$ of knots in QSRs.
d) The normalized cumulative distributions of X-ray luminosities for knots in QSRs.
e) The distributions of $\rho$ of knots in QSRs.
f) The normalized cumulative distributions of luminosity ratios for knots in QSRs.
}
\label{fig:qsr_ks}
\end{figure*}

\begin{figure*}
\begin{tabular}{cc}
\includegraphics[height=6.cm,width=6.3cm,angle=0]{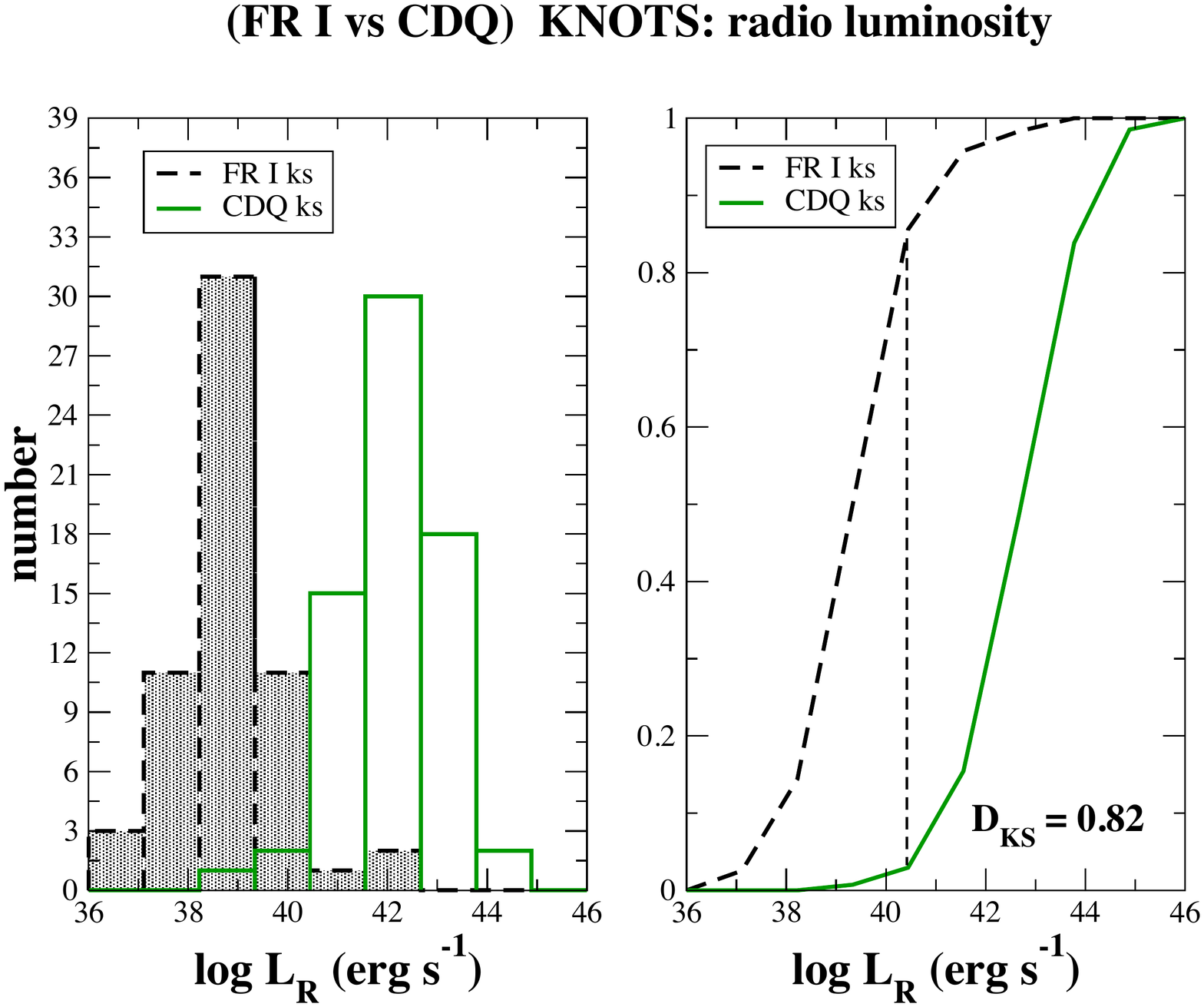}
\includegraphics[height=6.cm,width=6.3cm,angle=0]{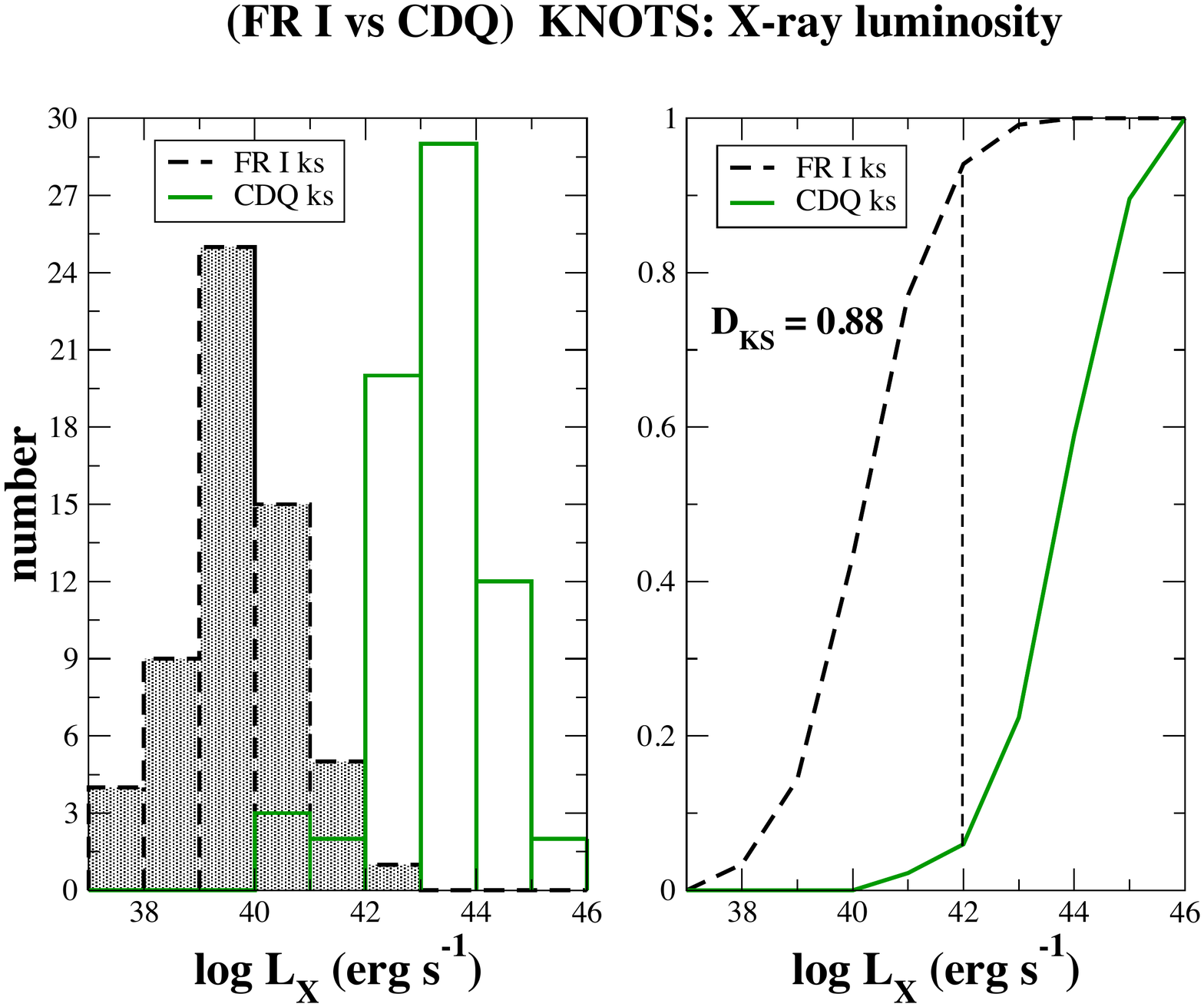}
\includegraphics[height=6.cm,width=6.3cm,angle=0]{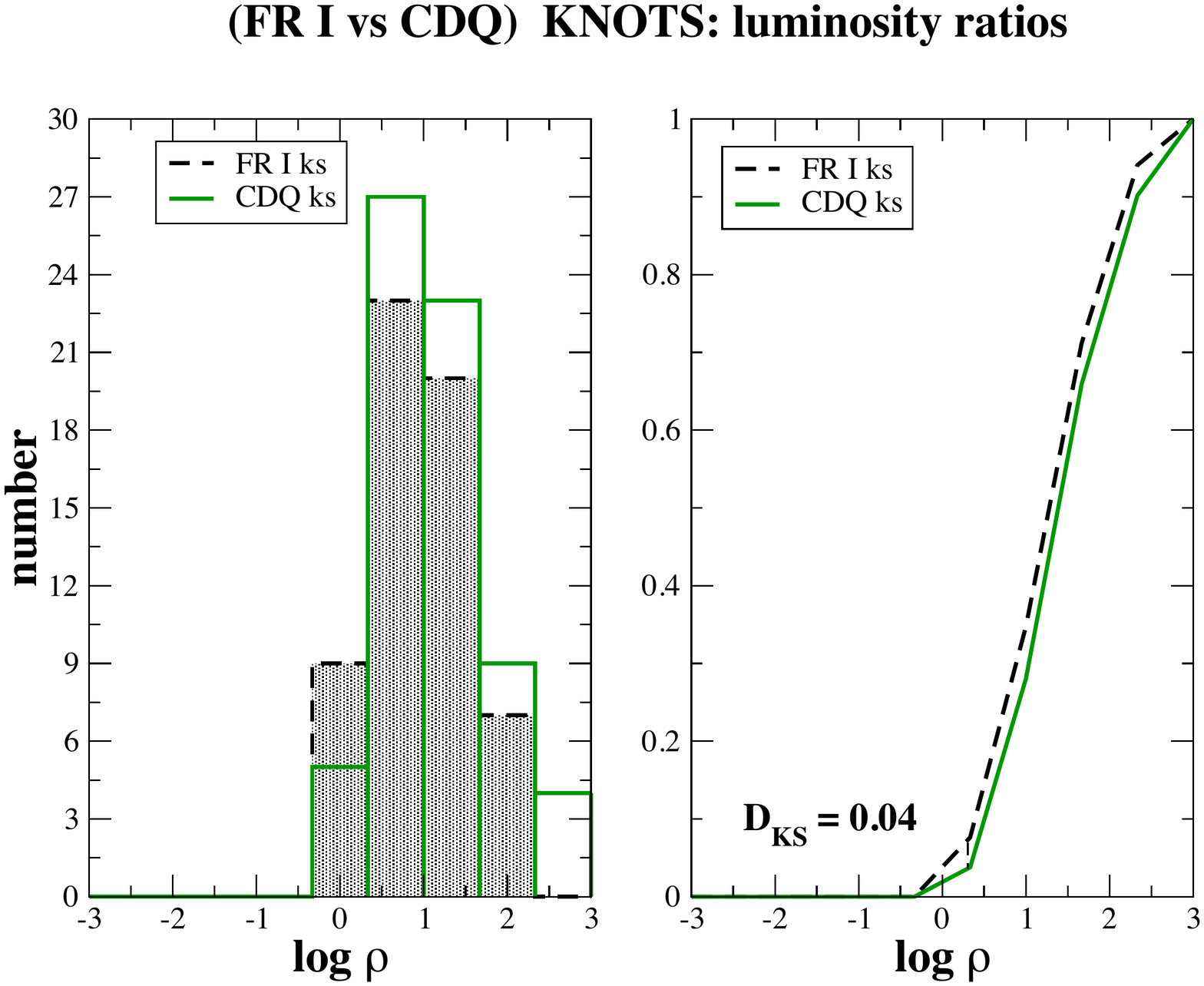}
\end{tabular}
\caption{
a) The distributions of radio luminosities $L_{\rm R}$ of knots in FR\,Is and CDQs.
b) The normalized cumulative distributions of radio luminosities for knots in FR\,Is and CDQs.
c) The distributions of X-ray luminosities $L_{\rm X}$ of knots in FR\,Is and CDQs.
d) The normalized cumulative distributions of X-ray luminosities for knots in FR\,Is and CDQs.
e) The distributions of $\rho$ of knots in FR\,Is and CDQs.
f) The normalized cumulative distributions of luminosity ratios for knots in FR\,Is and CDQs.
}
\label{fig:k1}
\end{figure*}

\begin{figure*}
\begin{tabular}{cc}
\includegraphics[height=6.cm,width=6.3cm,angle=0]{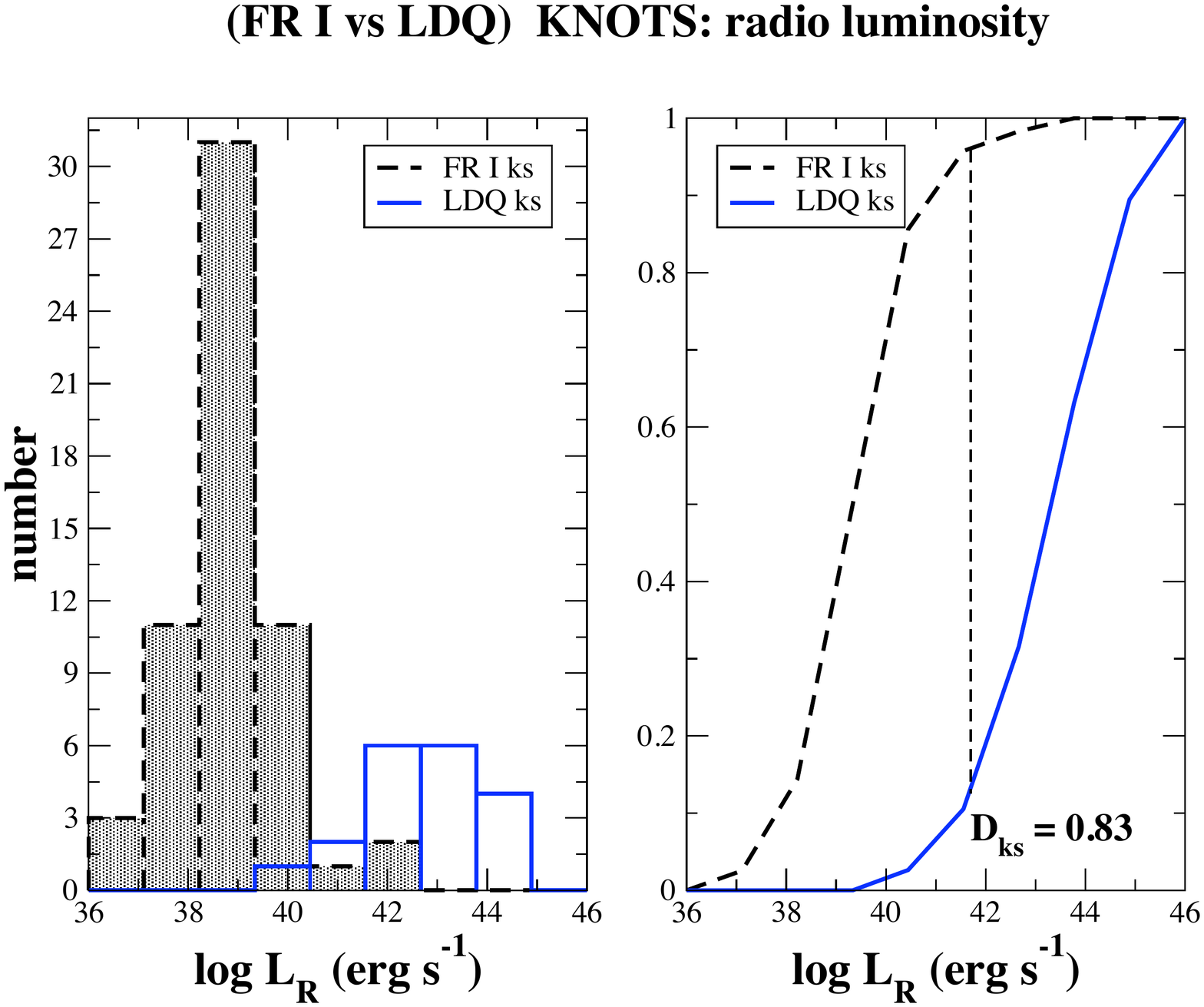}
\includegraphics[height=6.cm,width=6.3cm,angle=0]{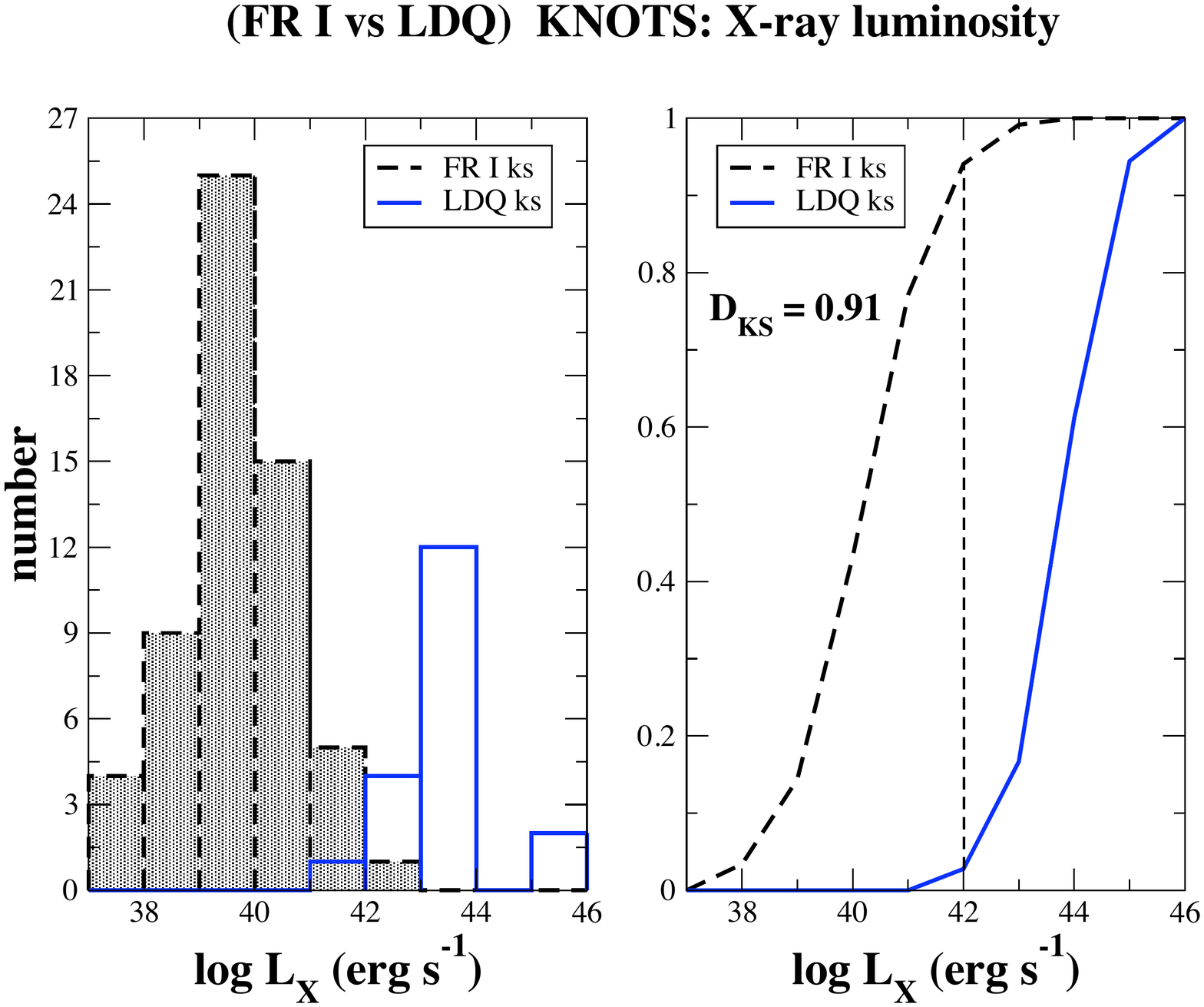}
\includegraphics[height=6.cm,width=6.3cm,angle=0]{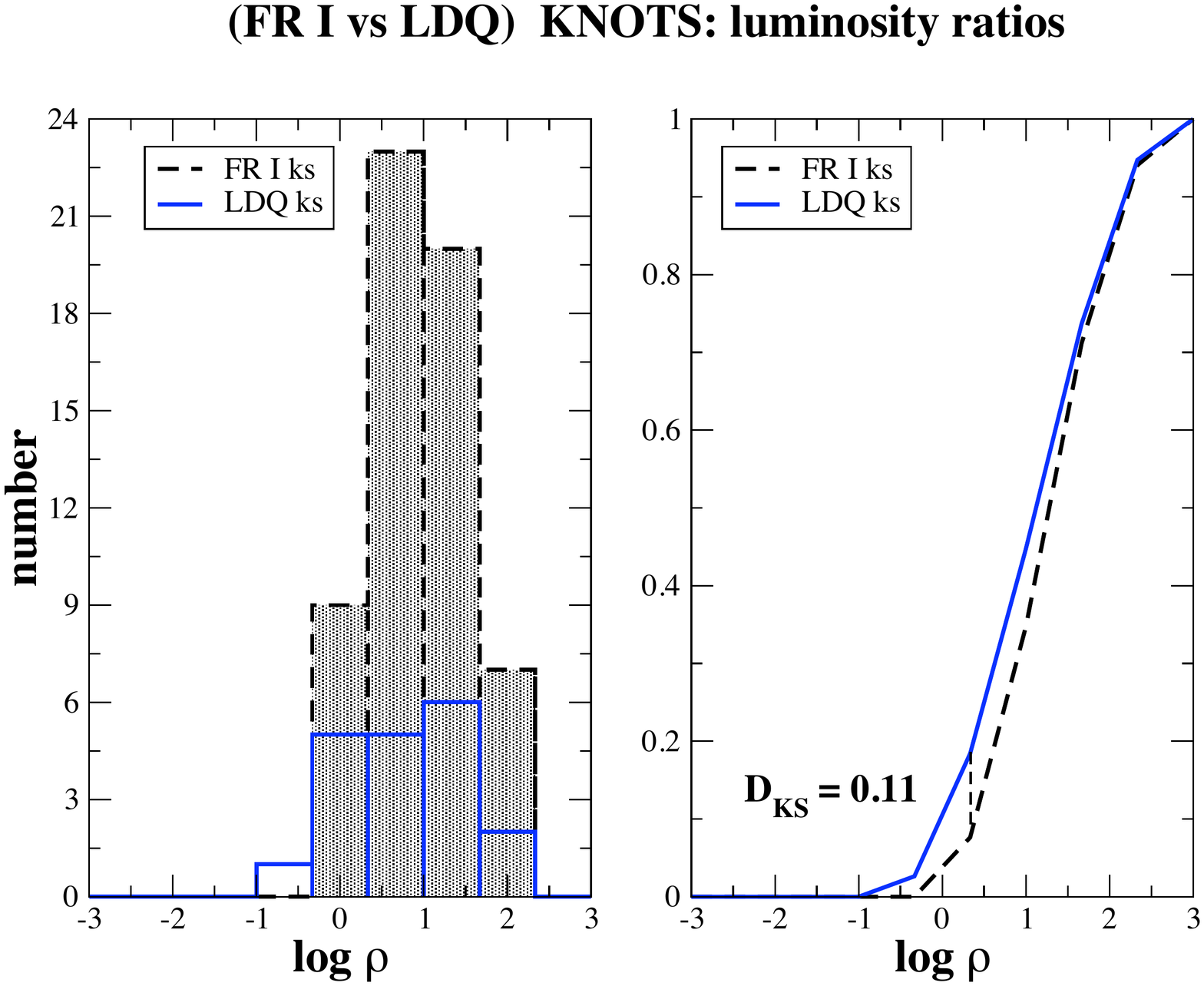}
\end{tabular}
\caption{
a) The distributions of radio luminosities $L_{\rm R}$ of knots in FR\,Is and LDQs.
b) The normalized cumulative distributions of radio luminosities for knots in FR\,Is and LDQs.
c) The distributions of X-ray luminosities $L_{\rm X}$ of knots in FR\,Is and LDQs.
d) The normalized cumulative distributions of X-ray luminosities for knots in FR\,Is and LDQs.
e) The distributions of $\rho$ of knots in FR\,Is and LDQs.
f) The normalized cumulative distributions of luminosity ratios for knots in FR\,Is and LDQs.
}
\label{fig:k3}
\end{figure*}

\begin{figure*}
\begin{tabular}{cc}
\includegraphics[height=6.cm,width=6.3cm,angle=0]{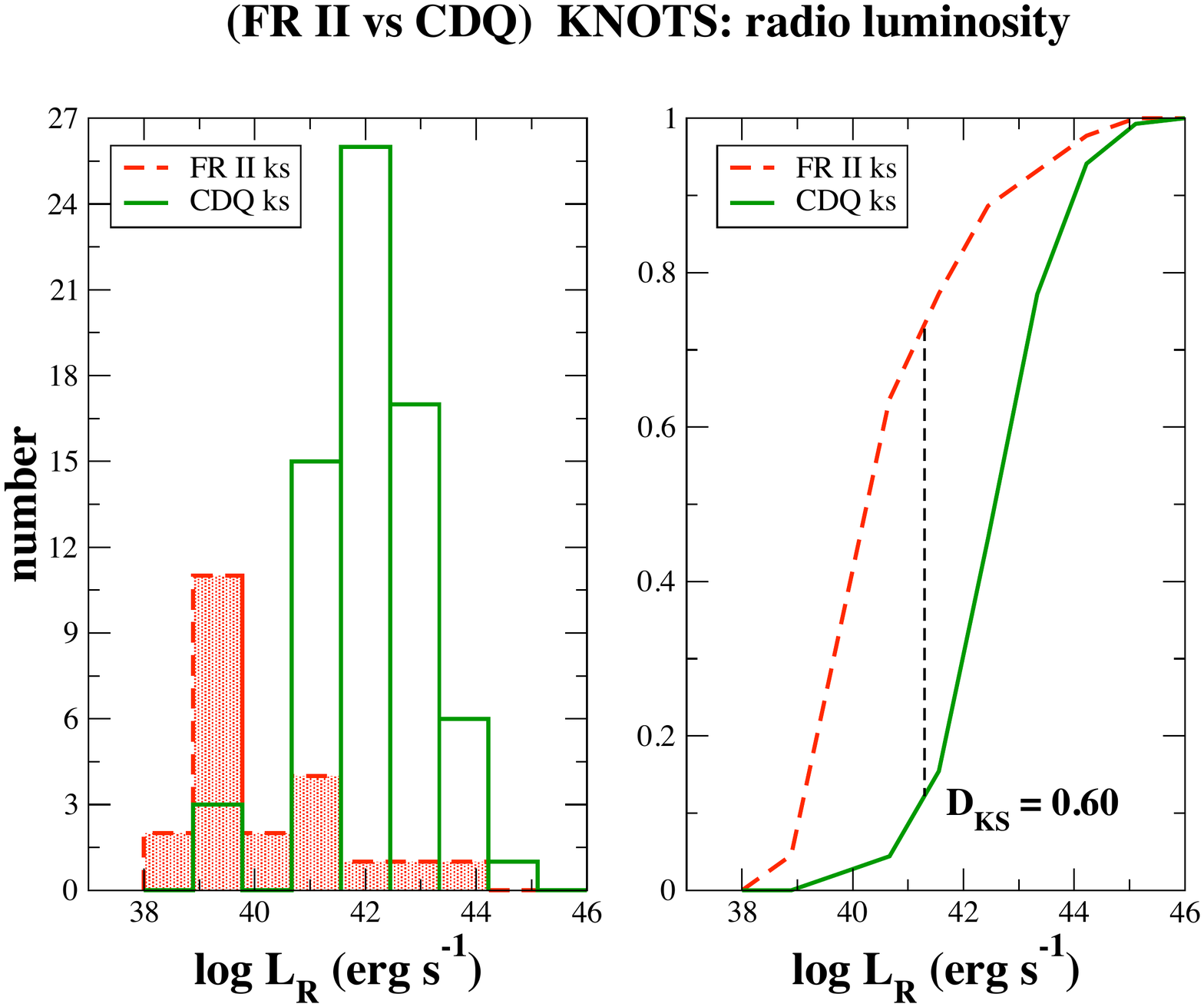}
\includegraphics[height=6.cm,width=6.3cm,angle=0]{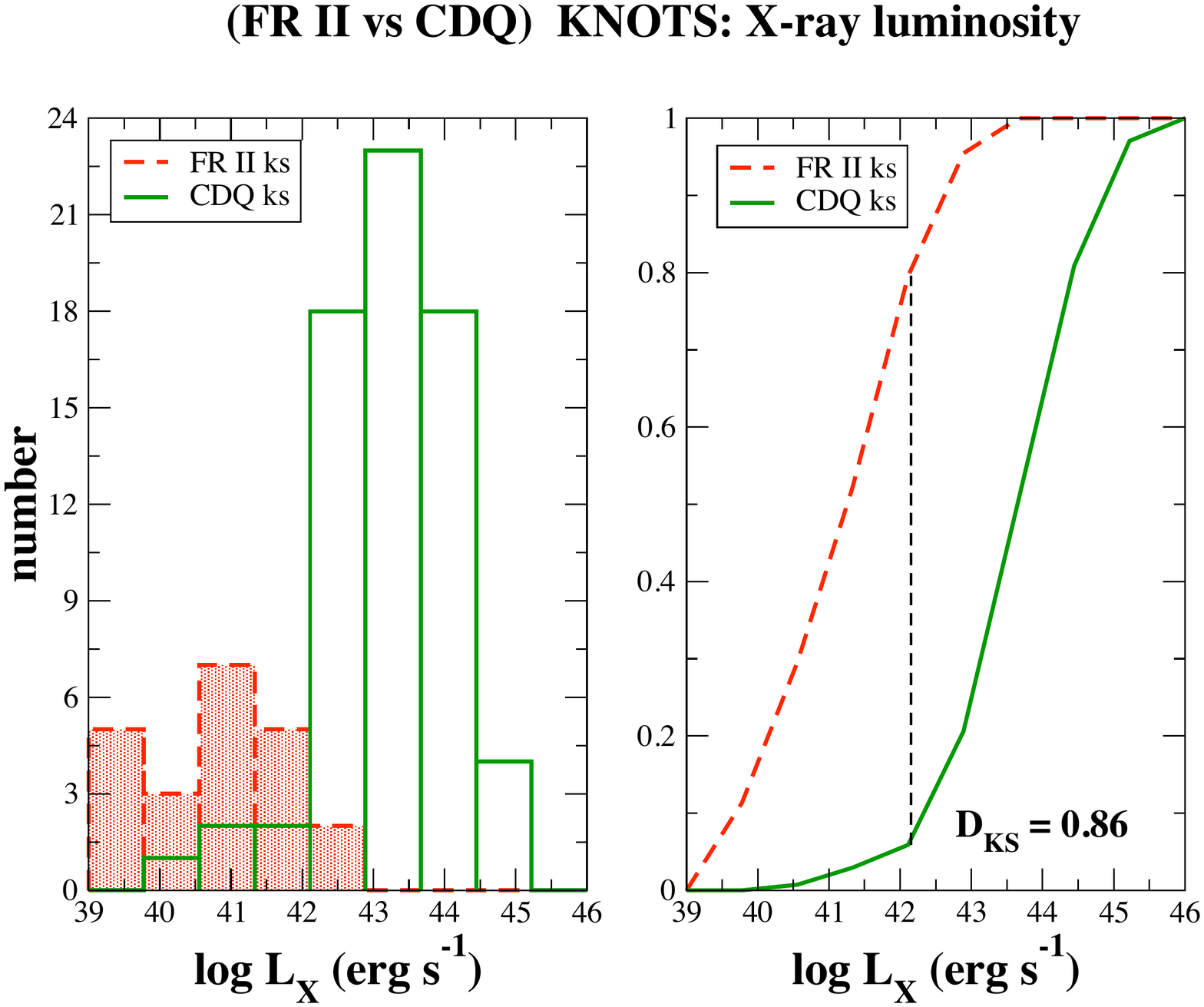}
\includegraphics[height=6.cm,width=6.3cm,angle=0]{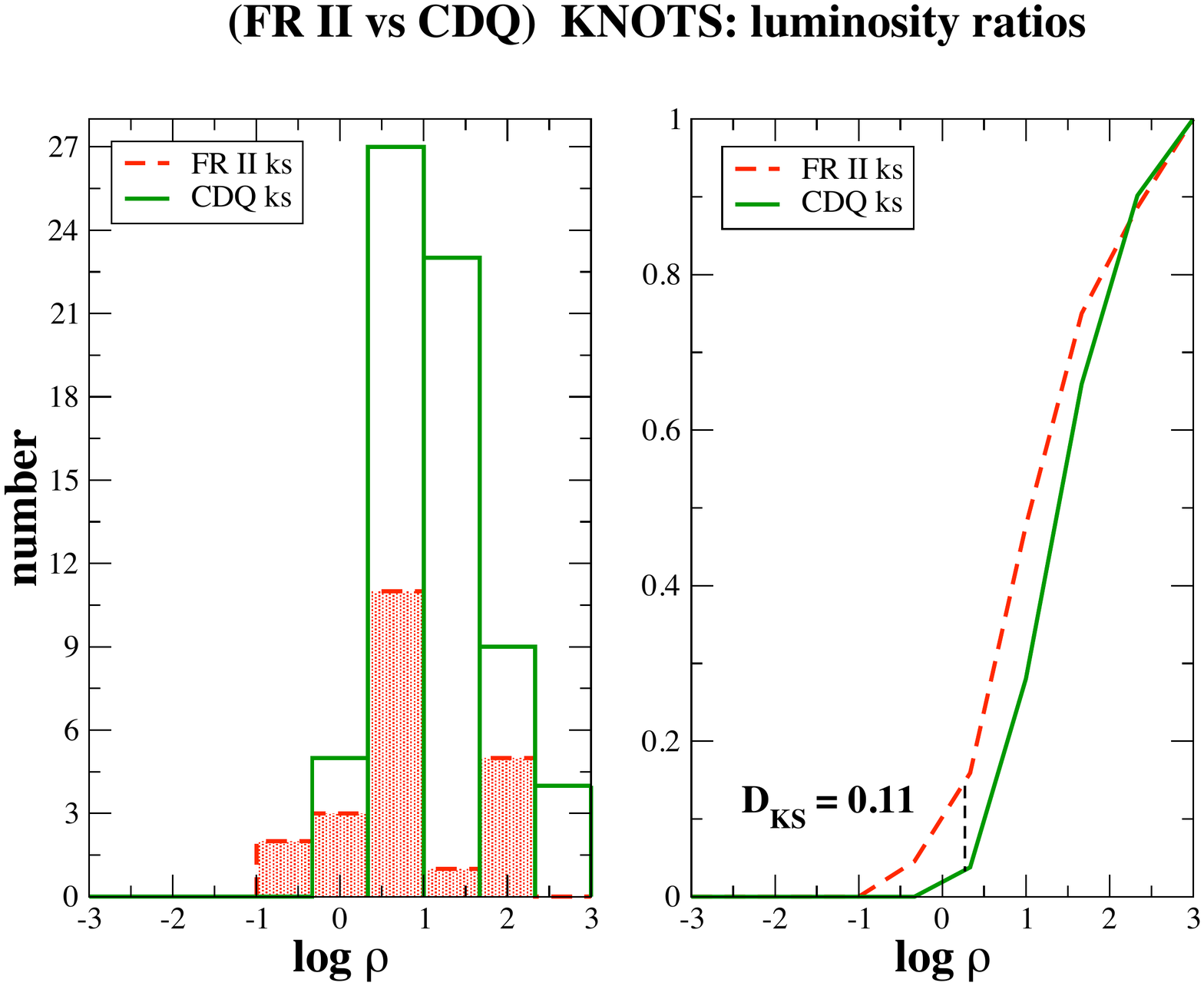}
\end{tabular}
\caption{
a) The distributions of radio luminosities $L_{\rm R}$ of knots in FR\,IIs and CDQs.
b) The normalized cumulative distributions of radio luminosities for knots in FR\,IIs and CDQs.
c) The distributions of X-ray luminosities $L_{\rm X}$ of knots in FR\,IIs and CDQs.
d) The normalized cumulative distributions of X-ray luminosities for knots in FR\,IIs and CDQs.
e) The distributions of $\rho$ of knots in FR\,Is and CDQs.
f) The normalized cumulative distributions of luminosity ratios for knots in FR\,IIs and CDQs.
}
\label{fig:k4}
\end{figure*}

\begin{figure*}
\begin{tabular}{cc}
\includegraphics[height=6.cm,width=6.3cm,angle=0]{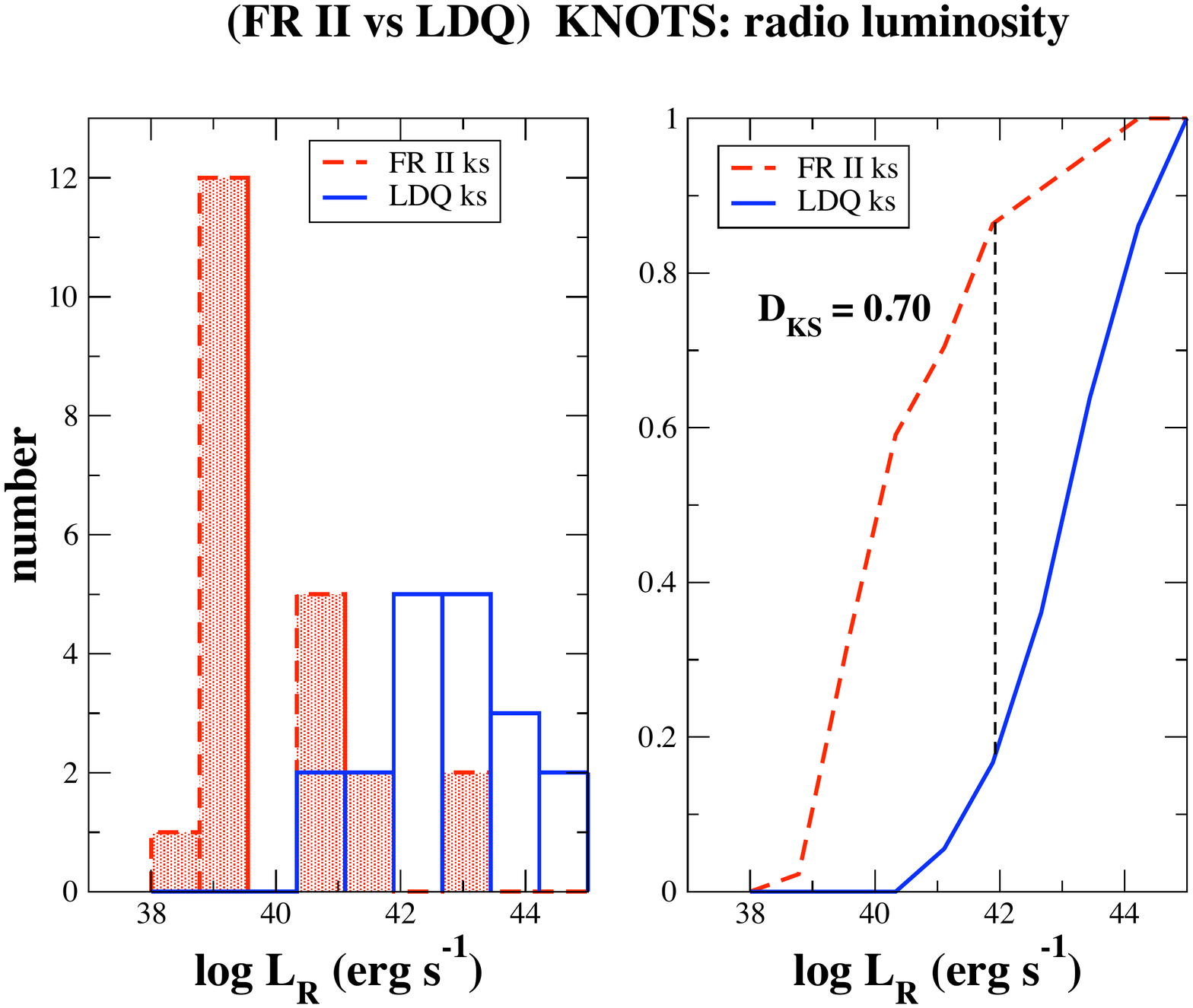}
\includegraphics[height=6.cm,width=6.3cm,angle=0]{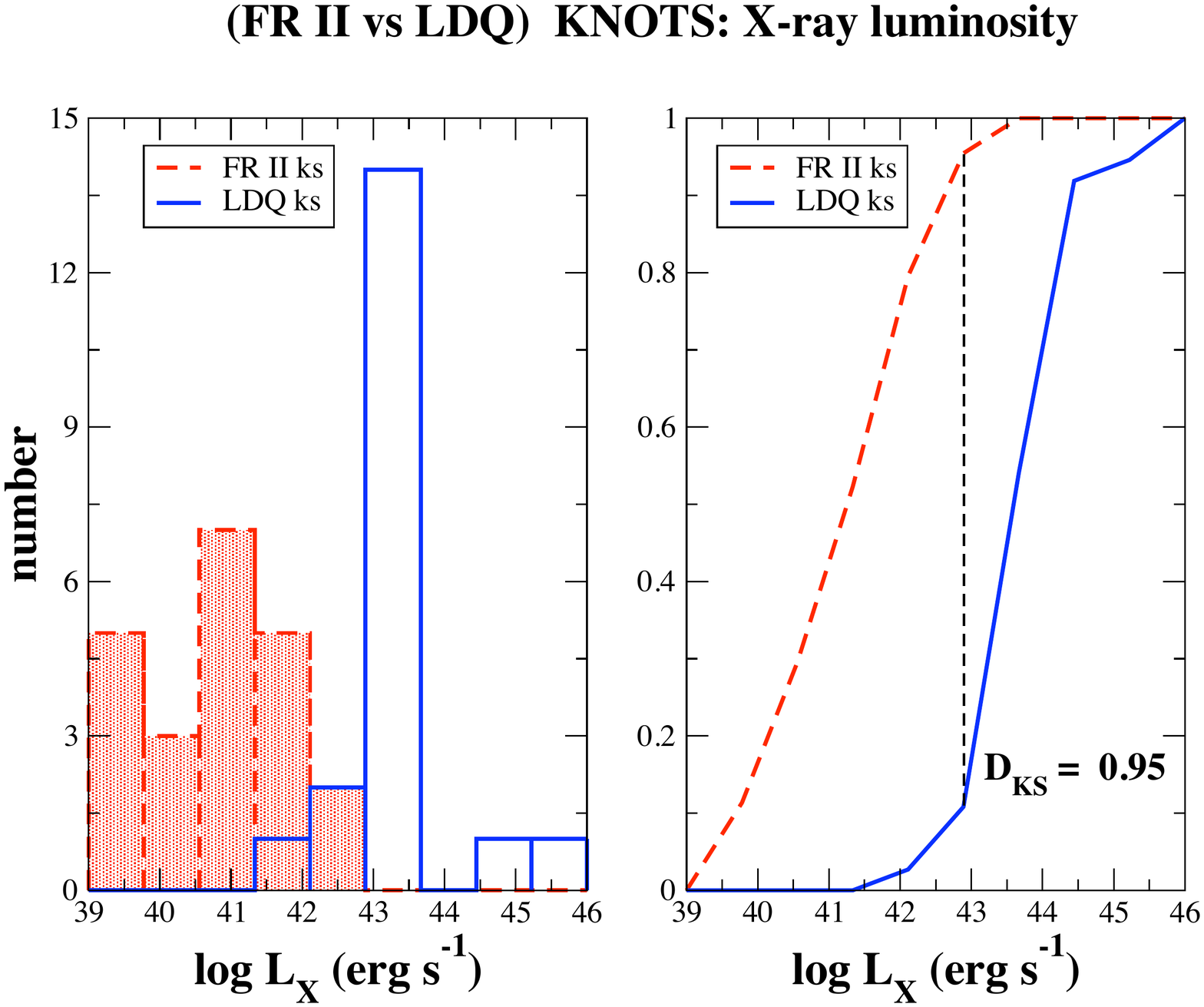}
\includegraphics[height=6.cm,width=6.3cm,angle=0]{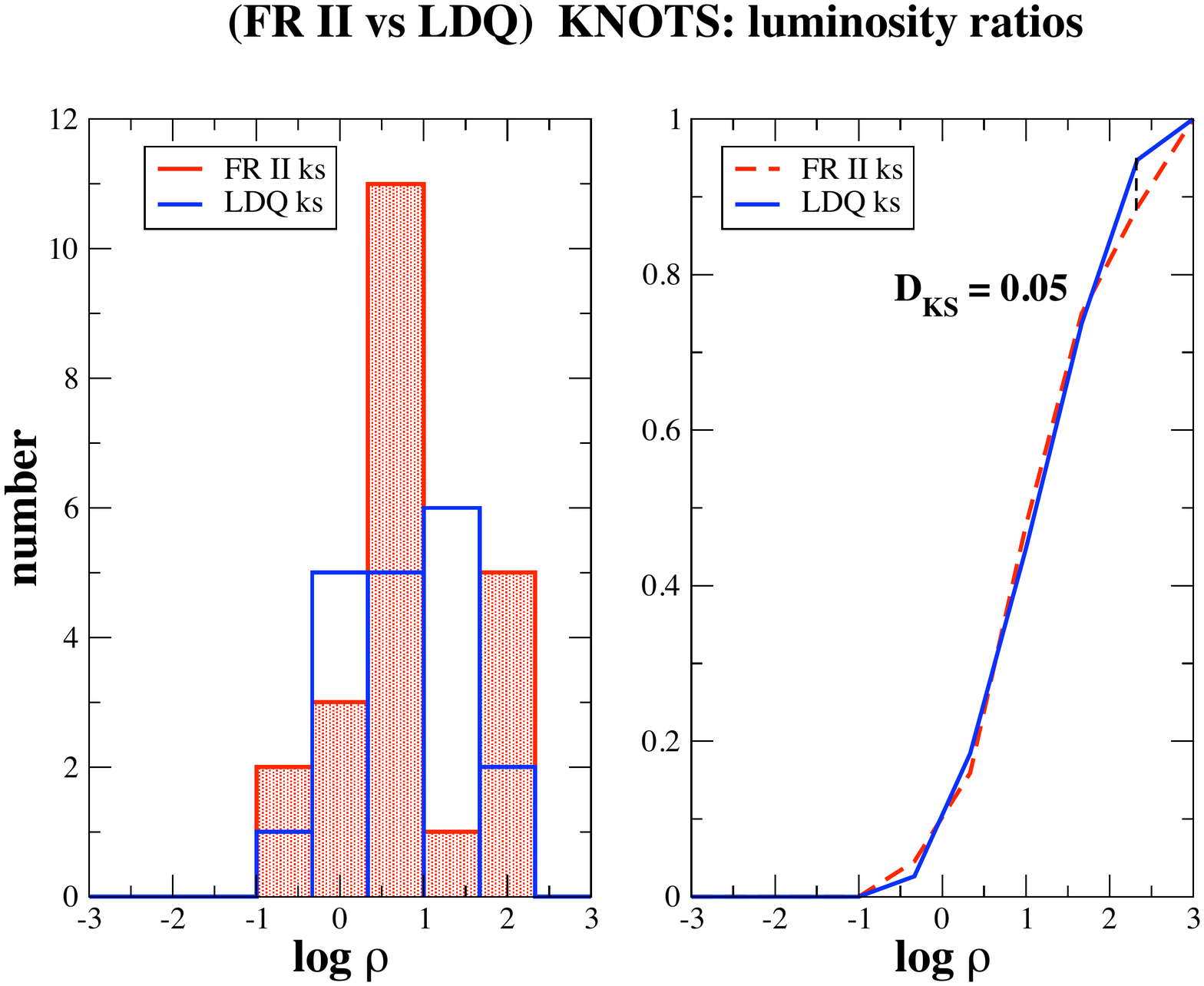}
\end{tabular}
\caption{
a) The distributions of radio luminosities $L_{\rm R}$ of knots in FR\,IIs and LDQs.
b) The normalized cumulative distributions of radio luminosities for knots in FR\,IIs and LDQs.
c) The distributions of X-ray luminosities $L_{\rm X}$ of knots in FR\,IIs and LDQs.
d) The normalized cumulative distributions of X-ray luminosities for knots in FR\,IIs and LDQs.
e) The distributions of $\rho$ of knots in FR\,IIs and LDQs.
f) The normalized cumulative distributions of luminosity ratios for knots in FR\,IIs and LDQs.
}
\label{fig:k2}
\end{figure*}

\begin{figure*}
\begin{tabular}{cc}
\includegraphics[height=6.cm,width=6.3cm,angle=0]{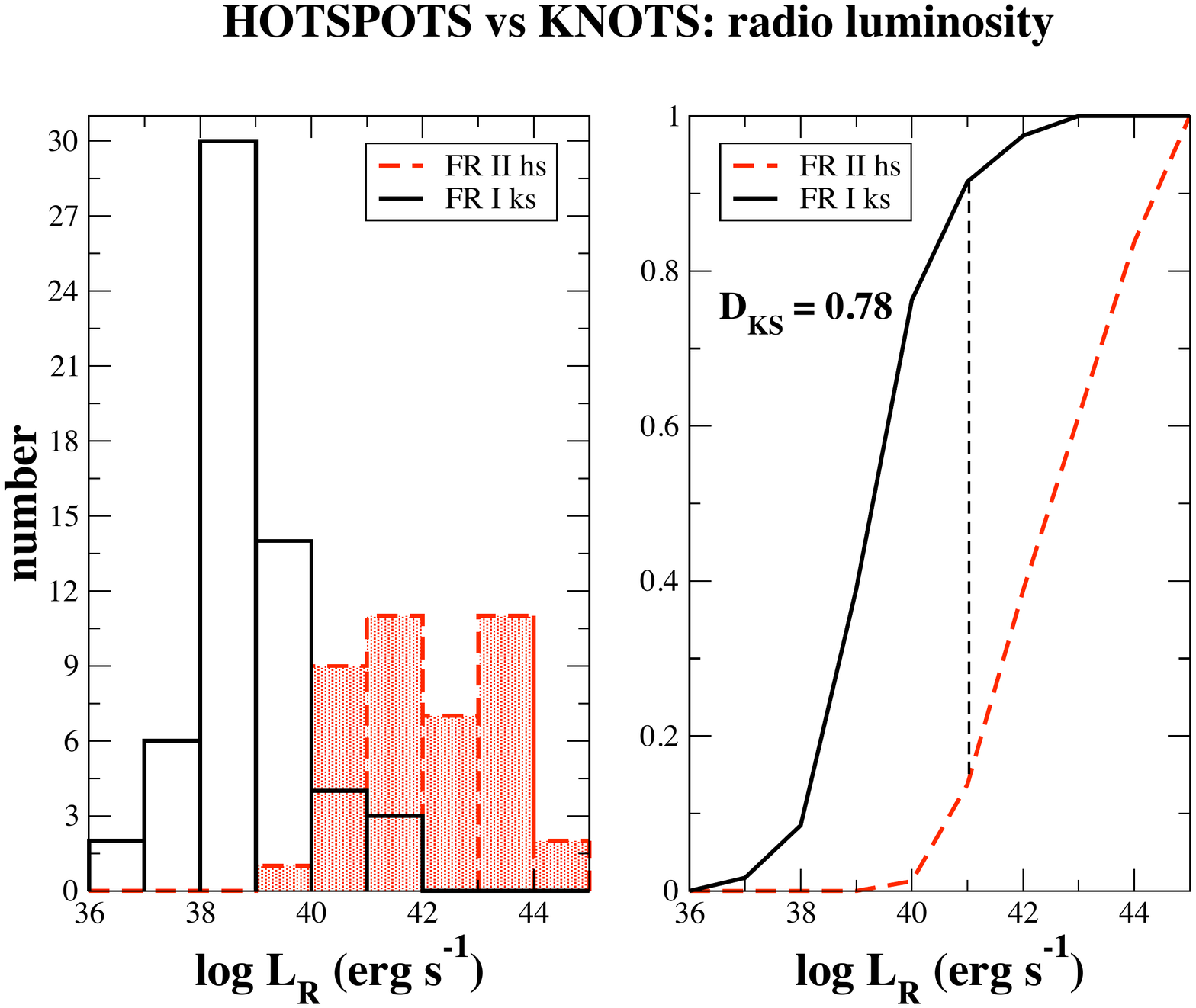}
\includegraphics[height=6.cm,width=6.3cm,angle=0]{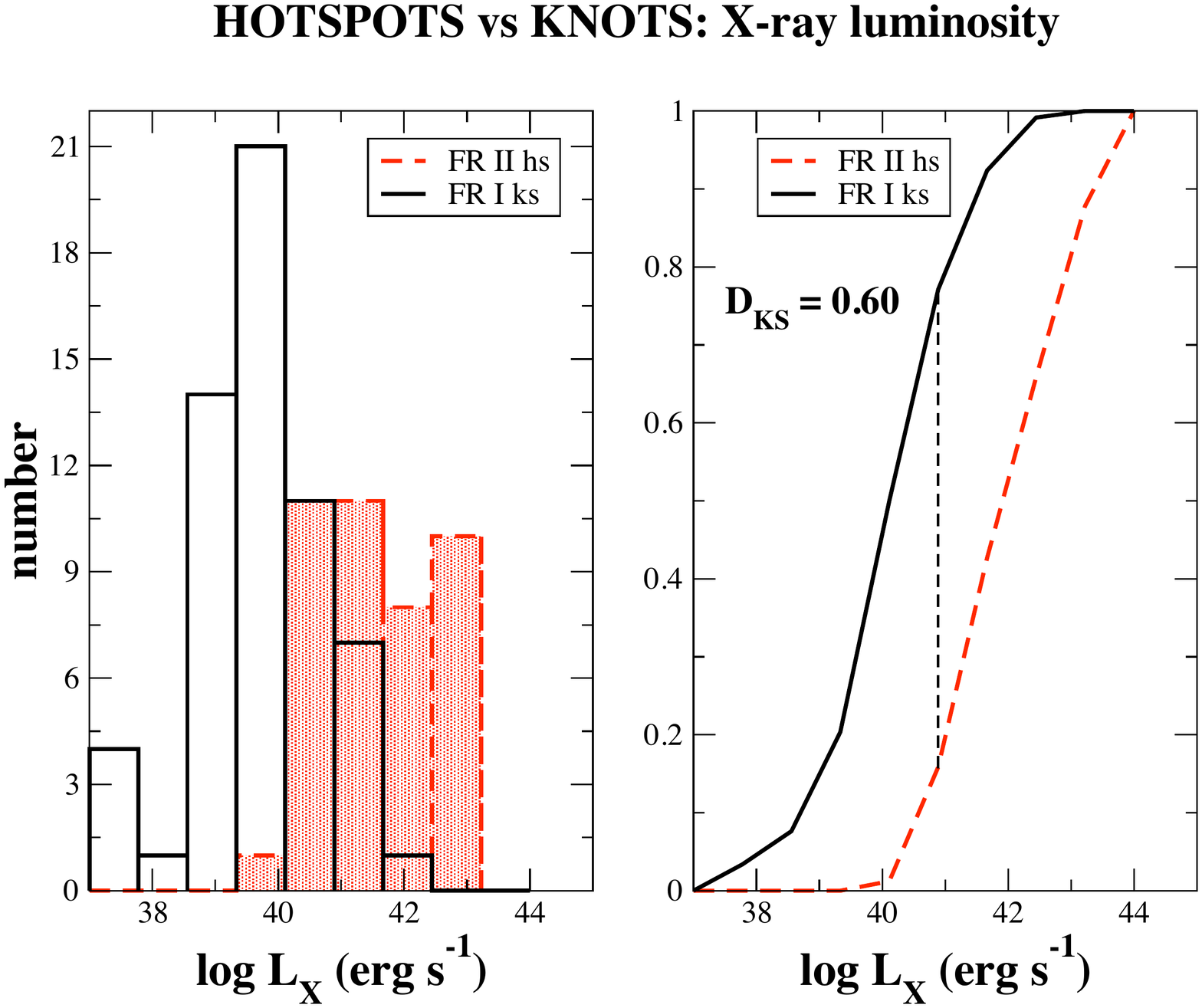}
\includegraphics[height=6.cm,width=6.3cm,angle=0]{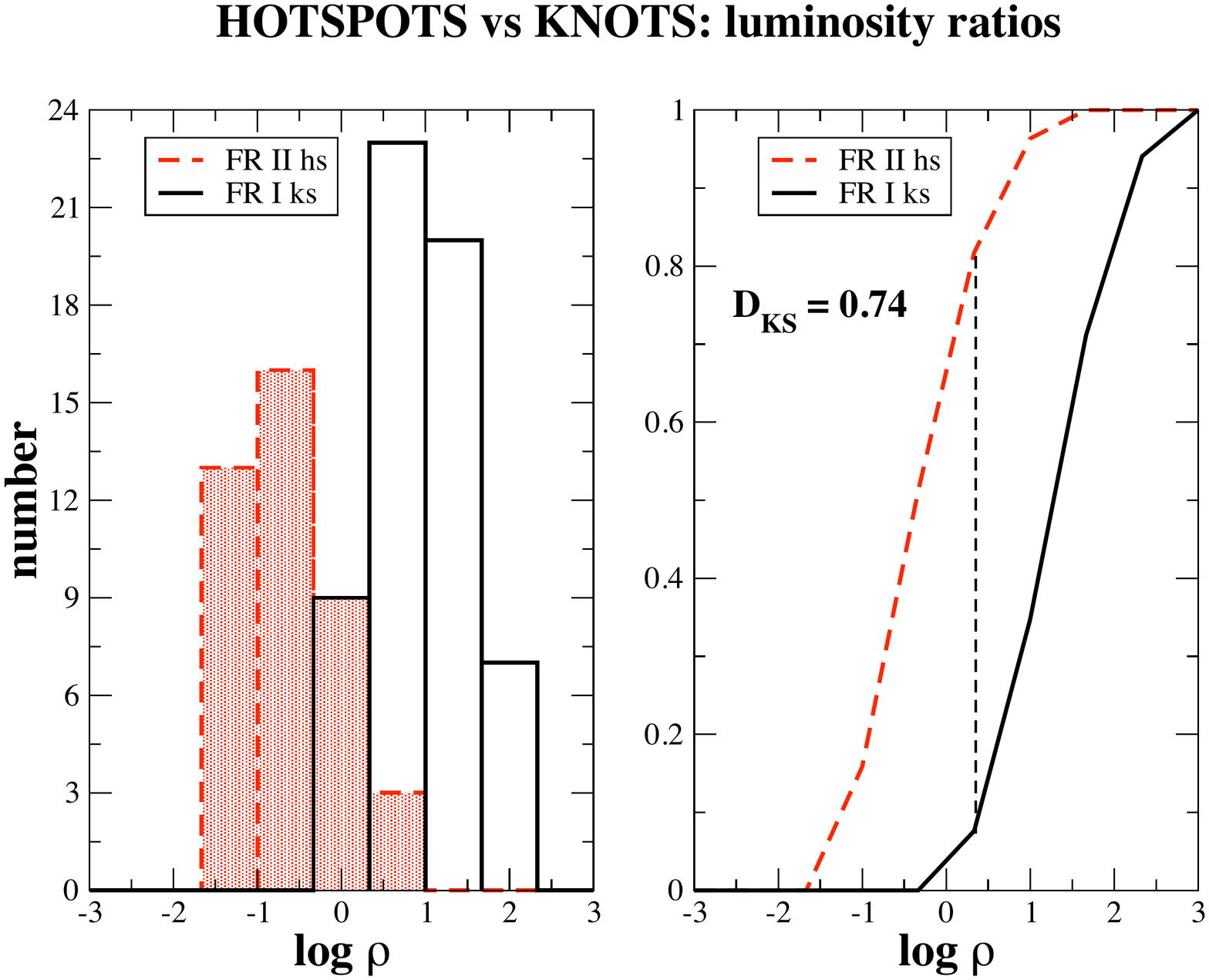}
\end{tabular}
\caption{
a) The distributions of radio luminosities $L_{\rm R}$ of hotspots and knots in FR\,IIs and FR\,Is.
b) The normalized cumulative distributions of radio luminosities for hotspots and knots in FR\,IIs and FR\,Is.
c) The distributions of X-ray luminosities $L_{\rm X}$ of hotspots and knots in FR\,IIs and FR\,Is.
d) The normalized cumulative distributions of X-ray luminosities for hotspots and knots in FR\,IIs and FR\,Is.
e) The distributions of $\rho$ of hotspots and knots in FR\,IIs and FR\,Is.
f) The normalized cumulative distributions of luminosity ratios for hotspots and knots in FR\,IIs and FR\,Is.
}
\label{fig:hk1}
\end{figure*}

\begin{figure*}
\begin{tabular}{cc}
\includegraphics[height=6.cm,width=6.3cm,angle=0]{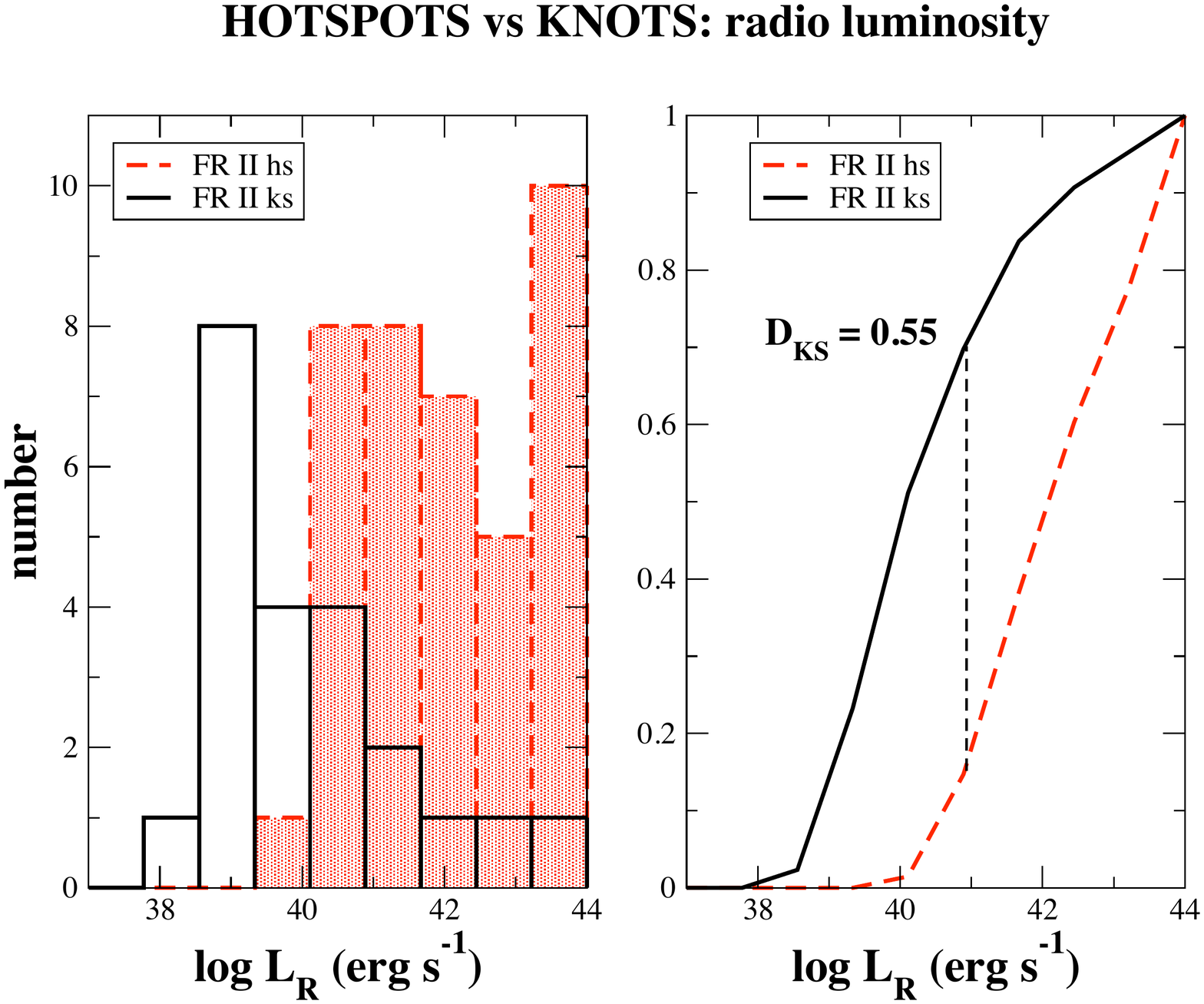}
\includegraphics[height=6.cm,width=6.3cm,angle=0]{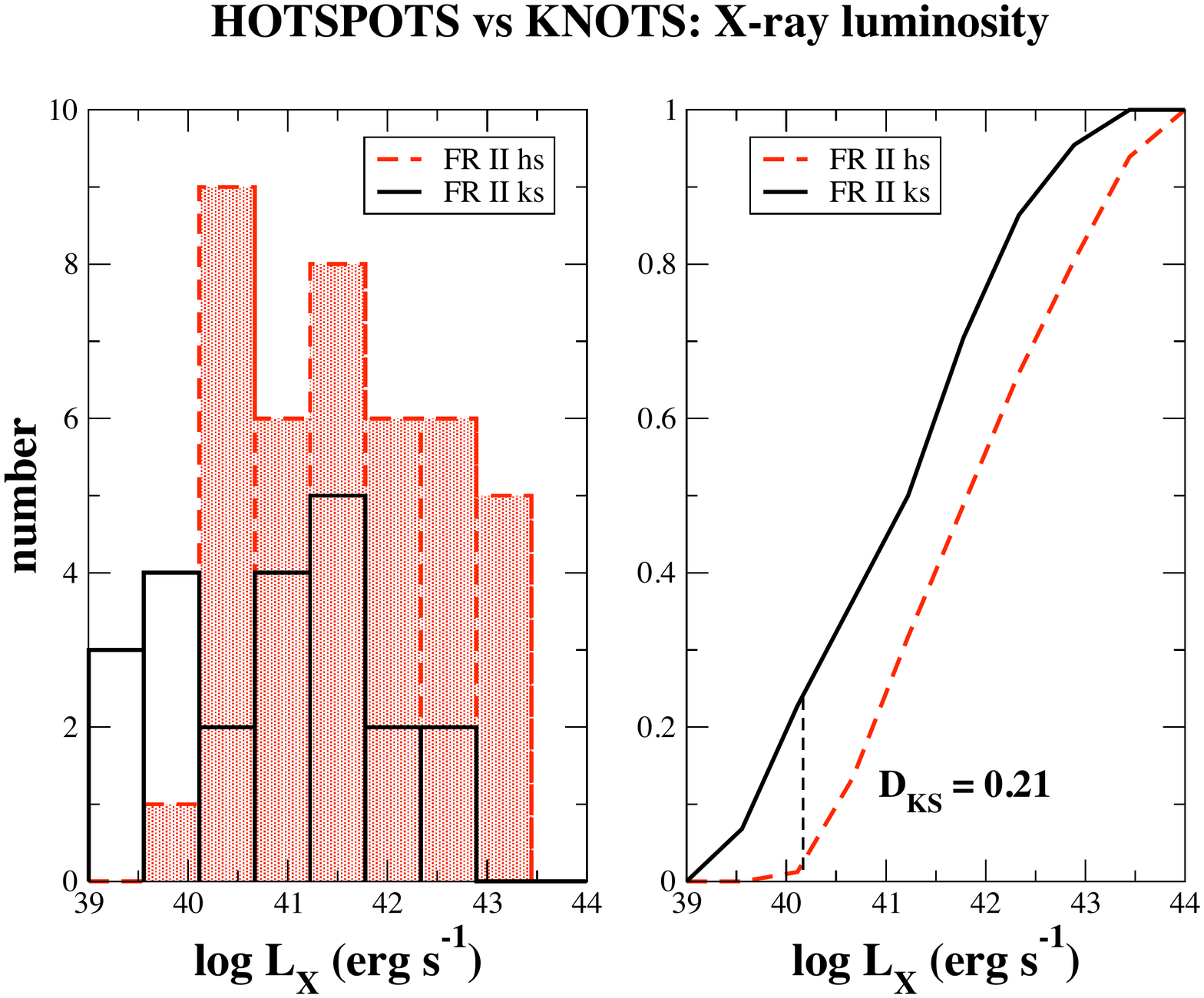}
\includegraphics[height=6.cm,width=6.3cm,angle=0]{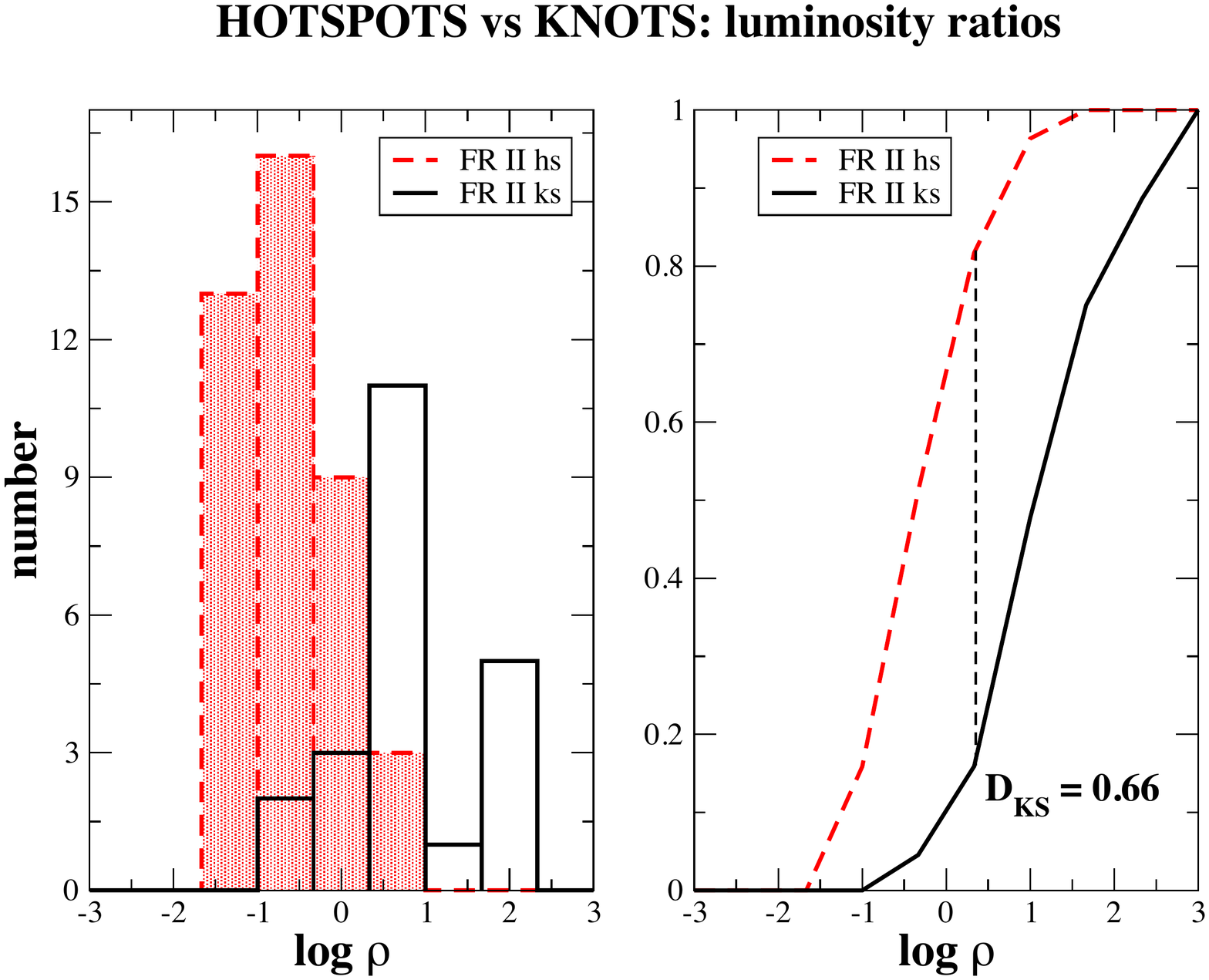}
\end{tabular}
\caption{
a) The distributions of radio luminosities $L_{\rm R}$ of hotspots and knots in FR\,IIs.
b) The normalized cumulative distributions of radio luminosities for hotspots and knots in FR\,IIs.
c) The distributions of X-ray luminosities $L_{\rm X}$ of hotspots and knots in FR\,IIs.
d) The normalized cumulative distributions of X-ray luminosities for hotspots and knots in FR\,IIs.
e) The distributions of $\rho$ of hotspots and knots in FR\,IIs.
f) The normalized cumulative distributions of luminosity ratios for hotspots and knots in FR\,IIs.
(Note the different color convention adopted for this figure).
}
\label{fig:hk2}
\end{figure*}

\begin{figure*}
\begin{tabular}{cc}
\includegraphics[height=6.cm,width=6.3cm,angle=0]{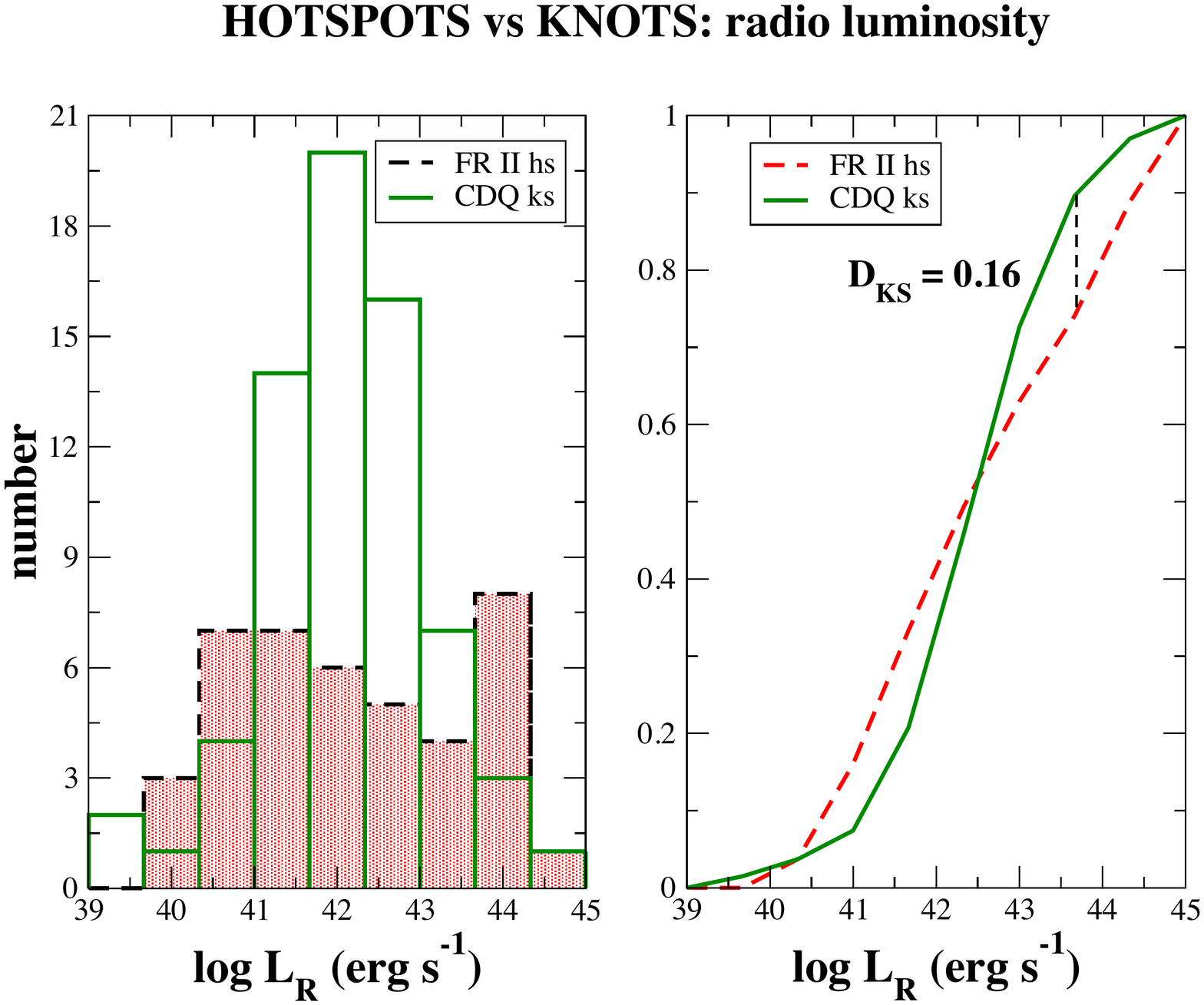}
\includegraphics[height=6.cm,width=6.3cm,angle=0]{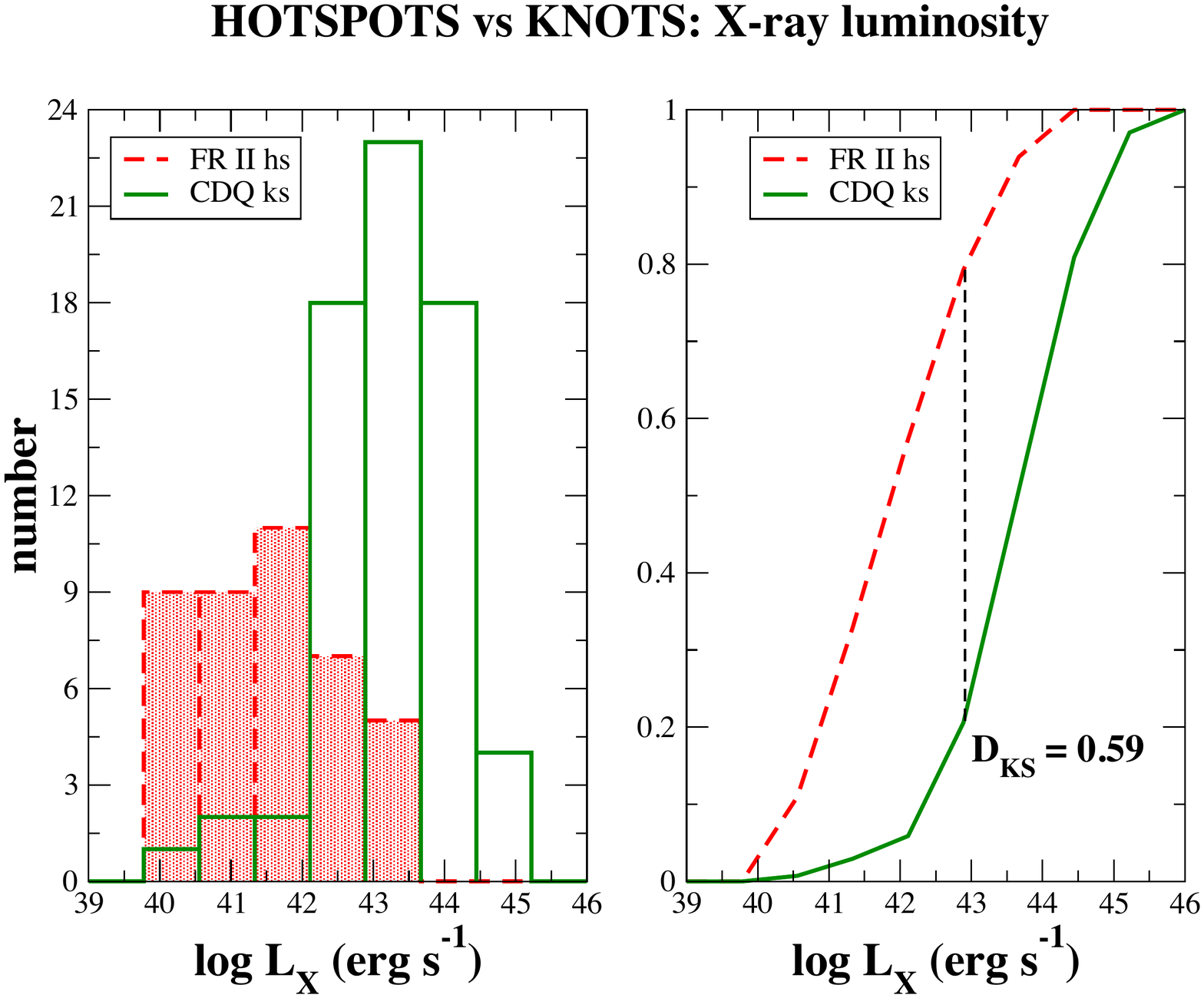}
\includegraphics[height=6.cm,width=6.3cm,angle=0]{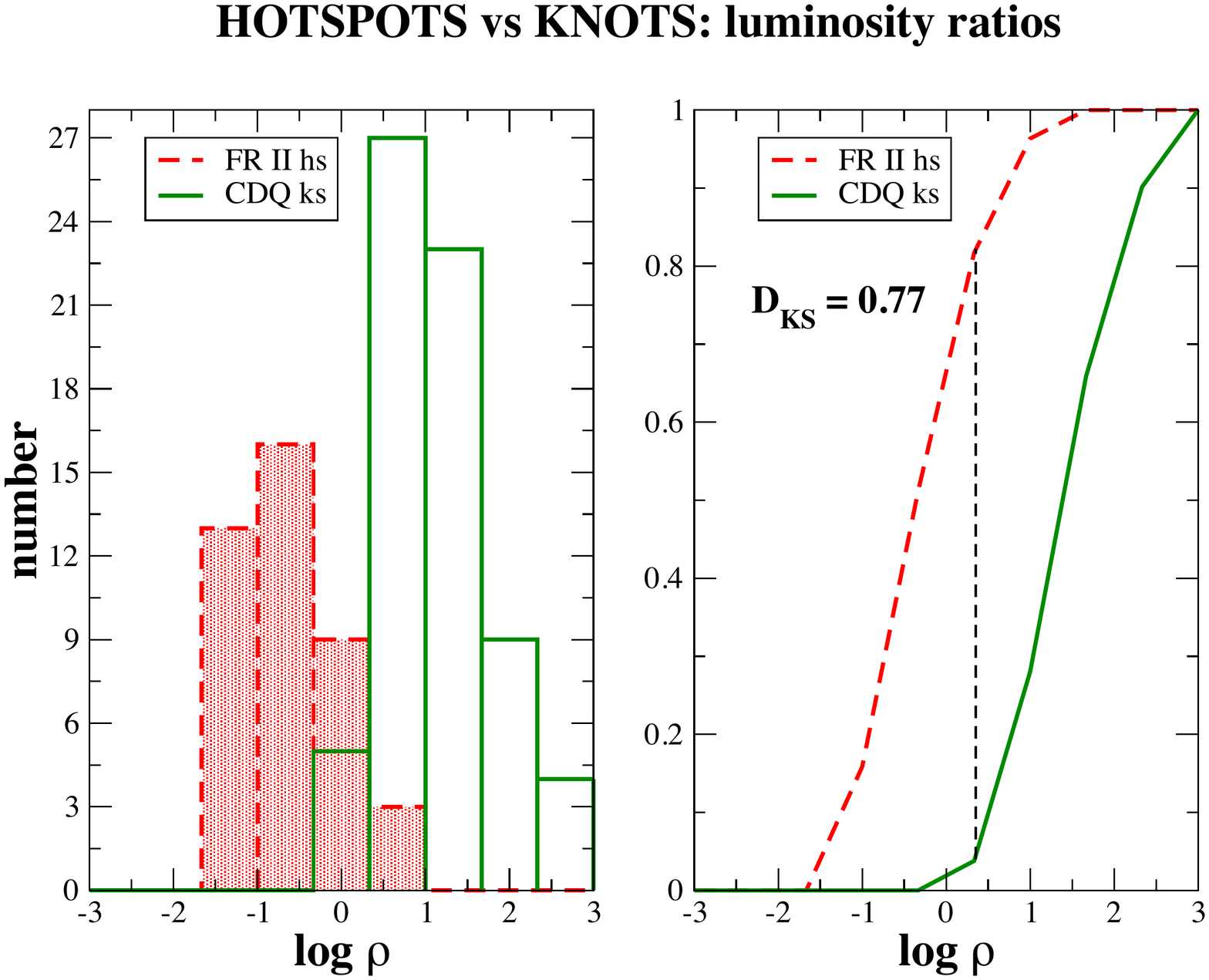}
\end{tabular}
\caption{
a) The distributions of radio luminosities $L_{\rm R}$ of hotspots and knots in FR\,IIs and CDQs.
b) The normalized cumulative distributions of radio luminosities for hotspots and knots in FR\,IIs and CDQs.
c) The distributions of X-ray luminosities $L_{\rm X}$ of hotspots and knots in FR\,IIs and CDQs.
d) The normalized cumulative distributions of X-ray luminosities for hotspots and knots in FR\,IIs and CDQs.
e) The distributions of $\rho$ of hotspots and knots in FR\,IIs and CDQs.
f) The normalized cumulative distributions of luminosity ratios for hotspots and knots in FR\,IIs and CDQs.
}
\label{fig:hk3}
\end{figure*}

\begin{figure*}
\begin{tabular}{cc}
\includegraphics[height=6.cm,width=6.3cm,angle=0]{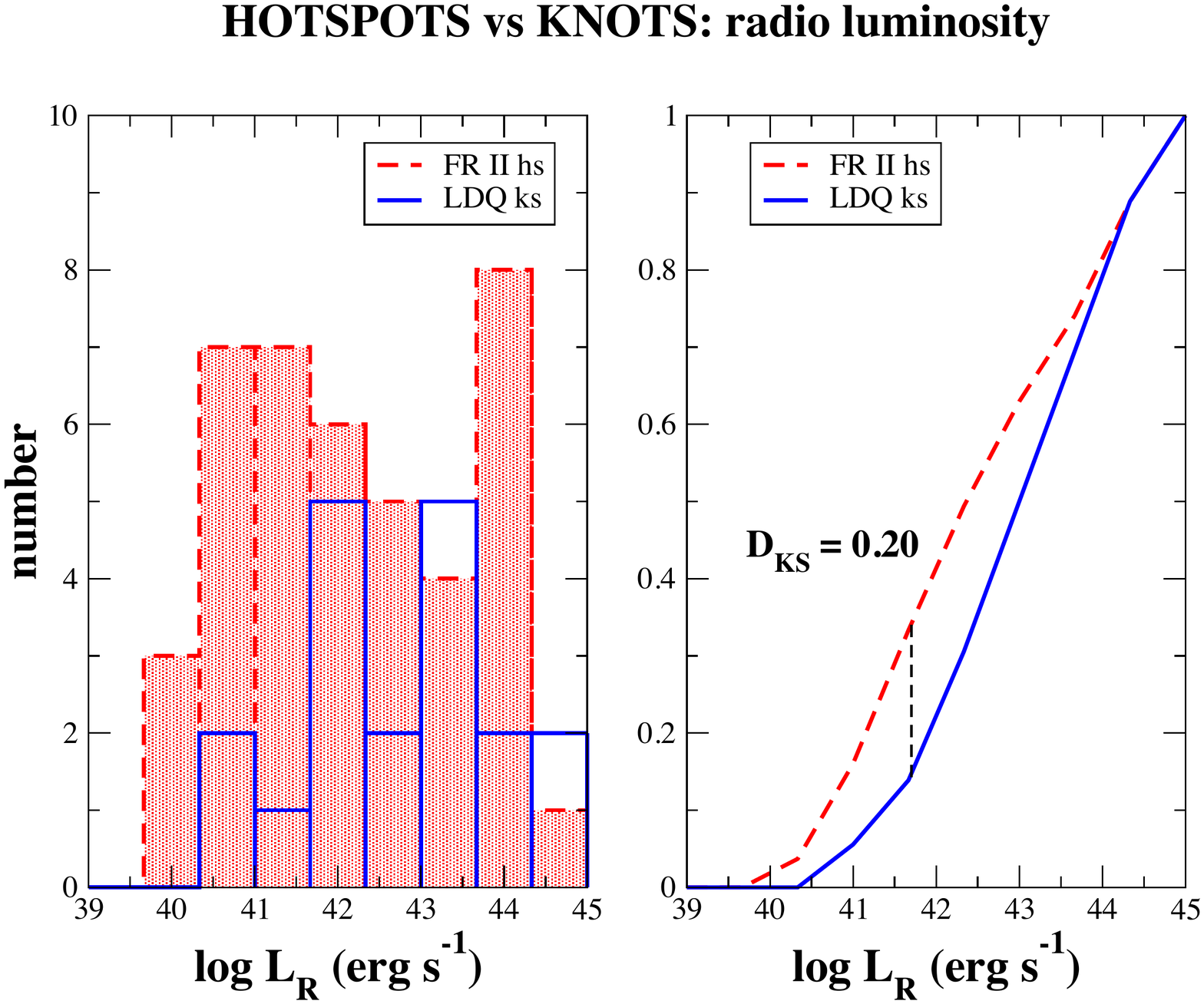}
\includegraphics[height=6.cm,width=6.3cm,angle=0]{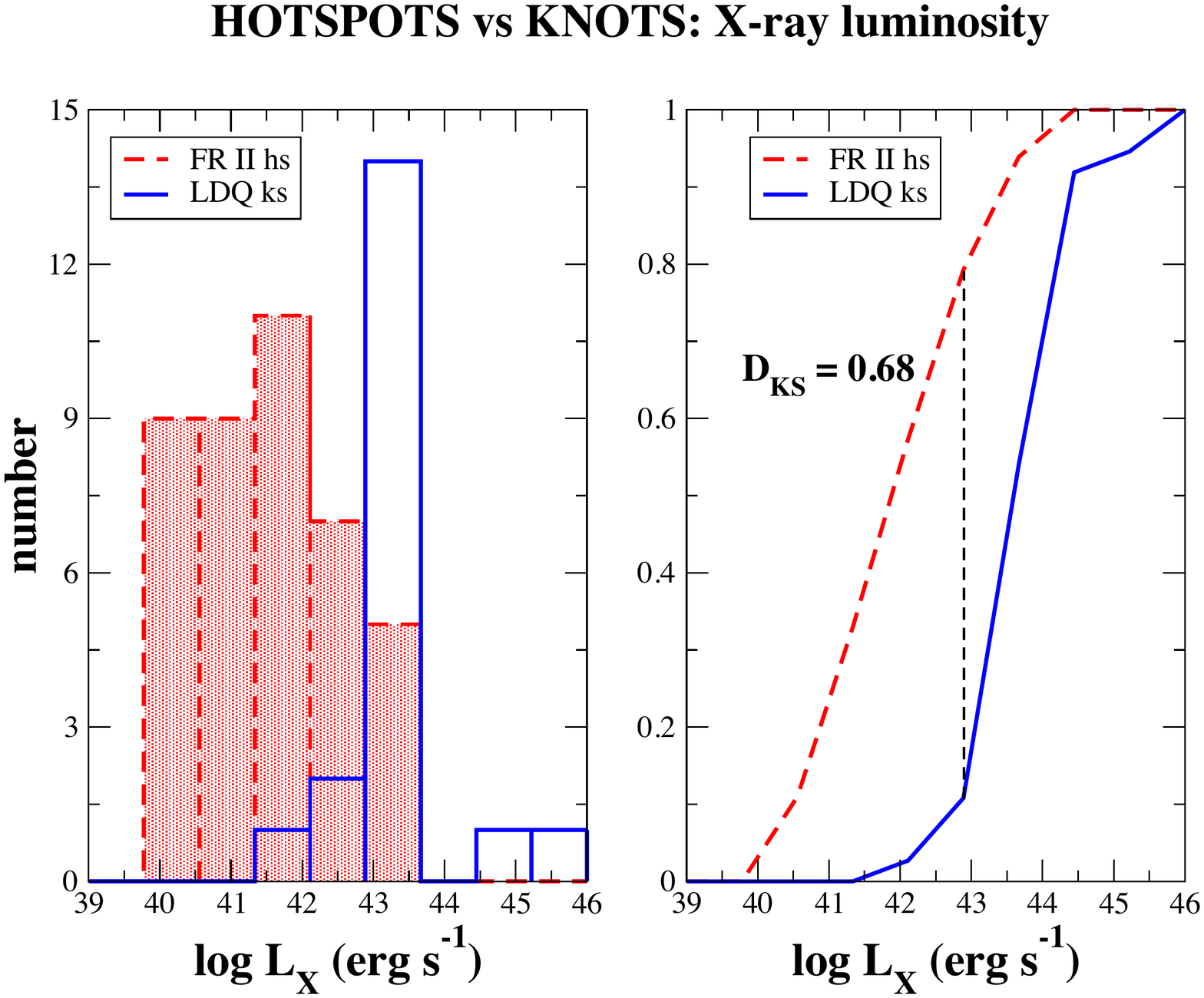}
\includegraphics[height=6.cm,width=6.3cm,angle=0]{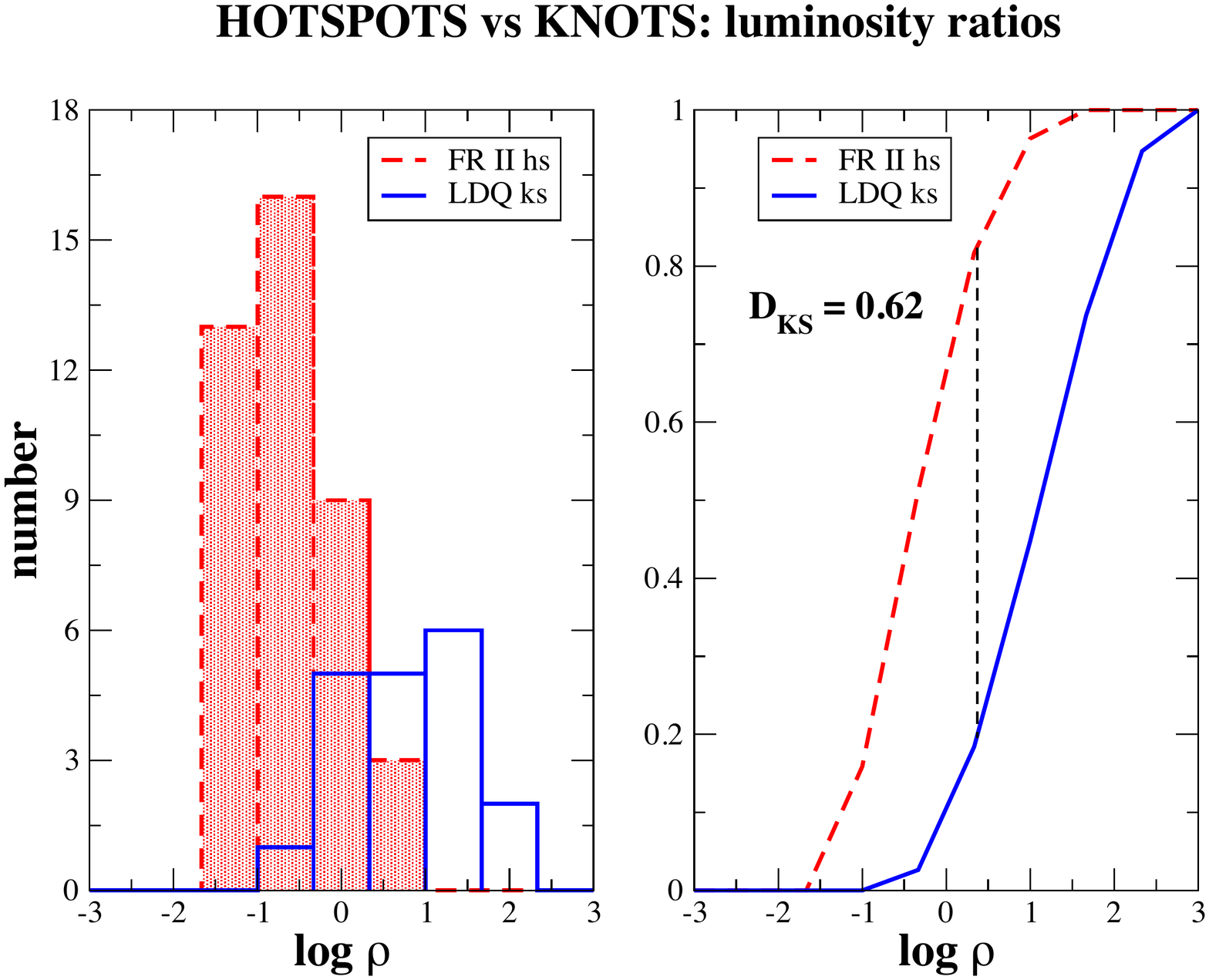}
\end{tabular}
\caption{
a) The distributions of radio luminosities $L_{\rm R}$ of hotspots and knots in FR\,IIs and LDQs.
b) The normalized cumulative distributions of radio luminosities for hotspots and knots in FR\,IIs and LDQs.
c) The distributions of X-ray luminosities $L_{\rm X}$ of hotspots and knots in FR\,IIs and LDQs.
d) The normalized cumulative distributions of X-ray luminosities for hotspots and knots in FR\,IIs and LDQs.
e) The distributions of $\rho$ of hotspots and knots in FR\,IIs and LDQs.
f) The normalized cumulative distributions of luminosity ratios for hotspots and knots in FR\,IIs and LDQs.
}
\label{fig:hk4}
\end{figure*}

\begin{figure*}
\begin{tabular}{cc}
\includegraphics[height=6.cm,width=6.3cm,angle=0]{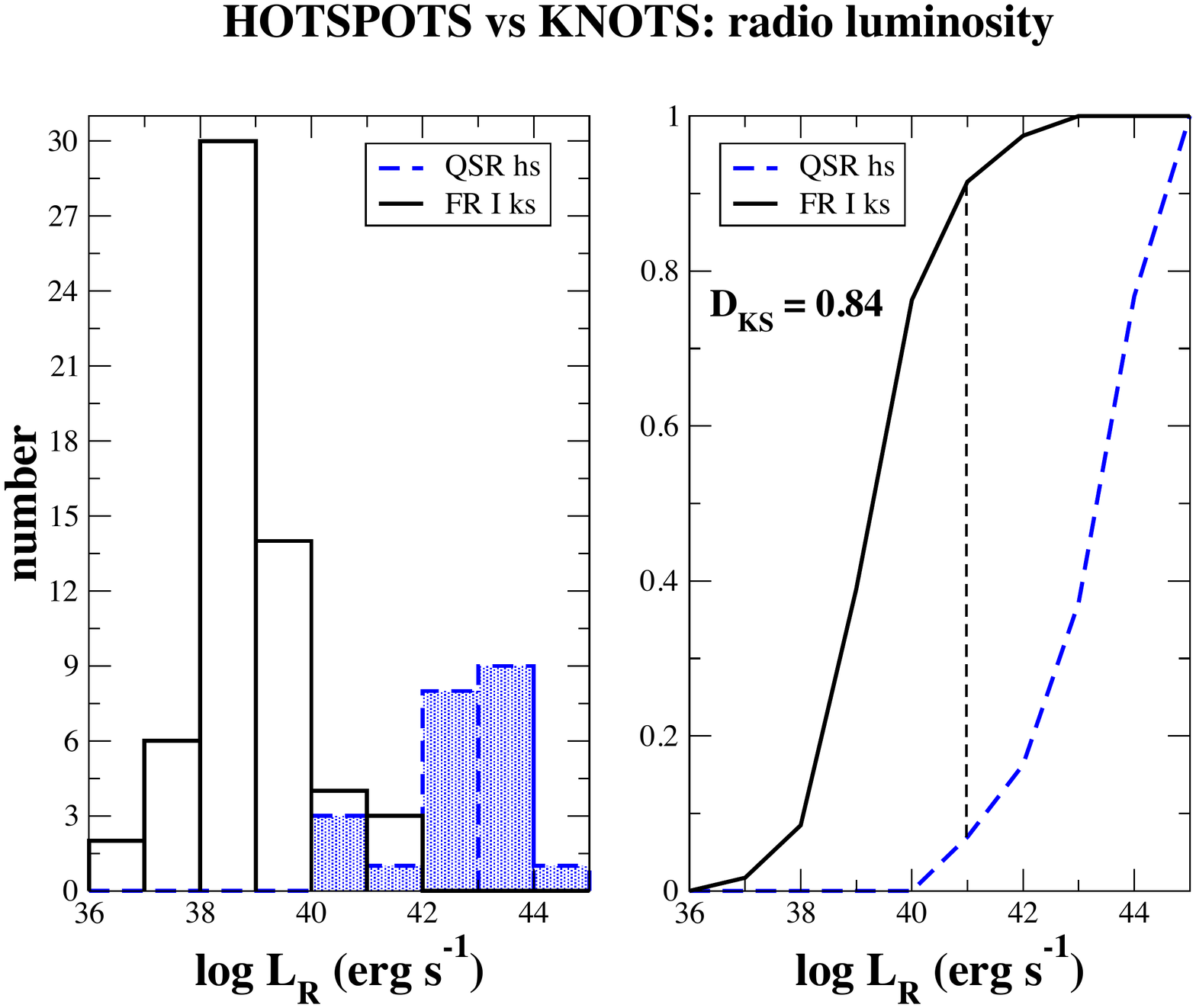}
\includegraphics[height=6.cm,width=6.3cm,angle=0]{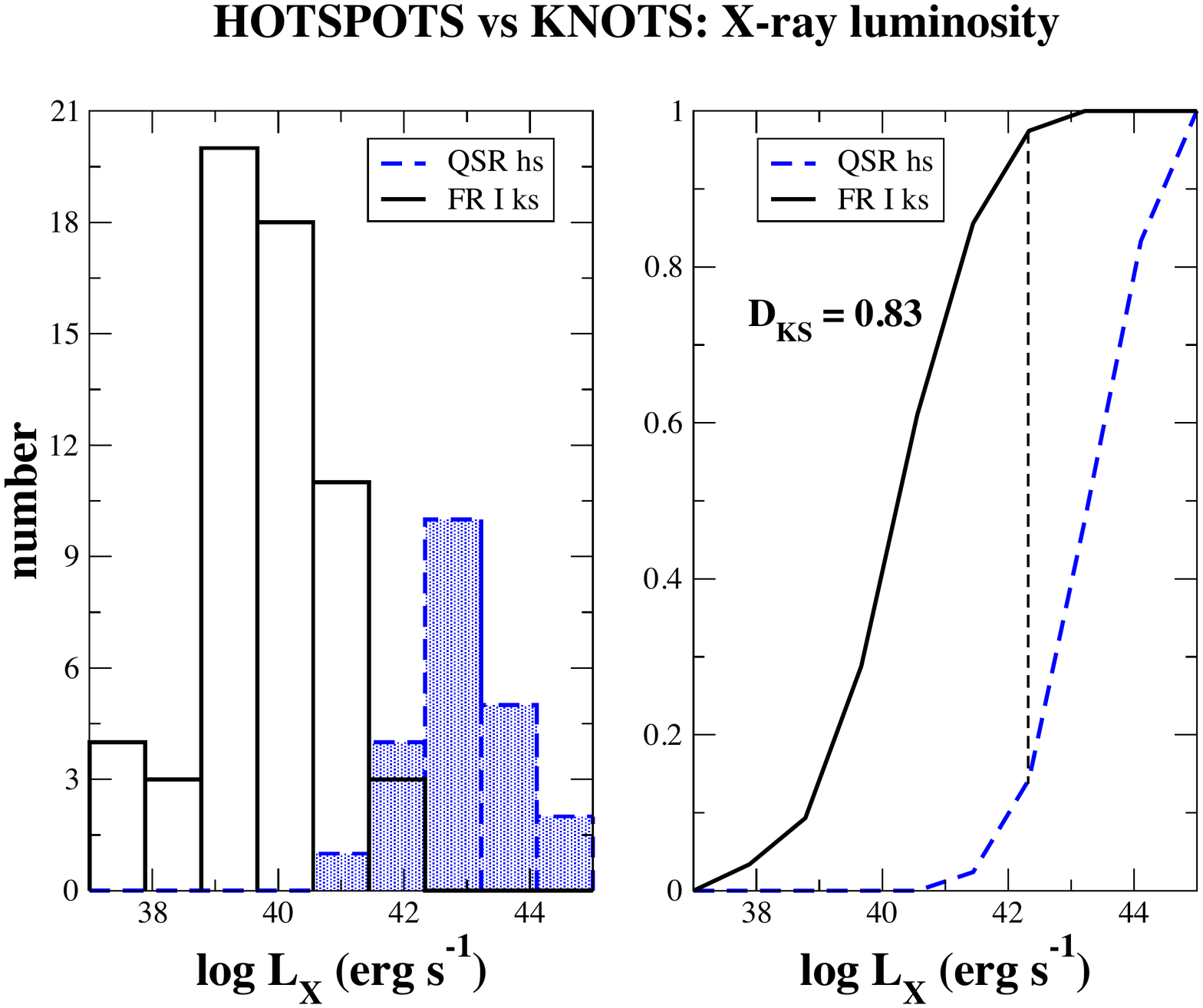}
\includegraphics[height=6.cm,width=6.3cm,angle=0]{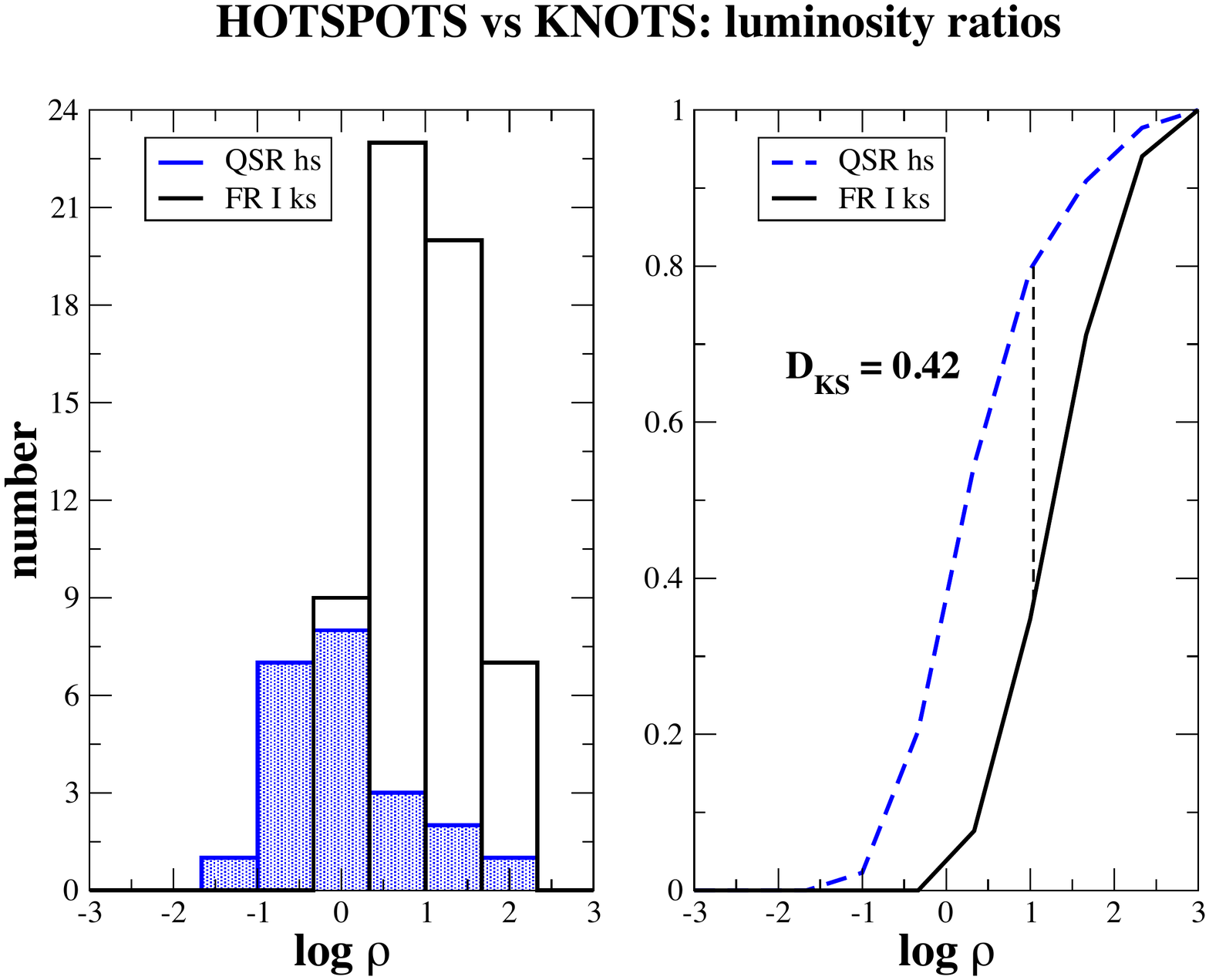}
\end{tabular}
\caption{
a) The distributions of radio luminosities $L_{\rm R}$ of hotspots and knots in LDQs and FR\,Is.
b) The normalized cumulative distributions of radio luminosities for hotspots and knots in LDQs and FR\,Is.
c) The distributions of X-ray luminosities $L_{\rm X}$ of hotspots and knots in LDQs and FR\,Is.
d) The normalized cumulative distributions of X-ray luminosities for hotspots and knots in LDQs and FR\,Is.
e) The distributions of $\rho$ of hotspots and knots in LDQs and FR\,Is.
f) The normalized cumulative distributions of luminosity ratios for hotspots and knots in LDQs and FR\,Is.
}
\label{fig:hk5}
\end{figure*}

\begin{figure*}
\begin{tabular}{cc}
\includegraphics[height=6.cm,width=6.3cm,angle=0]{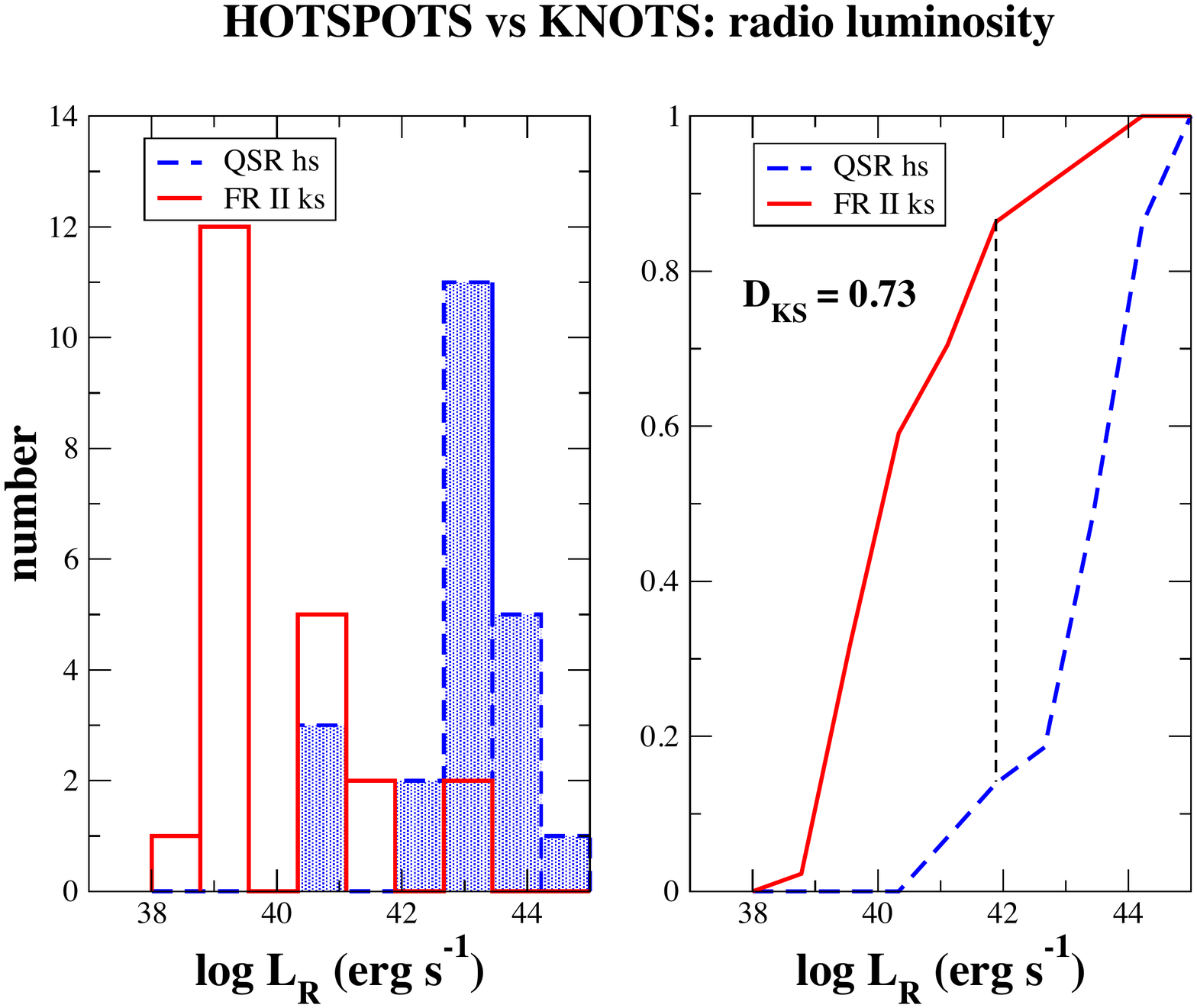}
\includegraphics[height=6.cm,width=6.3cm,angle=0]{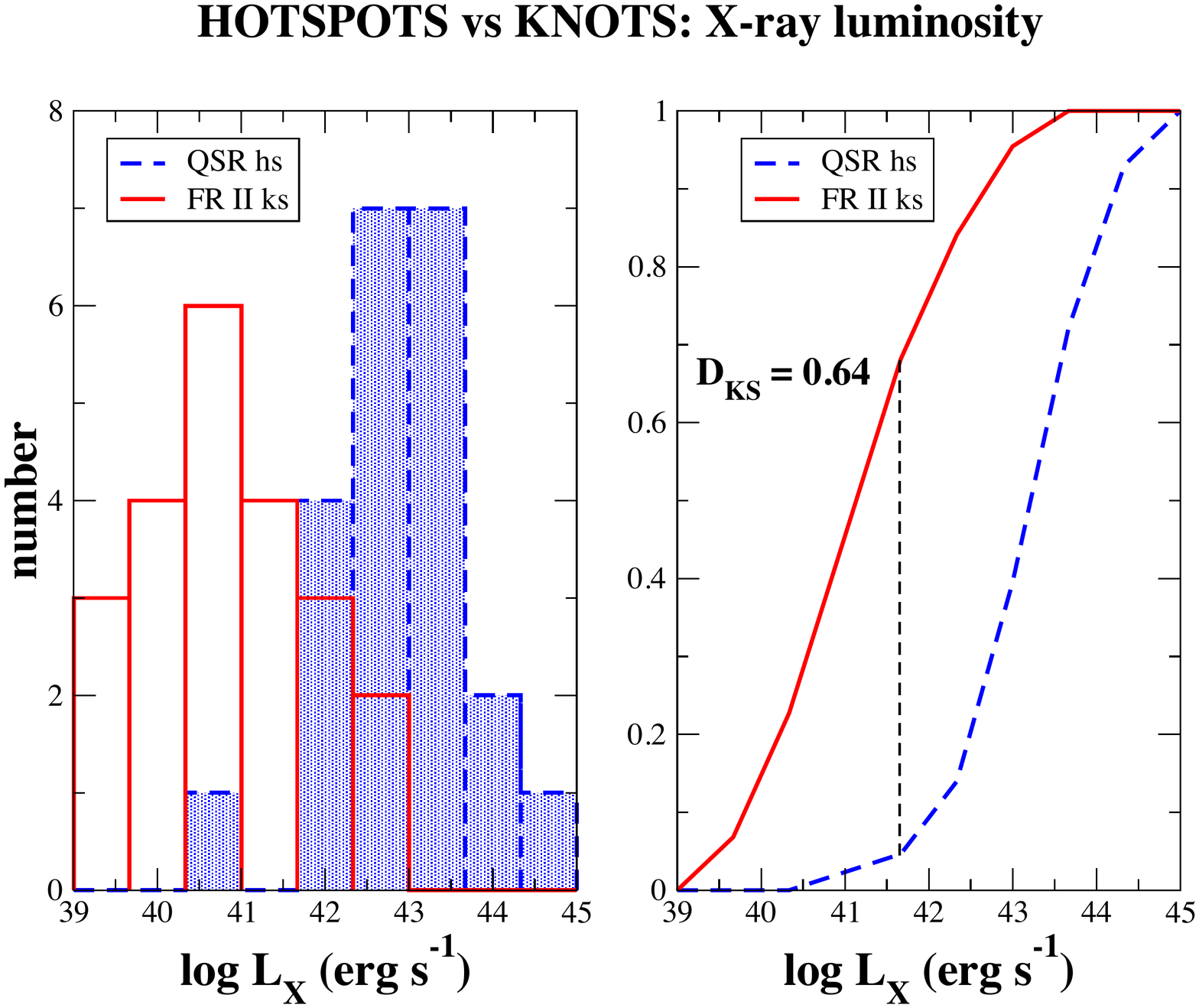}
\includegraphics[height=6.cm,width=6.3cm,angle=0]{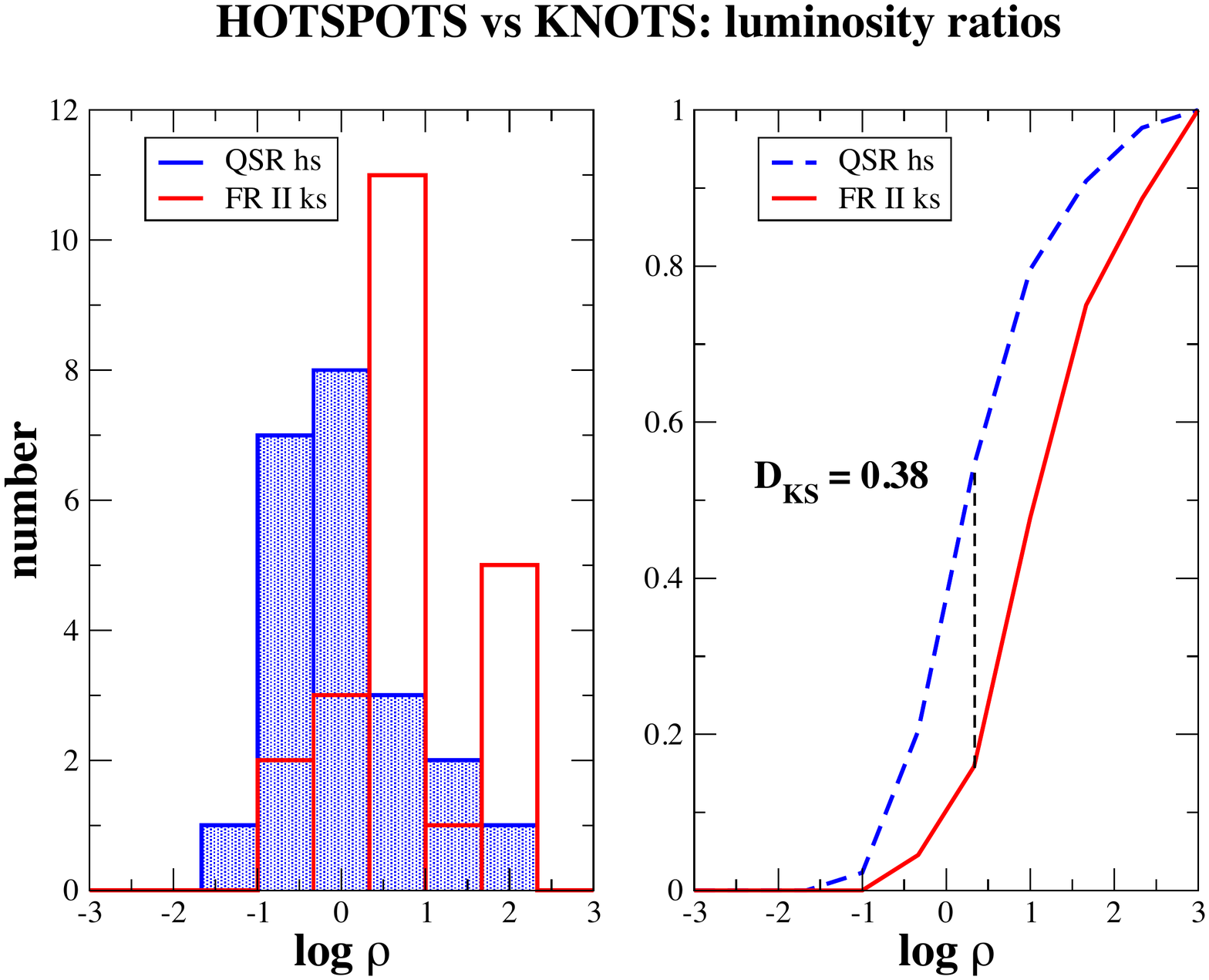}
\end{tabular}
\caption{
a) The distributions of radio luminosities $L_{\rm R}$ of hotspots and knots in LDQs and FR\,IIs.
b) The normalized cumulative distributions of radio luminosities for hotspots and knots in LDQs and FR\,IIs.
c) The distributions of X-ray luminosities $L_{\rm X}$ of hotspots and knots in LDQs and FR\,IIs.
d) The normalized cumulative distributions of X-ray luminosities for hotspots and knots in LDQs and FR\,IIs.
e) The distributions of $\rho$ of hotspots and knots in LDQs and FR\,IIs.
f) The normalized cumulative distributions of luminosity ratios for hotspots and knots in LDQs and FR\,IIs.
}
\label{fig:hk6}
\end{figure*}

\begin{figure*}
\begin{tabular}{cc}
\includegraphics[height=6.cm,width=6.3cm,angle=0]{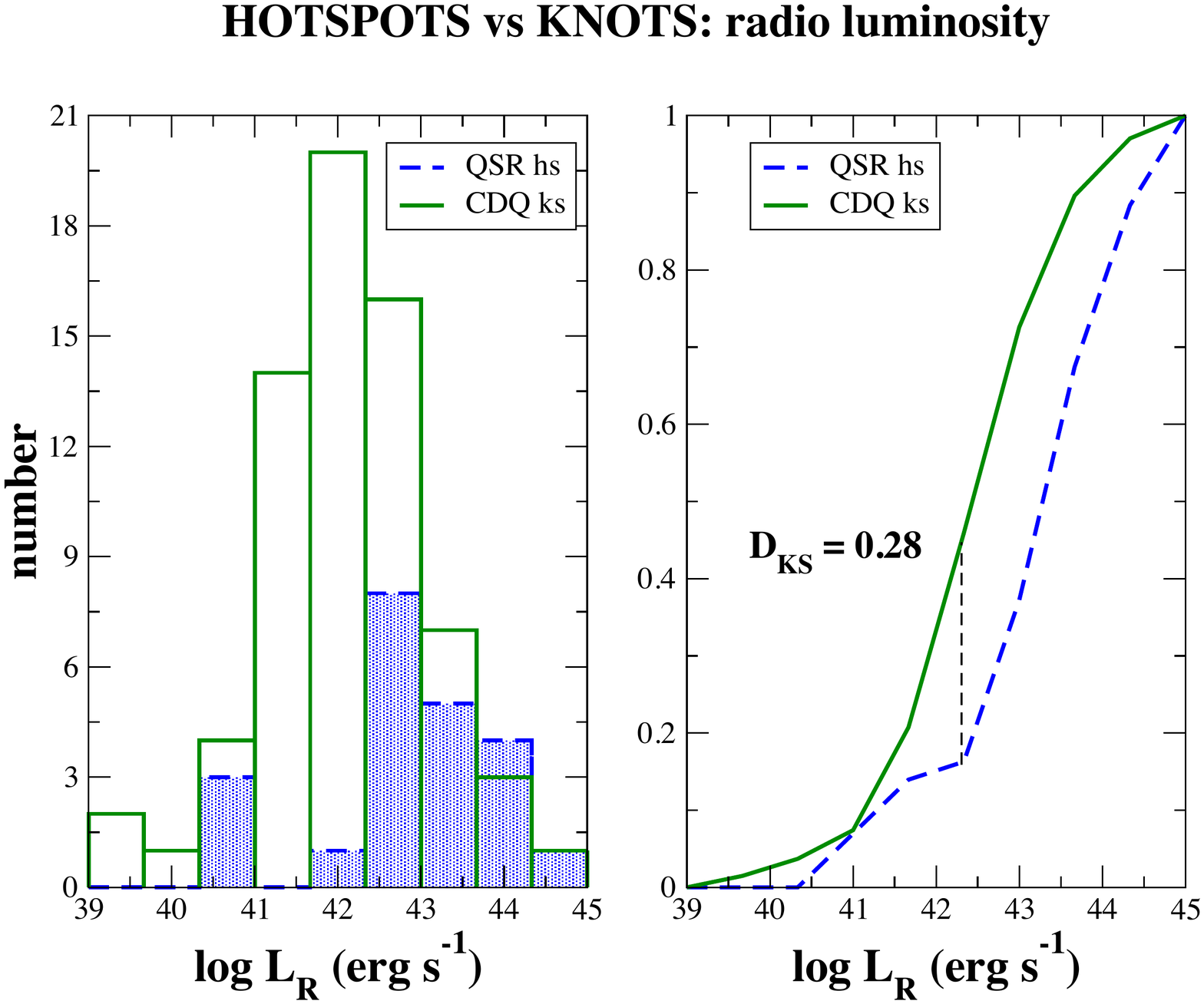}
\includegraphics[height=6.cm,width=6.3cm,angle=0]{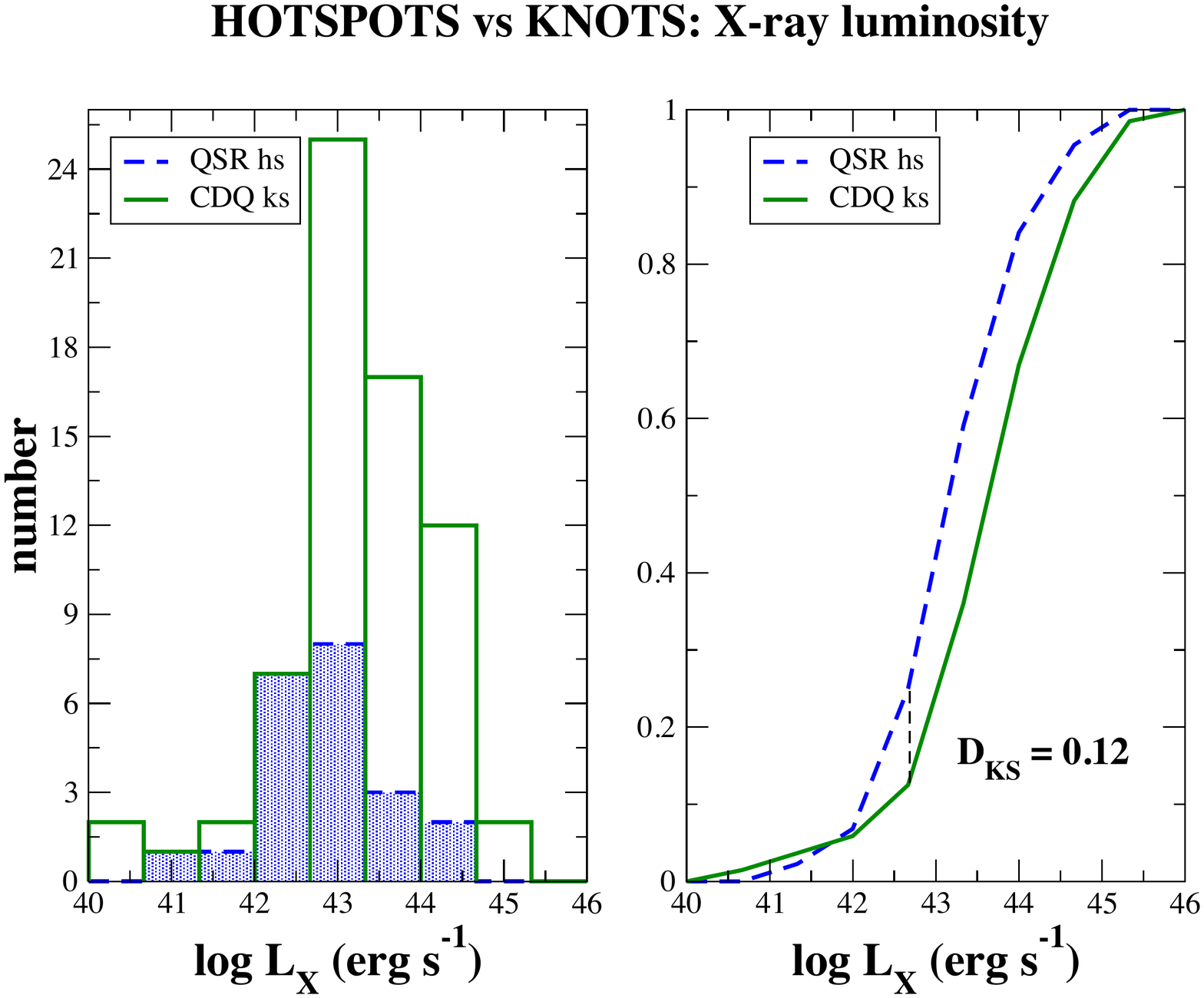}
\includegraphics[height=6.cm,width=6.3cm,angle=0]{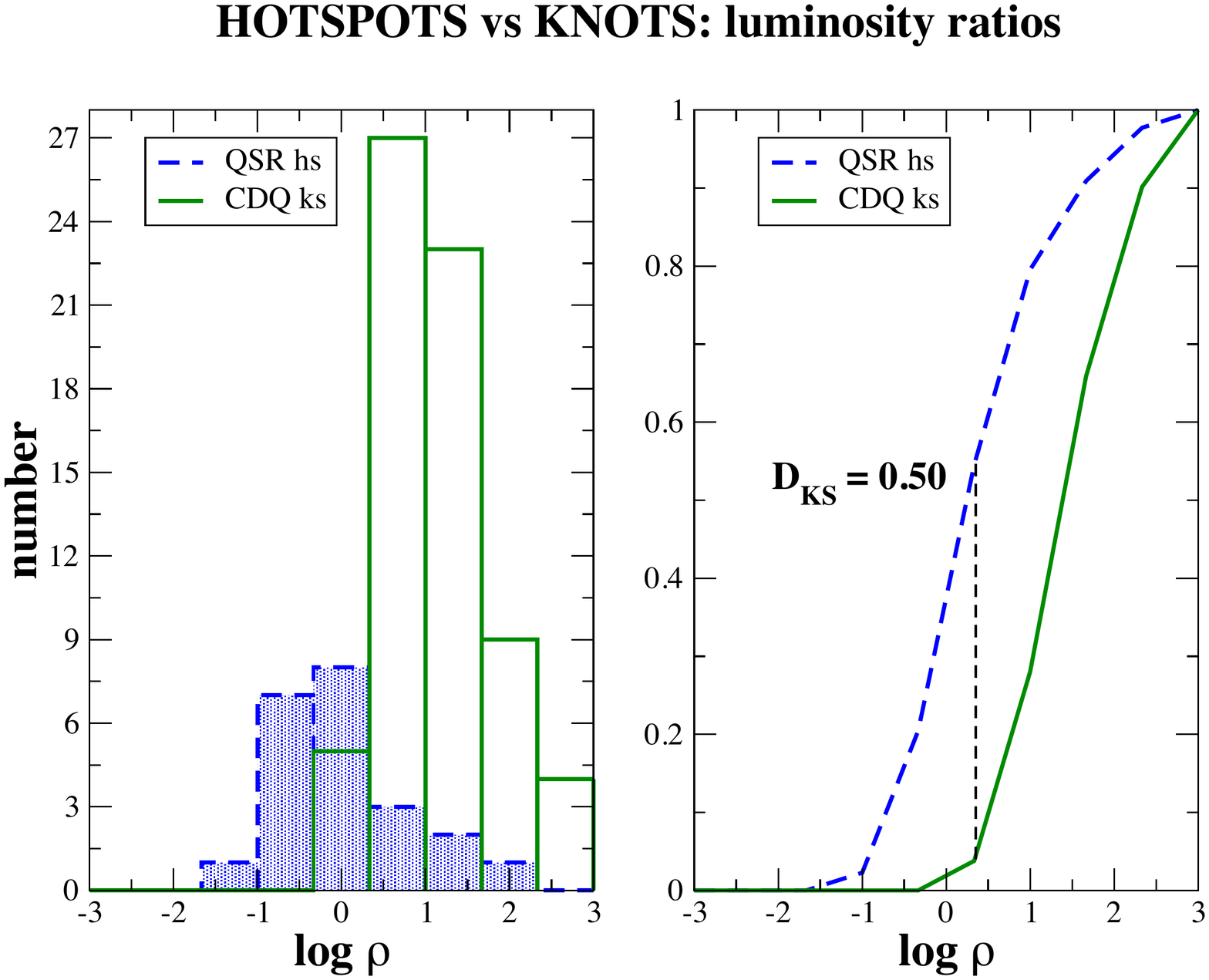}
\end{tabular}
\caption{
a) The distributions of radio luminosities $L_{\rm R}$ of hotspots and knots in LDQs and CDQs.
b) The normalized cumulative distributions of radio luminosities for hotspots and knots in LDQs and CDQs.
c) The distributions of X-ray luminosities $L_{\rm X}$ of hotspots and knots in LDQs and CDQs.
d) The normalized cumulative distributions of X-ray luminosities for hotspots and knots in LDQs and CDQs.
e) The distributions of $\rho$ of hotspots and knots in LDQs and CDQs.
f) The normalized cumulative distributions of luminosity ratios for hotspots and knots in LDQs and CDQs.
}
\label{fig:hk7}
\end{figure*}

\begin{figure*}
\begin{tabular}{cc}
\includegraphics[height=6.cm,width=6.3cm,angle=0]{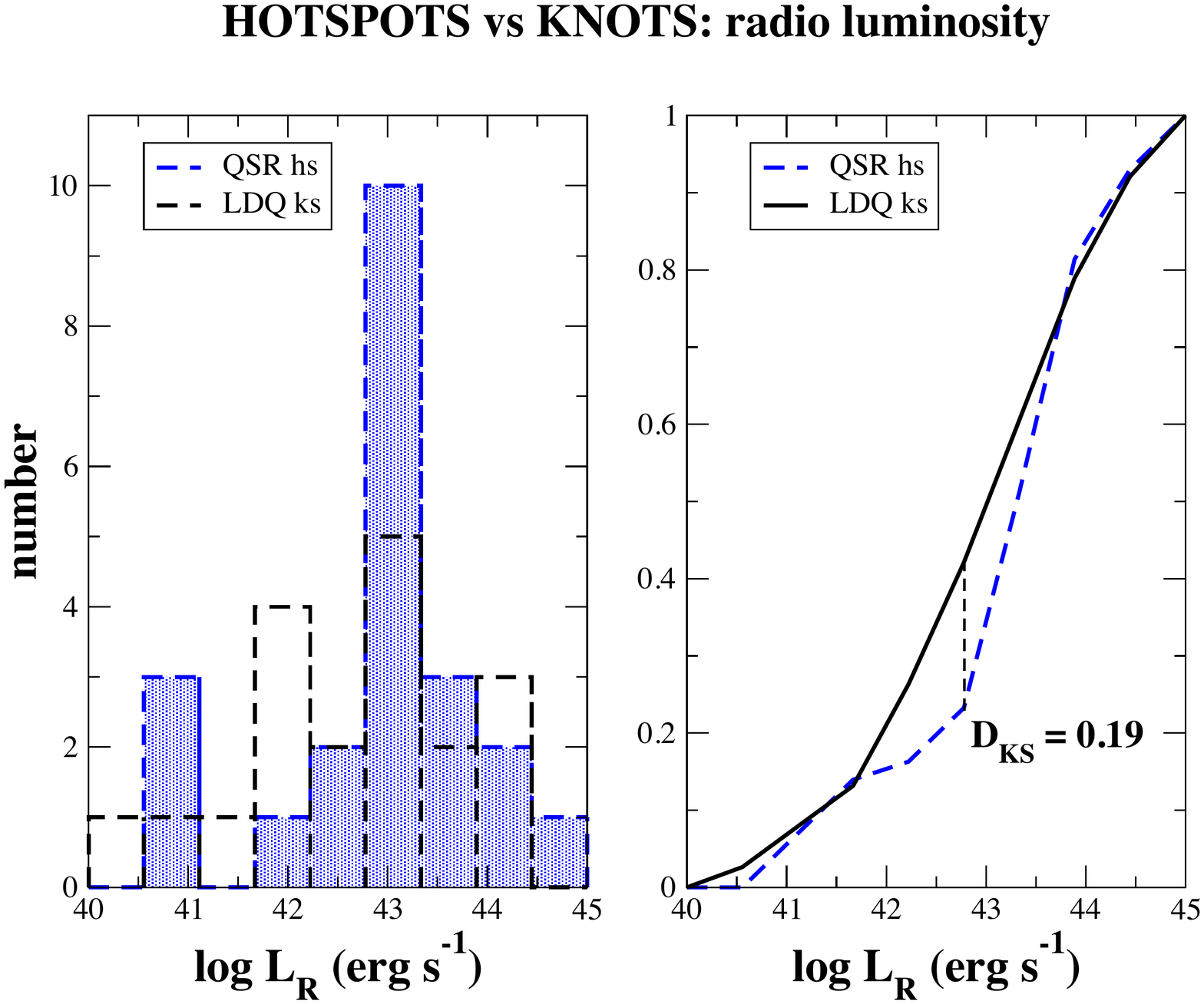}
\includegraphics[height=6.cm,width=6.3cm,angle=0]{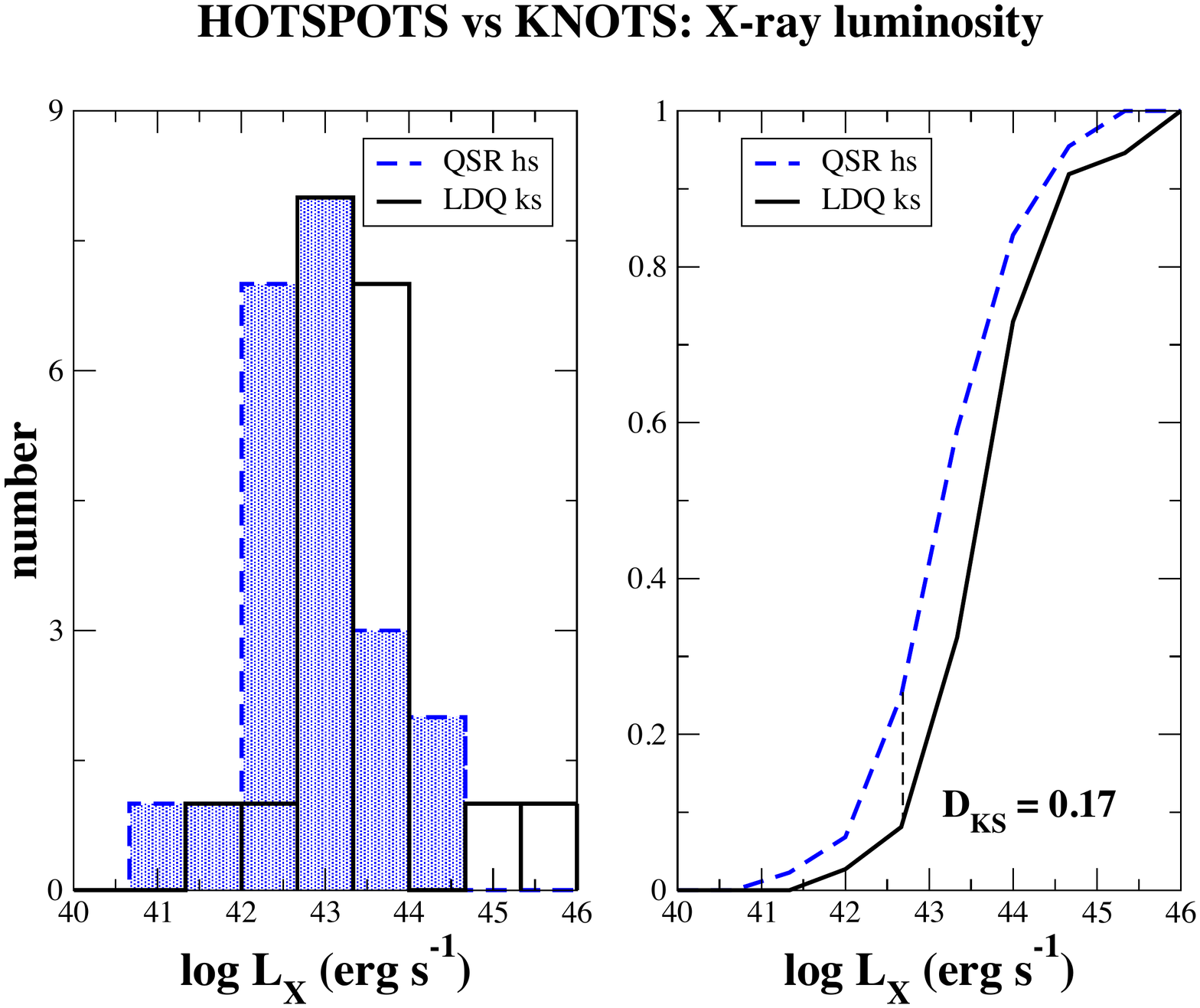}
\includegraphics[height=6.cm,width=6.3cm,angle=0]{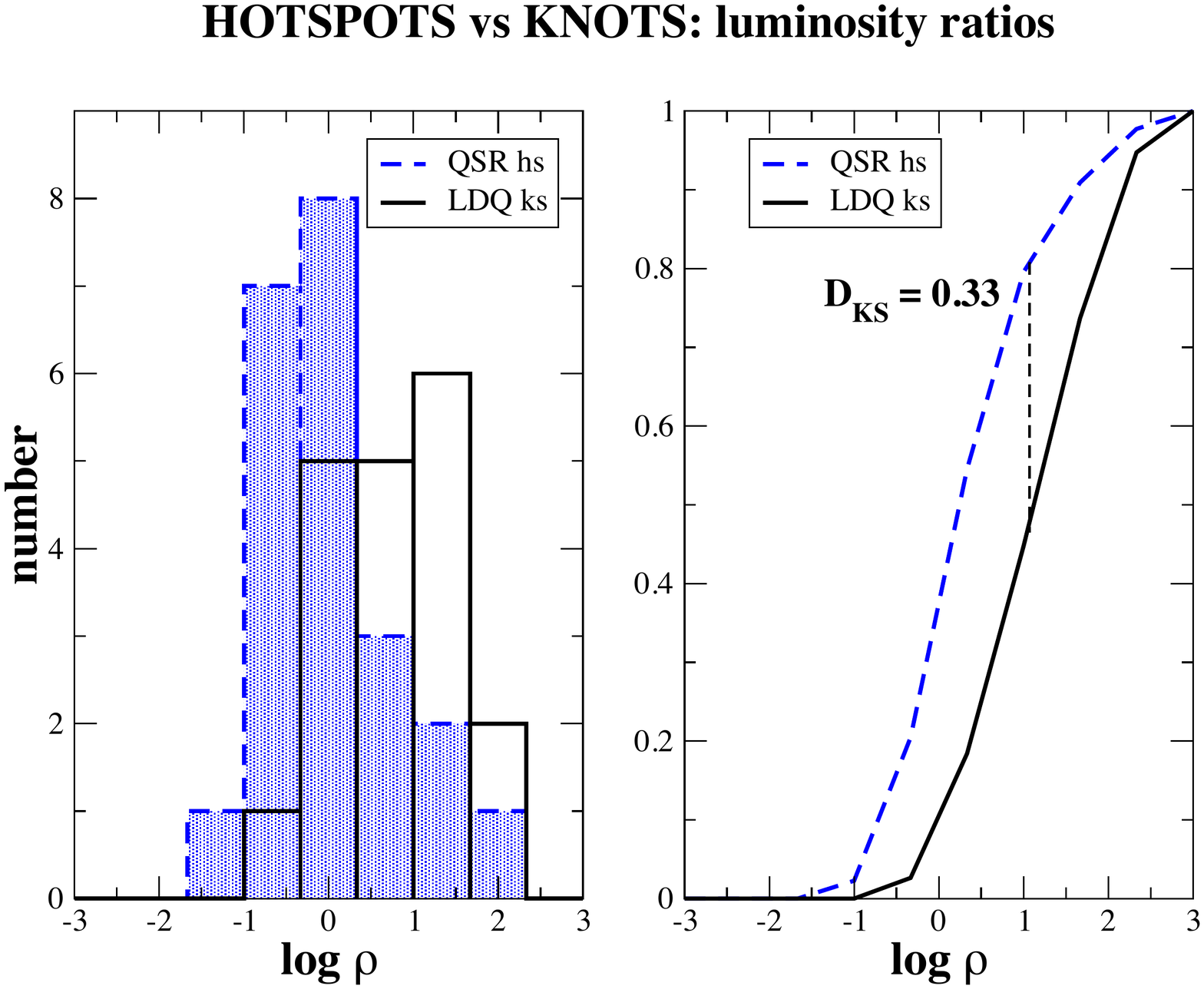}
\end{tabular}
\caption{
a) The distributions of radio luminosities $L_{\rm R}$ of hotspots and knots in LDQs.
b) The normalized cumulative distributions of radio luminosities for hotspots and knots in LDQs.
c) The distributions of X-ray luminosities $L_{\rm X}$ of hotspots and knots in LDQs.
d) The normalized cumulative distributions of X-ray luminosities for hotspots and knots in LDQs.
e) The distributions of $\rho$ of hotspots and knots in LDQs.
f) The normalized cumulative distributions of luminosity ratios for hotspots and knots in LDQs.
(Note the different color convention adopted for this figure).
}
\label{fig:hk8}
\end{figure*}

All the distributions of the observed parameters to compare their properties are shown in Figures~\ref{fig:rg_hs}, \ref{fig:rg_ks}, \ref{fig:qsr_ks}
\ref{fig:k1}, \ref{fig:k3}, \ref{fig:k4}, \ref{fig:k2}, \ref{fig:hk1}, \ref{fig:hk2}, \ref{fig:hk3}, \ref{fig:hk4}, \ref{fig:hk5}, \ref{fig:hk6}, \ref{fig:hk7}, 
and \ref{fig:hk8}.

\begin{figure}
\includegraphics[height=7.cm,width=8.5cm,angle=0]{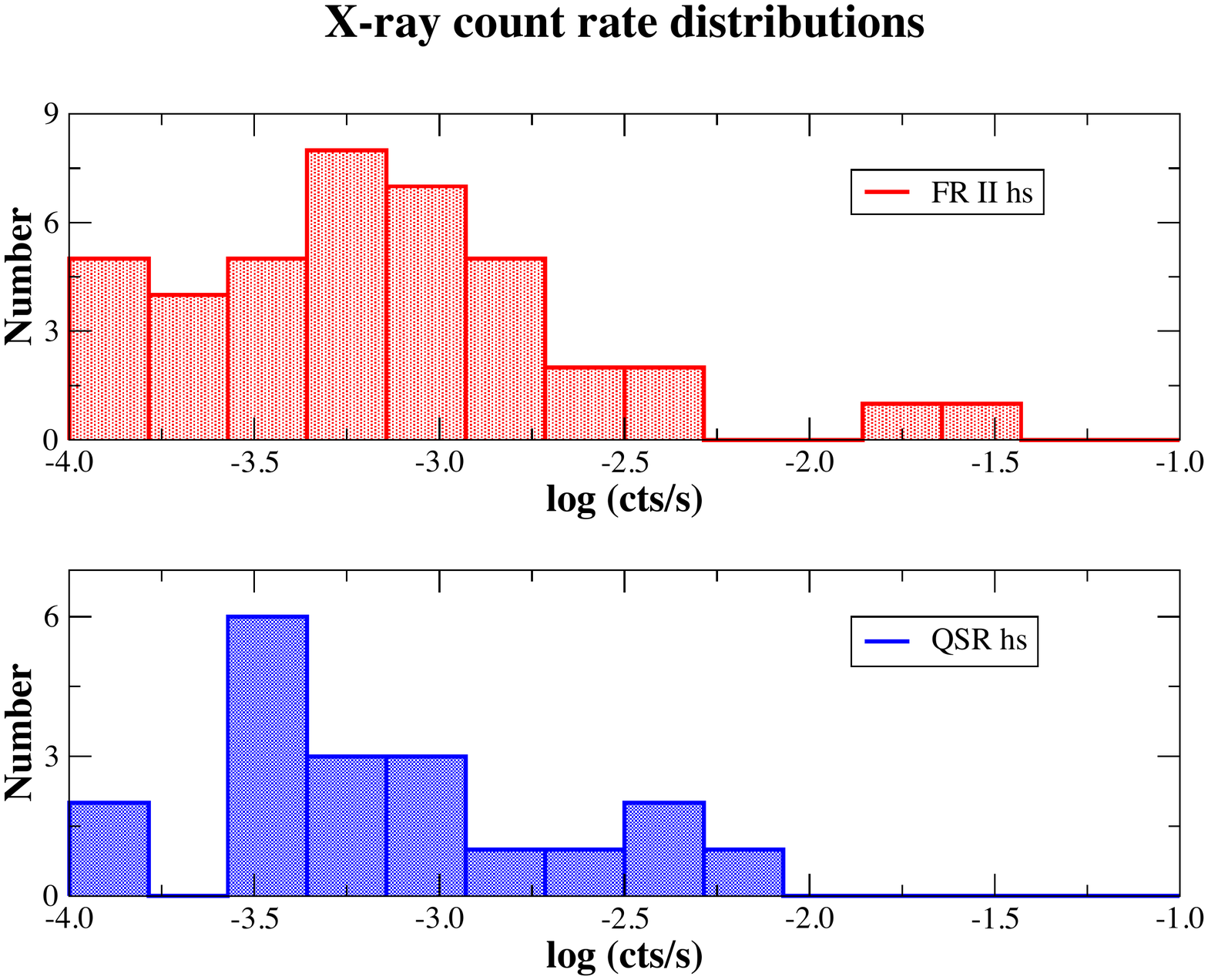}
\includegraphics[height=7.cm,width=8.5cm,angle=0]{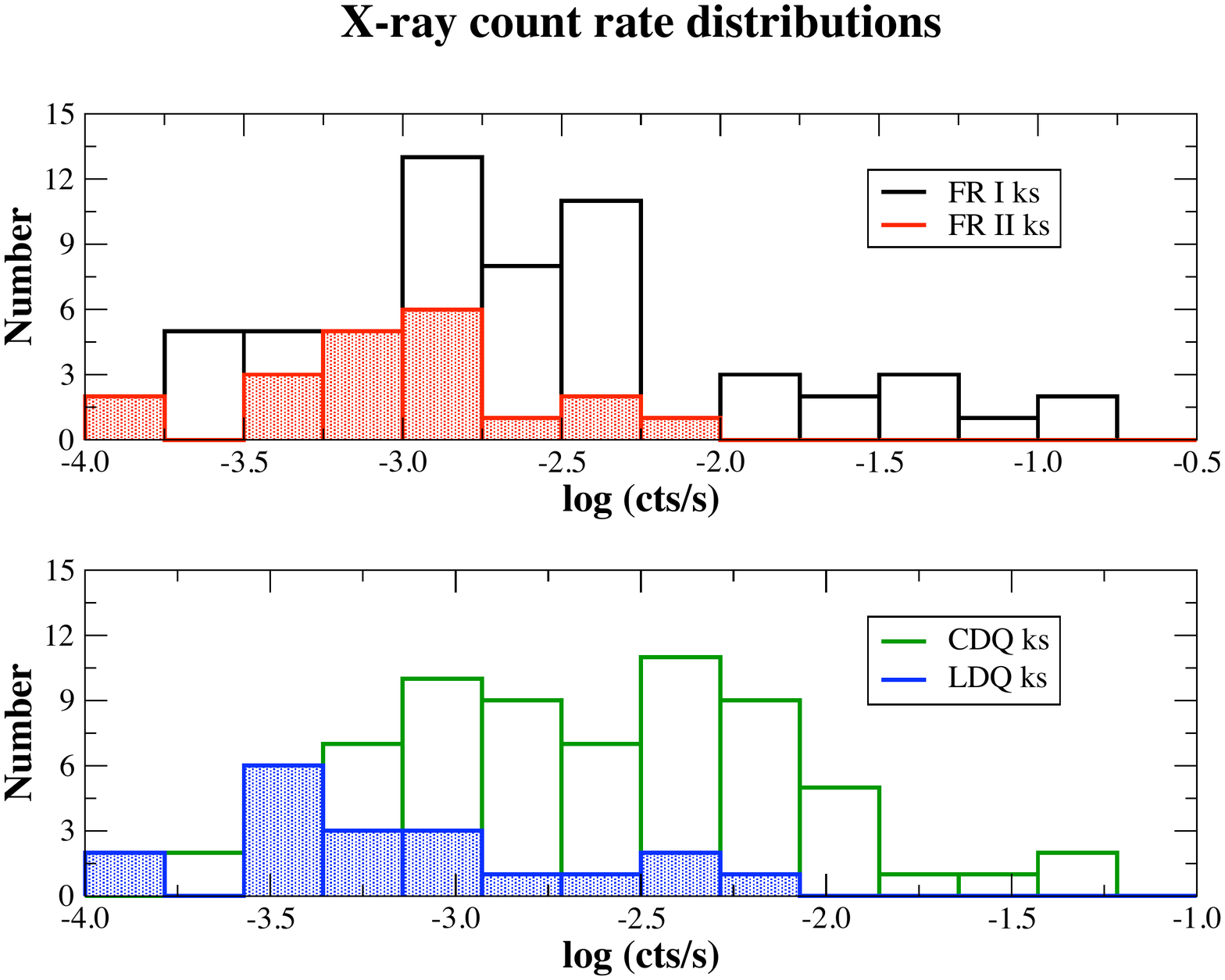}
\caption{The X-ray count rate distribution of hotspots (upper panel) and knots (lower panel).}
\label{fig:counts}
\end{figure}

Finally, we also report the distribution of the X-ray count rate for the different components in Figure~\ref{fig:counts}.

\begin{table}
\caption{}
\begin{tabular}{|lccccccccc|}
\hline
Component  & type & region & $\nu_{\rm R}$ & $S_{\nu}(\nu_{\rm R})$ &$F_{\rm soft}$     &   $F_{\rm med}$   &   $F_{\rm hard}$  & $F_{\rm tot}$     & $\rho$       \\
      &      &(arcsec)&   GHz         &     mJy                &10$^{-15}$ cgs     &10$^{-15}$ cgs     &10$^{-15}$ cgs     &10$^{-15}$ cgs     &              \\   
\hline
\noalign{\smallskip}
{\bf 3C\,6.1}        &     &              &      &        &             &             &             &             &              \\
n14.0                  & hs2 & c(1.5x1.5)   & 8.46 & 255.2  & 0.26(0.15)  & 0.54(0.22)  & 1.68(0.84)  & 2.48(0.88)  & 0.14(0.04)   \\
s12.0                  & hs2 & c(2.0x2.0)   & 8.46 & 219.5  & 0.00(0.00)  & 0.59(0.24)  & 0.42(0.88)  & 1.01(0.91)  & 0.06(0.05)   \\
\hline
\noalign{\smallskip}
{\bf 3C\,9}            &     &              &      &        &             &             &             &             &              \\
n5.0                   & kql & c(1.0x1.0)   & 8.44 &  19.1  & 0.22(0.16)  & 0.27(0.21)  & 0.85(0.72)  & 1.34(0.76)  & 0.87(0.48)   \\
s7.3                   & kql & c(1.0x1.0)   & 8.44 &  30.5  & 0.28(0.20)  & 0.19(0.19)  & 0.00(0.00)  & 0.47(0.27)  & 0.21(0.12)   \\
\hline
\noalign{\smallskip}
{\bf 3C\,15}           &     &              &      &        &             &             &             &             &              \\
n4.0                   & k1  & c(1.2x1.2)   & 8.46 & 101.5  & 1.86(0.38)  & 1.67(0.46)  & 4.78(1.48)  & 8.31(1.60)  & 1.00(0.19)   \\
\hline
\noalign{\smallskip}
{\bf 3C\,17}           &     &              &      &        &             &             &             &             &              \\
s3.7                   & k2  & e(0.9x0.6)   & 4.86 &  29.9  & 2.92(1.31)  & 2.39(0.98)  & 1.67(1.67)  & 6.98(2.34)  & 5.07(1.67)   \\
s11.3                  & k2  & c(0.74x0.74) & 4.86 &  85.3  & 0.40(0.40)  & 0.00(0.00)  & 5.92(3.49)  & 6.32(3.51)  & 1.54(0.85)   \\
\hline
\noalign{\smallskip}
{\bf NGC\,315}         &     &              &      &        &             &             &             &             &              \\
n6.3                   & k1  & c(2.0x2.0)   & 1.43 &  26.3  & 2.18(0.68)  & 4.42(0.59)  & 3.02(1.07)  & 9.62(1.40)  & 28.65(4.15)  \\
n10.5                  & k1  & c(2.0x2.0)   & 1.43 &  43.1  & 3.24(0.49)  & 3.31(0.47)  & 4.78(1.31)  &11.33(1.47)  & 20.78(2.53)  \\
n15.3                  & k1  & c(2.0x2.0)   & 1.43 &  34.7  & 2.98(0.45)  & 3.82(0.49)  & 4.66(1.18)  &11.46(1.35)  & 25.95(2.89)  \\
\hline
\noalign{\smallskip}
{\bf 3C\,31}           &     &              &      &        &             &             &             &             &              \\
n2.1                   & k1  & c(0.8x0.8)   & 8.46 &  2.1   & 4.16(0.59)  & 5.08(0.67)  & 4.58(1.25)  & 13.82(1.53) & 86.50(9.09)  \\
n4.2                   & k1  & r(2.95x1.97) & 8.46 & 18.9   & 3.47(0.54)  & 3.97(0.62)  & 3.04(1.09)  & 10.48(1.37) &  7.48(0.92)  \\
n7.0                   & k1  & c(1.2x1.2)   & 8.46 & 10.3   & 0.50(0.22)  & 1.39(0.34)  & 2.74(0.82)  &  4.63(0.92) &  5.60(1.09)  \\
\hline
\noalign{\smallskip}
{\bf 0106+013}         &     &              &      &        &             &             &             &             &              \\
s4.4                   & kqc & c(2.0x2.0)   & 1.40 & 468.7  & 7.98(2.01)  & 10.2(1.96)  & 24.9(6.28)  & 43.08(6.88) & 6.77(1.06)   \\
\hline
\noalign{\smallskip}
{\bf 3C\,33}           &     &              &      &        &             &             &             &             &              \\
n140.0                 & hs2 & c(3.0x3.0)   & 4.89 & 216.1  & 0.98(0.40)  & 0.90(0.37)  & 2.03(2.26)  & 3.91(2.32)  & 0.39(0.22)   \\
s112.0                 & hs2 & c(4.0x4.0)   & 4.89 &1076.3  & 1.11(0.42)  & 0.36(0.33)  & 1.35(0.98)  & 2.82(1.12)  & 0.06(0.02)   \\
\hline
\noalign{\smallskip}
{\bf 3C\,47}           &     &              &      &        &             &             &             &             &              \\
s39.0                  & hsq & c(2.0x2.0)   & 4.89 & 319.0  & 0.53(0.17)  & 0.49(0.24)  & 1.49(0.61)  & 2.52(0.67)  & 0.17(0.04)   \\
\hline
\noalign{\smallskip}
{\bf 3C\,52}           &     &              &      &        &             &             &             &             &              \\
n34.0                  & hs2 & c(1.2x1.2)   & 8.44 & 475.1  & 0.97(0.69)  & 1.34(0.77)  & 0.00(0.00)  & 2.31(1.03)  & 0.10(0.05)   \\
\hline
\noalign{\smallskip}
{\bf 4C\,+35.03}       &     &              &      &        &             &             &             &             &              \\
n3.0                   & k1  & r(4.43x2.46) & 4.86 &  42.4  & 6.05(1.37)  & 4.27(1.38)  & 4.16(3.00)  & 14.48(3.58) & 8.37(1.87)   \\
\hline
\noalign{\smallskip}
{\bf PKS\,0208-512}    &     &              &      &        &             &             &             &             &              \\
s4.0                   & kqc & c(1.5x1.5)   & 8.64 &  33.5  & 5.10(0.64)  & 7.25(0.81)  & 17.90(2.37) & 30.25(2.58) & 10.70(0.90)  \\
\hline
\noalign{\smallskip}
{\bf 3C\,61.1}         &     &              &      &        &             &             &             &             &              \\
s102.0                 & hs2 & c(4.0x4.0)   & 4.86 & 465.7  & 1.27(0.73)  & 1.14(0.82)  & 5.01(2.89)  & 7.42(3.09)  & 0.36(0.14)   \\
\hline
\noalign{\smallskip}
{\bf 3C\,66B}          &     &              &      &        &             &             &             &             &              \\
n2.6                   & k1  & c(1.25x1.25) & 8.44 & 29.2   & 12.70(0.88) & 13.20(1.02) & 17.40(2.37) & 43.30(2.73) & 20.74(1.19)  \\
n5.5                   & k1  & c(1.25x1.25) & 8.44 & 16.8   & 2.26(0.42)  & 3.83(0.57)  & 4.23(1.14)  & 10.32(1.34) & 8.36(1.02)   \\
\hline
\noalign{\smallskip}
{\bf 3C\,83.1}         &     &              &      &        &             &             &             &             &              \\
e8.0                   & k1  & c(1.23x1.23) & 4.99 &  6.54  & 0.03(0.06)  & 0.32(0.11)  & 0.51(0.30)  & 0.86(0.33)  & 2.90(1.08)   \\
e13.0                  & k1  & c(1.23x1.23) & 4.99 &  5.98  & 0.11(0.07)  & 0.07(0.07)  & 0.24(0.31)  & 0.43(0.32)  & 1.90(1.17)   \\
w7.0                   & k1  & c(1.23x1.23) & 4.99 &  2.01  & 0.14(0.08)  & 0.14(0.09)  & 0.53(0.31)  & 0.81(0.34)  & 9.96(3.72)   \\
\hline
\noalign{\smallskip}
{\bf 3C\,105}          &     &              &      &        &             &             &             &             &              \\
s166.0                 & k2  & c(1.5x1.5)   & 8.41 &  40.0  & 1.81(0.90)  & 1.83(0.92)  & 10.10(4.12) & 13.74(4.32) & 4.64(1.36)   \\
s169.0                 & hs2 & c(1.5x1.5)   & 8.41 & 346.5  & 0.37(0.37)  & 0.55(0.55)  & 4.96(2.87)  & 5.87(2.94)  & 0.22(0.10)   \\
\hline
\noalign{\smallskip}
{\bf PKS\,0405-123}    &     &              &      &        &             &             &             &             &              \\
n18.7                  & hsq & c(1.3x1.3)   & 4.86 & 197.7  & 3.59(1.00)  & 2.92(1.04)  & 5.13(2.57)  & 11.64(2.95) & 1.31(0.32)   \\
\hline
\noalign{\smallskip}
{\bf 3C\,109}          &     &              &      &        &             &             &             &             &              \\
n46.4                  & hs2 & c(1.5x1.5)   & 8.26 &  37.9  & 0.00(0.00)  & 0.06(0.09)  & 0.93(0.47)  & 0.99(0.48)  & 0.32(0.15)   \\
s48.4                  & hs2 & c(1.5x1.5)   & 8.26 & 175.8  & 0.05(0.05)  & 0.24(0.14)  & 0.22(0.22)  & 0.51(0.27)  & 0.04(0.02)   \\
\hline
\noalign{\smallskip}
{\bf PKS\,0413-210}    &     &              &      &        &             &             &             &             &              \\
s1.8                   & kqc & c(1.63x1.63) & 4.86 & 375.8  & 2.23(1.54)  & 6.99(2.36)  & 23.70(7.89) & 32.92(8.38) & 1.83(0.46)   \\
\hline
\noalign{\smallskip}
{\bf 3C\,111}          &     &              &      &        &             &             &             &             &              \\
e9.0                   & k2  & c(1.5x1.5)   & 1.44 & 8.6    & 0.00(0.00)  & 1.68(1.11)  & 12.20(5.50) & 13.88(5.61) & 116.91(45.92)\\
e15.8                  & k2  & r(5.0x2.5)   & 1.44 & 12.9   & 0.00(0.00)  & 3.25(1.25)  & 9.44(4.70)  & 12.69(4.86) & 75.48(27.06) \\
e29.5                  & k2  & r(4.0x2.5)   & 1.44 & 12.9   & 0.00(0.00)  & 2.48(1.01)  & 15.20(4.59) & 17.68(4.70) & 100.56(25.87)\\
e43.6                  & k2  & r(4.0x2.5)   & 1.44 & 11.0   & 0.00(0.00)  & 0.00(0.00)  & 8.45(3.45)  & 8.45(3.45)  & 53.50(21.84) \\
e60.0                  & k2  & r(6.5x3.0)   & 1.44 & 15.2   & 0.00(0.00)  & 6.63(1.66)  & 22.90(6.21) & 29.53(6.43) & 146.61(30.21)\\
e98.0                  & k2  & c(2.0x2.0)   & 1.44 & 39.6   & 0.00(0.00)  & 1.51(0.75)  & 4.10(2.36)  & 5.61(2.48)  & 10.91(4.54)  \\
e121.0                 & hs2 & c(3.0x3.0)   & 1.44 & 1771.1 & 0.51(0.51)  & 1.67(0.84)  & 4.20(2.10)  & 6.38(2.32)  & 0.34(0.13)   \\
\noalign{\smallskip}
\hline
\end{tabular}\\
\label{tab:app1}
Note: Col. (2) reports the class component accordingly with the definition of Section~\ref{sec:sample}. In Col. (3), where the region size for the 
flux measurements is shown, the letters `c', `r', and `e' indicate a circular, a rectangular and an elliptical region, respectively. Col. (10) reports the value of the
ratios as defined in Section~\ref{sec:ratios}.  
\end{table}

\begin{table}
\caption{}
\begin{tabular}{|lccccccccc|}
\hline
Component & type & region & $\nu_{\rm R}$ & $S_{\nu}(\nu_{\rm R})$ &$F_{\rm soft}$     &   $F_{\rm med}$   &   $F_{\rm hard}$  & $F_{\rm tot}$     & $\rho$       \\
      &      &(arcsec)&   GHz         &     mJy                &10$^{-15}$ cgs     &10$^{-15}$ cgs     &10$^{-15}$ cgs     &10$^{-15}$ cgs     &              \\   
\hline
\noalign{\smallskip}
{\bf 3C\,120}          &     &              &      &        &             &             &             &             &              \\
w4.0                   & k1  & c(0.74x0.74) & 4.86 & 57.1   & 0.63(0.22)  & 2.03(0.39)  & 14.20(2.21) & 16.86(2.25) & 6.35(0.82)   \\
w25.0                  & k1  & c(2.0x2.0)   & 4.86 & 9.1    & 0.48(0.15)  & 2.47(0.39)  & 31.50(2.97) & 34.45(3.00) & 79.40(6.81)  \\
\hline
\noalign{\smallskip}
{\bf 3C\,123}          &     &              &      &        &             &             &             &             &              \\
e9.0                   & k2  & c(1.5x1.5)   & 1.49 &12593.8 & 0.82(0.23)  & 4.76(0.71)  & 16.20(2.53) & 21.78(2.64) & 0.12(0.01)   \\
\hline
\noalign{\smallskip}
{\bf 3C\,129}          &     &              &      &        &             &             &             &             &              \\
n2.0                   & k1  & c(0.8x0.8)   & 7.99 & 1.4    & 0.00(0.00)  & 0.44(0.28)  & 3.03(1.26)  & 3.47(1.29)  & 33.45(11.83) \\
n5.0                   & k1  & c(0.8x0.8)   & 7.99 & 2.3    & 0.00(0.00)  & 0.32(0.18)  & 0.52(1.02)  & 0.84(1.04)  & 6.22(5.93)   \\
\hline
\noalign{\smallskip}
{\bf Pictor A}         &     &              &      &        &             &             &             &             &              \\
w250.0                 & hs2  & c(5.0x5.0)   & 4.87 & 1709.0 & 91.8(2.70)  & 121.0(3.74) & 201.00(9.88)&413.80(10.90)& 5.21(0.13)   \\
\hline
\noalign{\smallskip}
{\bf PKS\,0521-365}    &     &              &      &        &             &             &             &             &              \\
w1.9                   & kbl & c(0.8x0.8)   & 4.86 & 139.1  & 21.40(2.65) & 27.10(3.67) & 27.60(7.76) & 76.10(8.98) & 12.12(1.37)  \\
e8.6                   & kbl & c(2.0x2.0)   & 4.86 & 1526.7 & 1.36(0.81)  & 2.07(1.12)  & 1.12(1.77)  & 4.55(2.25)  & 0.07(0.03)   \\
\hline
\noalign{\smallskip}
{\bf 0529+075}         &     &              &      &        &             &             &             &             &              \\
w2.8                   & kqc & r(2.0x1.8)   & 1.40 & 50.3   & 1.16(0.67)  & 3.82(1.21)  & 12.90(4.60) & 17.88(4.80) & 29.12(7.27)  \\
\hline
\noalign{\smallskip}
{\bf PKS\,0605-085}    &     &              &      &        &             &             &             &             &              \\
e2.1                   & kqc & c(0.8x0.8)   & 4.86 & 10.4   & 2.97(1.00)  & 4.62(1.41)  & 23.10(6.52) & 30.69(6.75) & 72.94(14.41) \\
e4.2                   & kqc & c(0.8x0.8)   & 4.86 & 22.3   & 3.09(1.03)  & 4.23(1.30)  & 19.00(5.28) & 26.32(5.53) & 29.90(5.67)  \\
\hline
\noalign{\smallskip}
{\bf 3C\,173.1}        &     &              &      &        &             &             &             &             &              \\
s31.0                  & hs2 & c(2.5x2.5)   & 1.48 & 443.3  & 0.11(0.11)  & 0.16(0.16)  & 1.94(0.97)  & 2.20(0.99)  & 0.34(0.15)   \\
\hline
\noalign{\smallskip}
{\bf 3C\,179}          &     &              &      &        &             &             &             &             &              \\
w4.5                   & kql & c(0.8x0.8)   & 4.87 & 48.1   & 0.65(0.46)  & 1.14(0.66)  & 0.91(0.91)  & 2.70(1.21)  & 1.26(0.55)   \\
w6.5                   & hsq & c(1.2x1.2)   & 4.87 & 69.7   & 1.18(0.59)  & 2.79(0.99)  & 3.18(2.29)  & 7.15(2.56)  & 2.25(0.77)   \\
\hline
\noalign{\smallskip}
{\bf B2\,0738+313}     &     &              &      &        &             &             &             &             &              \\
s31.0                  & hsq & c(2.5x2.5)   & 4.99 & 0.83   & 0.52(0.24)  & 0.91(0.33)  & 0.72(0.67)  & 2.15(0.78)  & 56.77(19.68) \\
s36.0                  & hsq & c(2.5x2.5)   & 4.99 & 1.21   & 0.38(0.19)  & 0.81(0.33)  & 0.00(0.00)  & 1.19(0.38)  & 22.24(7.09)  \\
\hline
\noalign{\smallskip}
{\bf 3C\,189}          &     &              &      &        &             &             &             &             &              \\
e1.3                   & k1  & c(0.5x0.5)   & 1.66 & 27.8   & 3.99(1.33)  & 5.99(1.63)  & 11.30(4.39) & 21.28(4.87) & 49.25(10.86) \\
\hline
\noalign{\smallskip}
{\bf 0827+243}         &     &              &      &        &             &             &             &             &              \\
s2.6                   & kqc & c(0.8x0.8)   & 4.86 & 0.24   & 0.47(0.39)  & 1.74(0.69)  & 3.70(2.25)  & 5.91(2.39)  &521.17(206.90)\\ 
s4.7                   & kqc & c(0.8x0.8)   & 4.86 & 0.54   & 2.18(0.59)  & 1.39(0.58)  & 8.06(2.45)  & 11.63(2.59) & 456.79(99.10)\\
s5.9                   & kqc & c(0.8x0.8)   & 4.86 & 1.98   & 0.51(0.30)  & 0.91(0.47)  & 0.47(3.08)  & 4.50(1.49)  & 48.13(15.68) \\
\hline
\noalign{\smallskip}
{\bf 4C\,+29.30}       &     &              &      &        &             &             &             &             &              \\
n11.3                  & k1  & c(1.5x1.5)   & 4.87 & 17.5   & 2.20(0.84)  & 0.34(0.34)  & 0.00(0.00)  & 2.54(0.91)  & 3.71(1.33)   \\
s17.7                  & k1  & c(1.5x1.5)   & 4.87 & 18.2   & 2.18(0.82)  & 2.13(0.87)  & 10.30(4.21) & 14.61(4.38) & 17.23(4.99)  \\
\hline
\noalign{\smallskip}
{\bf 3C\,207}          &     &              &      &        &             &             &             &             &              \\
e4.3                   & kql & c(1.0x1.0)   & 8.46 & 92.5   & 2.78(0.44)  & 4.98(0.68)  & 14.50(2.43) & 22.26(2.56) & 2.97(0.33)   \\
e6.3                   & kql & c(1.0x1.0)   & 8.46 & 79.0   & 0.98(0.28)  & 0.98(0.33)  & 3.05(1.34)  & 5.01(1.41)  & 0.80(0.21)   \\
\hline
\noalign{\smallskip}
{\bf 3C\,212}          &     &              &      &        &             &             &             &             &              \\
n5.0                   & hsq & c(1.0x1.0)   & 8.46 & 61.1   & 0.00(0.00)  & 0.12(0.28)  & 0.70(0.70)  & 0.82(0.75)  & 0.16(0.15)   \\
s4.7                   & hsq & c(1.0x1.0)   & 8.46 & 166.1  & 0.00(0.00)  & 0.64(0.45)  & 0.00(0.00)  & 0.64(0.45)  & 0.05(0.03)   \\
\hline
\noalign{\smallskip}
{\bf PKS\,0903-573}    &     &              &      &        &             &             &             &             &              \\
n2.6                   & kqc & c(1.0x1.0)   & 8.64 & 321.1 & 0.53(0.84)  & 5.13(2.14)  & 4.10(4.10)  & 9.76(4.70)  & 0.48(0.22)   \\
\hline
\noalign{\smallskip}
{\bf 3C\,219}          &     &              &      &        &             &             &             &             &              \\
s9.3                   & k2  & c(3.0x3.0)   & 4.56 & 17.1   & 0.22(0.25)  & 0.17(0.44)  & 3.08(1.74)  & 3.47(1.81)  & 4.45(2.32)   \\
s17.6                  & k2  & c(3.0x3.0)   & 4.56 & 26.0   & 0.20(0.26)  & 0.59(0.41)  & 5.73(2.87)  & 6.53(2.91)  & 5.53(2.46)   \\
\hline
\noalign{\smallskip}
{\bf PKS\,0920-397}    &     &              &      &        &             &             &             &             &              \\
s2.5                   & kqc & c(1.0x1.0)   & 8.64 & 42.0   & 2.24(0.76)  & 3.48(1.05)  & 11.90(3.50) & 17.62(3.73) & 6.06(1.16)   \\
s4.6                   & kqc & c(1.0x1.0)   & 8.64 & 14.8   & 1.71(0.61)  & 2.72(0.79)  & 1.67(1.21)  & 6.10(1.57)  & 7.38(1.72)   \\
s8.6                   & hsq & c(1.0x1.0)   & 8.64 & 50.1   & 0.19(0.19)  & 2.07(0.65)  & 2.13(1.26)  & 4.39(1.43)  & 1.20(0.36)   \\
\hline
\noalign{\smallskip}
{\bf 3C\,227}          &     &              &      &        &             &             &             &             &              \\
w109.0                 & hs2 & c(2.5x2.5)   & 1.43 & 261.6  & 1.77(0.43)  & 2.36(0.52)  & 2.85(1.01)  & 6.98(1.21)  & 1.94(0.33)   \\
w117.0                 & hs2 & c(3.0x3.0)   & 1.43 & 98.3   & 0.37(0.21)  & 0.84(0.32)  & 3.33(1.27)  & 4.54(1.33)  & 3.27(0.95)   \\
\hline
\noalign{\smallskip}
{\bf Q\,0957+561}     &     &              &      &        &             &             &             &             &              \\
e3.5                  & kql & c(0.8x0.8)   & 4.86 & 23.4   & 0.34(0.17)  & 0.36(0.20)  & 1.75(0.73)  & 2.46(0.77)  & 2.18(0.68)   \\
\hline
\noalign{\smallskip}
{\bf 4C\,+13.41}      &     &              &      &        &             &             &             &             &              \\
s9.0                  & kql & c(2.0x2.0)   & 8.26 & 1.6    & 0.47(0.18)  & 0.58(0.24)  & 0.43(0.56)  & 1.47(0.63)  & 12.37(5.01)  \\
\noalign{\smallskip}
\hline
\end{tabular}\\
\label{tab:app2}
Note: Col. (2) reports the class component accordingly with the definition of Section~\ref{sec:sample}. In Col. (3), where the region size for the 
flux measurements is shown, the letters `c', `r', and `e' indicate a circular, a rectangular and an elliptical region, respectively. Col. (10) reports the value of the
ratios as defined in Section~\ref{sec:ratios}.  
\end{table}

\begin{table}
\caption{}
\begin{tabular}{|lccccccccc|}
\hline
Component & type & region & $\nu_{\rm R}$ & $S_{\nu}(\nu_{\rm R})$ &$F_{\rm soft}$     &   $F_{\rm med}$   &   $F_{\rm hard}$  & $F_{\rm tot}$     & $\rho$       \\
      &      &(arcsec)&   GHz         &     mJy                &10$^{-15}$ cgs     &10$^{-15}$ cgs     &10$^{-15}$ cgs     &10$^{-15}$ cgs     &              \\   
\hline
\noalign{\smallskip}
{\bf 1045-188}        &     &              &      &        &             &             &             &             &              \\
e3.4                  & kqc & c(1.0x1.0)   & 1.43 & 25.3   & 1.76(0.79)  & 2.11(0.96)  & 8.00(3.58)  & 11.87(3.79) & 33.92(10.59) \\
e5.3                  & kqc & c(1.0x1.0)   & 1.43 & 24.4   & 0.00(0.00)  & 2.40(0.98)  & 0.49(1.11)  & 2.89(1.48)  & 8.49(4.30)   \\
e7.3                  & kqc & c(1.0x1.0)   & 1.43 & 78.3   & 0.60(0.59)  & 1.40(0.81)  & 2.73(2.71)  & 4.73(2.89)  & 4.36(2.61)   \\
\hline
\noalign{\smallskip}
{\bf PKS\,1046-406}   &     &              &      &        &             &             &             &             &              \\
s3.0                  & kqc & r(3.43x1.8)  & 8.64 & 20.5   & 5.19(2.36)  & 3.24(2.36)  & 8.24(7.17)  & 16.67(7.91) & 11.11(4.76)  \\
\hline
\noalign{\smallskip}
{\bf 4C\,+20.24}      &     &              &      &        &             &             &             &             &              \\
n4.2                  & kql & c(1.2x1.2)   & 1.43 & 10.9   & 0.14(0.10)  & 0.35(0.14)  & 0.93(0.33)  & 1.42(0.37)  & 9.27(2.43)   \\
n11.1                 & kql & c(1.2x1.2)   & 1.43 & 9.1    & 0.21(0.08)  & 0.35(0.10)  & 0.98(0.36)  & 1.53(0.38)  & 11.98(2.97)  \\
n14.7                 & kql & c(1.2x1.2)   & 1.43 & 6.3    & 0.50(0.12)  & 0.64(0.14)  & 1.38(0.40)  & 2.52(0.44)  & 28.64(4.97)  \\
n18.8                 & kql & c(1.2x1.2)   & 1.43 & 21.8   & 1.02(0.15)  & 1.53(0.19)  & 2.91(0.53)  & 5.46(0.59)  & 18.00(1.92)  \\
n21.3                 & hsq & c(1.2x1.2)   & 1.43 & 936.0  & 1.85(0.20)  & 2.09(0.22)  & 2.72(0.52)  & 6.66(0.60)  & 0.52(0.05)   \\
\hline
\noalign{\smallskip}
{\bf 3C\,254}         &     &              &      &        &             &             &             &             &              \\
w11.8                 & hsq & c(0.5x0.5)   & 4.89 & 154.7  & 0.39(0.17)  & 0.19(0.13)  & 0.26(0.26)  & 0.84(0.34)  & 0.12(0.05)   \\
\hline
\noalign{\smallskip}
{\bf PKS\,1127-145}   &     &              &      &        &             &             &             &             &              \\
n3.0                  & kqc & c(1.2x1.2)   & 8.46 & 1.1    & 2.41(0.33)  & 2.64(0.41)  & 7.28(1.42)  & 12.33(1.51) & 138.58(16.43)\\
n6.5                  & kqc & c(1.5x1.5)   & 1.43 & 2.4    & 0.87(0.19)  & 1.32(0.25)  & 1.82(0.73)  & 4.01(0.79)  & 123.08(23.42)\\
n11.0                 & kqc & r(5.0x2.0)   & 8.46 & 1.3    & 1.12(0.18)  & 1.51(0.24)  & 3.49(0.70)  & 6.12(0.76)  & 58.77(7.06)  \\
n19.0                 & kqc & r(3.0x2.0)   & 8.46 & 7.1    & 0.65(0.14)  & 0.58(0.14)  & 1.16(0.48)  & 2.39(0.52)  & 4.22(0.88)   \\
n28.0                 & kqc & c(2.5x2.5)   & 8.46 & 13.6   & 0.31(0.10)  & 0.36(0.10)  & 0.66(0.33)  & 1.33(0.36)  & 1.22(0.32)   \\
\hline
\noalign{\smallskip}
{\bf PKS\,1136-135}   &     &              &      &        &             &             &             &             &              \\
w4.9                  & kql & c(1.0x1.0)   & 4.86 & 1.4    & 1.94(0.28)  & 1.96(0.33)  & 6.30(1.21)  & 10.20(1.28) & 157.63(19.22)\\
w7.0                  & kql & c(1.0x1.0)   & 4.86 & 37.3   & 4.69(0.43)  & 5.96(0.54)  & 11.50(1.50) & 22.15(1.65) & 12.85(0.93)  \\
\hline
\noalign{\smallskip}
{\bf 3C\,263}         &     &              &      &        &             &             &             &             &              \\
e16.3                 & hsq & c(1.5x1.5)   & 4.86 & 640.7  & 2.56(0.35)  & 2.04(0.35)  & 4.73(1.20)  & 9.33(1.30)  & 0.30(0.04)   \\
\hline
\noalign{\smallskip}
{\bf 3C\,264}         &     &              &      &        &             &             &             &             &              \\
n1.8                  & k1  & c(0.5x0.5)   & 4.95 & 14.9   & 2.23(0.76)  & 0.84(0.81)  & 2.47(1.82)  & 5.54(2.13)  & 7.83(2.93)   \\
\hline
\noalign{\smallskip}
{\bf 3C\,265}         &     &              &      &        &             &             &             &             &              \\
e31.0                 & hs2 & c(1.5x1.5)   & 4.85 & 313.5  & 0.47(0.15)  & 0.28(0.13)  & 0.85(0.50)  & 1.59(0.53)  & 0.11(0.04)   \\
w47.0                 & hs2 & c(1.2x1.2)   & 4.85 & 41.7   & 0.13(0.07)  & 0.14(0.08)  & 0.90(0.52)  & 1.17(0.53)  & 0.59(0.26)   \\
\hline
\noalign{\smallskip}
{\bf 4C\,+49.22}      &     &              &      &        &             &             &             &             &              \\
s2.3                  & kqc & r(1.7x1.2)   & 4.87 & 29.6   & 4.12(1.28)  & 12.80(2.31) & 30.70(7.46) & 47.62(7.91) & 33.62(5.53)  \\
s4.5                  & kqc & r(2.5x1.3)   & 4.87 & 24.7   & 2.21(0.93)  & 4.18(1.26)  & 5.33(3.47)  & 11.72(3.81) & 10.04(3.19)  \\
s8.0                  & hsq & r(2.5x1.5)   & 4.87 & 47.9   & 0.80(0.46)  & 2.20(0.91)  & 8.32(3.72)  & 11.32(3.86) & 4.92(1.66)   \\
\hline
\noalign{\smallskip}
{\bf PKS\,1202-262}   &     &              &      &        &             &             &             &             &              \\
n2.2                  & kqc & r(2.5x2.5)   & 4.86 & 40.8   & 8.85(1.07)  & 15.40(1.48) & 52.00(5.37) & 76.25(5.67) & 41.52(2.95)  \\
n5.5                  & hsq & c(1.5x1.5)   & 4.86 & 48.4   & 11.30(1.02) & 17.70(1.34) & 46.80(4.36) & 75.80(4.67) & 35.42(2.06)  \\
\hline
\noalign{\smallskip}
{\bf 3C\,270}         &     &              &      &        &             &             &             &             &              \\
w9.6                  & k1  & c(2.5x2.5)   & 4.86 & 4.4    & 0.00(0.00)  & 0.09(0.32)  & 3.33(1.20)  & 3.42(1.24)  & 15.85(5.76)  \\
w15.8                 & k1  & c(2.5x2.5)   & 4.86 & 7.8    & 0.84(0.46)  & 1.46(0.45)  & 3.42(1.32)  & 5.72(1.47)  & 15.49(3.95)  \\
\hline
\noalign{\smallskip}
{\bf PG\,1222-216}    &     &              &      &        &             &             &             &             &              \\
e2.6                  & kqc & c(0.8x0.8)   & 4.86 & 13.8   & 0.91(0.42)  & 1.15(0.52)  & 1.54(2.13)  & 3.60(2.23)  & 5.59(3.35)   \\
e4.7                  & kqc & c(0.8x0.8)   & 4.86 & 4.3    & 0.62(0.31)  & 0.00(0.00)  & 0.41(0.92)  & 1.02(0.97)  & 5.33(4.76)   \\
n1.0                  & kqc & c(0.5x0.5)   & 4.86 & 60.5   & 11.10(1.93) & 17.30(2.56) & 27.80(8.70) & 56.20(9.27) & 19.74(3.17)  \\
n1.8                  & kqc & c(0.5x0.5)   & 4.86 & 47.8   & 7.09(1.09)  & 8.15(1.29)  & 14.40(3.26) & 29.64(3.67) & 13.23(1.60)  \\
\hline
\noalign{\smallskip}
{\bf M\,84}           &     &              &      &        &             &             &             &             &              \\
n2.7                  & k1  & c(0.5x0.5)   & 4.89 & 3.0    & 0.39(0.35)  & 0.04(0.23)  & 1.04(0.61)  & 1.47(0.74)  & 10.40(5.21)  \\
n3.5                  & k1  & c(0.5x0.5)   & 4.89 & 10.1   & 0.33(0.32)  & 0.47(0.22)  & 0.49(0.46)  & 1.29(0.60)  & 2.76(1.29)   \\
n5.6                  & k1  & c(0.5x0.5)   & 4.89 & 11.5   & 0.35(0.23)  & 0.00(0.00)  & 0.47(0.33)  & 0.81(0.40)  & 1.56(0.76)   \\
\hline
\noalign{\smallskip}
{\bf 3C\,273}         &     &              &      &        &             &             &             &             &              \\
s12.9                 & kqc & c(1.0x1.0)   & 1.5 & 50.3    & 54.70(2.15) & 78.10(2.66) & 125.00(6.74)& 257.80(7.56)& 352.07(10.15)\\
s15.0                 & kqc & c(1.0x1.0)   & 1.5 & 94.2    & 41.40(1.85) & 41.30(1.94) & 77.70(5.21) & 160.40(5.86)& 117.34(4.20) \\
s16.8                 & kqc & c(0.8x0.8)   & 1.5 & 95.3    & 8.78(0.84)  & 7.96(0.88)  & 17.60(2.46) & 34.34(2.74) & 24.79(1.94)  \\
s18.9                 & kqc & c(0.8x0.8)   & 1.5 & 242.9   & 9.33(0.87)  & 11.30(1.01) & 21.30(2.77) & 41.93(3.07) & 11.85(0.85)  \\
s20.4                 & kqc & c(0.8x0.8)   & 1.5 & 536.1   & 8.85(0.85)  & 11.00(1.01) & 11.70(1.96) & 31.55(2.36) & 4.07(0.30)   \\
\hline
\noalign{\smallskip}
{\bf PKS\,1030-357}   &     &              &      &        &             &             &             &             &              \\
s3.5                  & kqc & c(1.5x1.5)   & 8.64 & 16.7   & 0.59(0.30)  & 1.87(0.50)  & 2.27(1.16)  & 4.73(1.30)  & 3.54(0.93)   \\
s6.6                  & kqc & c(1.5x1.5)   & 8.64 & 6.8    & 0.46(0.32)  & 1.44(0.44)  & 1.37(0.81)  & 3.27(0.97)  & 6.08(1.76)   \\
s13.3                 & kqc & c(2.0x2.0)   & 8.64 & 29.7   & 5.01(0.82)  & 6.74(0.89)  & 13.60(2.38) & 25.35(2.67) & 10.8(1.09)   \\
s13.5                 & kqc & c(2.0x2.0)   & 8.64 & 14.8   & 2.09(0.57)  & 4.12(0.70)  & 7.10(1.73)  & 13.30(1.95) & 11.30(1.60)  \\
\noalign{\smallskip}
\hline
\end{tabular}\\
\label{tab:app3}
Note: Col. (2) reports the class component accordingly with the definition of Section~\ref{sec:sample}. In Col. (3), where the region size for the 
flux measurements is shown, the letters `c', `r', and `e' indicate a circular, a rectangular and an elliptical region, respectively. Col. (10) reports the value of the
ratios as defined in Section~\ref{sec:ratios}.  
\end{table}

\begin{table}
\caption{}
\begin{tabular}{|lccccccccc|}
\hline
Component & type & region & $\nu_{\rm R}$ & $S_{\nu}(\nu_{\rm R})$ &$F_{\rm soft}$     &   $F_{\rm med}$   &   $F_{\rm hard}$  & $F_{\rm tot}$     & $\rho$       \\
      &      &(arcsec)&   GHz         &     mJy                &10$^{-15}$ cgs     &10$^{-15}$ cgs     &10$^{-15}$ cgs     &10$^{-15}$ cgs     &              \\   
\hline
\noalign{\smallskip}
{\bf PKS\,1030-357}   &     &              &      &        &             &             &             &             &              \\
s3.5                  & kqc & c(1.5x1.5)   & 8.64 & 16.7   & 0.59(0.30)  & 1.87(0.50)  & 2.27(1.16)  & 4.73(1.30)  & 3.54(0.93)   \\
s6.6                  & kqc & c(1.5x1.5)   & 8.64 & 6.8    & 0.46(0.32)  & 1.44(0.44)  & 1.37(0.81)  & 3.27(0.97)  & 6.08(1.76)   \\
s13.3                 & kqc & c(2.0x2.0)   & 8.64 & 29.7   & 5.01(0.82)  & 6.74(0.89)  & 13.60(2.38) & 25.35(2.67) & 10.8(1.09)   \\
s13.5                 & kqc & c(2.0x2.0)   & 8.64 & 14.8   & 2.09(0.57)  & 4.12(0.70)  & 7.10(1.73)  & 13.30(1.95) & 11.30(1.60)  \\
\hline
\noalign{\smallskip}
{\bf M\,87}           &     &              &      &        &             &             &             &             &              \\
w0.9                  & k1  & r(2.46x1.23) & 4.89 & 211.8  & 37.20(1.09) & 120.00(2.17)& 562.00(8.65)& 719.20(8.98)& 70.06(0.87)  \\
w2.7                  & k1  & r(2.46x1.23) & 4.89 & 220.6  & 66.90(1.38) & 90.30(1.81) & 112.00(3.86)& 269.20(4.48)& 25.80(0.42)  \\
w3.9                  & k1  & r(1.72x1.18) & 4.89 & 151.2  & 20.50(0.79) & 24.50(1.09) & 25.90(2.09) & 70.90(2.49) & 9.96(0.34)   \\
w6.0                  & k1  & r(2.31x1.23) & 4.89 & 239.5  & 39.20(1.08) & 40.20(1.33) & 31.70(2.19) & 111.10(2.78)& 9.92(0.24)   \\
w8.5                  & k1  & r(1.72x1.23) & 4.89 & 288.3  & 25.20(0.84) & 18.60(0.92) & 11.20(1.36) & 55.00(1.85) & 4.12(0.13)   \\
w9.9                  & k1  & r(0.98x1.23) & 4.89 & 130.1  & 3.44(0.378) & 3.56(0.54)  & 2.42(0.68)  & 9.42(0.95)  & 1.55(0.15)   \\
w10.9                 & k1  & r(0.98x1.23) & 4.89 & 202.5  & 9.07(0.54)  & 8.47(0.69)  & 7.41(1.13)  & 24.95(1.43) & 2.64(0.15)   \\
w12.5                 & k1  & r(2.12x2.12) & 4.89 & 1837.5 & 173.00(2.10)& 180.00(2.48)& 189.00(4.87)& 542.00(5.85)& 6.28(0.07)   \\
w14.4                 & k1  & r(1.72x1.48) & 4.89 & 1093.2 & 33.40(0.97) & 33.40(1.18) & 25.70(1.92) & 92.50(2.45) & 1.81(0.05)   \\
w17.6                 & k1  & r(1.72x1.23) & 4.89 & 850.3  & 7.63(0.50)  & 9.18(0.74)  & 4.17(0.96)  & 20.98(1.31) & 0.53(0.03)   \\
w20.2                 & k1  & e(0.69x0.84) & 4.89 & 240.0  & 3.35(0.37)  & 5.22(0.63)  & 7.73(1.16)  & 16.30(1.37) & 1.43(0.12)   \\
w26.1                 & k1  & e(0.69x0.84) & 4.89 & 166.7  & 2.01(0.30)  & 4.12(0.58)  & 14.20(1.49) & 20.33(1.63) & 2.53(0.20)   \\
\hline
\noalign{\smallskip}
{\bf PKS\,1229-027}   &     &              &      &        &             &             &             &             &              \\
w2.0                  & kqc & c(0.8x0.8)   & 4.75 & 113.9  & 3.15(0.85)  & 5.12(1.18)  & 9.45(3.49)  & 17.72(3.78) & 3.37(0.70)   \\
\hline
\noalign{\smallskip}
{\bf 3C\,275.1}       &     &              &      &        &             &             &             &             &              \\
n3.5                  & kql & c(1.2x1.2)   & 8.44 & 1.4    & 0.91(0.37)  & 0.55(0.28)  & 2.50(1.27)  & 3.96(1.35)  & 34.89(11.71) \\
n9.5                  & hsq & c(1.2x1.2)   & 8.44 & 155.68 & 2.31(0.49)  & 2.54(0.59)  & 6.29(1.76)  & 11.14(1.92) & 0.87(0.15)   \\
\hline
\noalign{\smallskip}
{\bf 3C\,280}         &     &              &      &        &             &             &             &             &              \\
e5.0                  & hs2 & c(1.5x1.5)   & 4.89 & 339.1  & 0.40(0.13)  & 0.41(0.17)  & 1.17(0.60)  & 1.98(0.63)  & 0.12(0.04)   \\
w6.0                  & k2  & c(1.0x1.0)   & 4.89 & 46.2   & 0.04(0.08)  & 0.10(0.09)  & 0.74(0.43)  & 0.88(0.45)  & 0.39(0.20)   \\
w8.0                  & hs2 & c(1.0x1.0)   & 4.89 & 1089.8 & 0.96(0.19)  & 0.89(0.23)  & 1.15(0.59)  & 3.00(0.66)  & 0.06(0.01)   \\
\hline
\noalign{\smallskip}
{\bf 1317-520}        &     &              &      &        &             &             &             &             &              \\
e10.0                 & kqc & c(1.5x1.5)   & 1.50 & 41.9   & 4.49(0.81)  & 3.23(0.77)  & 7.86(2.54)  & 15.58(2.77) & 25.46(4.45)  \\
\hline
\noalign{\smallskip}
{\bf 3C\,281}         &     &              &      &        &             &             &             &             &              \\
n13.5                 & kql & c(1.5x1.5)   & 4.86 & 18.8   & 0.52(0.30)  & 0.00(0.00)  & 1.60(1.13)  & 2.12(1.17)  & 2.32(1.28)   \\
n16.5                 & hsq & c(1.5x1.5)   & 4.86 & 79.0   & 0.12(0.12)  & 0.28(0.28)  & 0.00(0.00)  & 0.40(0.31)  & 0.11(0.09)   \\
\hline
\noalign{\smallskip}
{\bf Centaurus A}     &     &              &      &        &             &             &             &             &              \\
e15.0                 & k1  & c(2.5x2.5)   & 4.80 & 315.7  & 20.40(1.40) & 80.10(1.99) & 188.00(6.11)& 288.50(6.58)& 20.30(0.45)  \\
e29.0                 & k1  & c(2.5x2.5)   & 4.80 & 262.4  & 3.24(0.63)  & 11.30(0.77) & 31.80(2.67) & 46.34(2.85) & 3.91(0.24)   \\
\hline
\noalign{\smallskip}
{\bf 4C\,+65.15}     &     &              &      &        &             &             &             &             &              \\
s4.5                 & kql & c(0.8x0.8)   & 8.46 & 24.4   & 0.17(0.12)  & 0.81(0.29)  & 0.99(0.57)  & 1.97(0.65)  & 0.97(0.32)   \\
w5.0                 & kql & c(0.8x0.8)   & 8.46 & 9.8    & 0.10(0.10)  & 0.37(0.19)  & 0.00(0.00)  & 0.47(0.21)  & 0.59(0.26)   \\
\hline
\noalign{\smallskip}
{\bf 3C\,287.1}      &     &              &      &        &             &             &             &             &              \\
w65.0                & hs2 & c(5.0x5.0)   & 4.86 & 91.3   & 1.85(0.92)  & 1.37(0.97)  & 4.62(3.27)  & 7.84(3.53)  & 1.82(0.80)   \\
\hline
\noalign{\smallskip}
{\bf Centaurus B}    &     &              &      &        &             &             &             &             &              \\
w3.4                 & k1  & c(1.0x1.0)   & 8.64 & 46.2   & 0.71(0.71)  & 0.25(0.56)  & 7.90(4.56)  & 8.86(4.65)  & 14.32(12.22) \\
w5.7                 & k1  & c(1.0x1.0)   & 8.64 & 52.1   & 0.75(0.75)  & 5.94(2.10)  & 8.68(6.21)  & 15.4(6.60)  & 17.09(11.52) \\
\hline
\noalign{\smallskip}
{\bf 4C\,+19.44}     &     &              &      &        &             &             &             &             &              \\
n16.0                & hsq & r(2.46x1.72) & 4.86 & 55.4   & 0.15(0.09)  & 0.23(0.12)  & 0.49(0.43)  & 0.87(0.45)  & 0.33(0.17)   \\
s4.0                 & kqc & r(1.54x1.72) & 4.86 & 17.5   & 0.48(0.20)  & 0.91(0.30)  & 1.06(0.89)  & 2.45(0.96)  & 3.01(1.15)   \\
s5.3                 & kqc & r(1.08x1.72) & 4.86 & 7.1    & 0.04(0.10)  & 0.29(0.16)  & 0.00(0.00)  & 0.33(0.19)  & 1.02(0.58)   \\
s6.6                 & kqc & r(1.52x1.48) & 4.86 & 10.8   & 0.58(0.19)  & 0.73(0.20)  & 1.57(0.57)  & 2.88(0.63)  & 5.67(1.22)   \\
s8.3                 & kqc & r(1.87x1.72) & 4.86 & 19.1   & 0.77(0.20)  & 0.76(0.22)  & 1.19(0.58)  & 2.72(0.65)  & 3.07(0.71)   \\
s10.0                & kqc & r(1.38x1.72) & 4.86 & 6.2    & 0.29(0.13)  & 0.58(0.19)  & 1.26(0.52)  & 2.13(0.57)  & 7.24(1.90)   \\
s11.2                & kqc & r(1.23x1.72) & 4.86 & 5.4    & 0.13(0.07)  & 0.20(0.10)  & 0.53(0.38)  & 0.86(0.40)  & 3.38(1.53)   \\
s12.9                & kqc & r(1.87x1.72) & 4.86 & 9.9    & 0.37(0.16)  & 1.22(0.25)  & 1.71(0.66)  & 3.30(0.72)  & 7.02(1.51)   \\
s14.6                & kqc & r(1.48x1.72) & 4.86 & 9.3    & 0.59(0.17)  & 0.77(0.19)  & 1.19(0.50)  & 2.55(0.56)  & 5.92(1.26)   \\
s15.9                & kqc & r(1.23x1.72) & 4.86 & 2.8    & 0.72(0.19)  & 0.95(0.23)  & 0.82(0.45)  & 2.49(0.54)  & 19.36(4.07)  \\
s17.7                & kqc & r(2.46x1.72) & 4.86 & 2.5    & 1.45(0.27)  & 1.04(0.25)  & 2.67(0.78)  & 5.16(0.86)  & 44.78(7.27)  \\
s25.7                & kqc & c(1.33x1.33) & 4.86 & 3.1    & 0.49(0.15)  & 0.25(0.11)  & 1.82(0.76)  & 2.56(0.78)  & 17.02(5.06)  \\
s28.0                & hsq & c(1.72x1.72) & 4.86 & 86.1   & 1.53(0.28)  & 2.15(0.33)  & 4.58(1.03)  & 8.26(1.12)  & 2.04(0.27)   \\
\hline
\noalign{\smallskip}
{\bf 3C\,294}        &     &              &      &        &             &             &             &             &              \\
s9.0                 & hs2 & c(1.0x1.0)   & 4.82 & 41.9   & 0.19(0.07)  & 0.29(0.09)  & 0.20(0.20)  & 0.68(0.23)  & 0.35(0.11)   \\
n6.0                 & hs2 & c(1.0x1.0)   & 4.82 & 170.5  & 0.11(0.06)  & 0.14(0.08)  & 0.43(0.30)  & 0.68(0.32)  & 0.08(0.04)   \\
\noalign{\smallskip}
\hline
\end{tabular}\\
\label{tab:app4}
Note: Col. (2) reports the class component accordingly with the definition of Section~\ref{sec:sample}. In Col. (3), where the region size for the 
flux measurements is shown, the letters `c', `r', and `e' indicate a circular, a rectangular and an elliptical region, respectively. Col. (10) reports the value of the
ratios as defined in Section~\ref{sec:ratios}.  
\end{table}

\begin{table}
\caption{}
\begin{tabular}{|lccccccccc|}
\hline
Component & type & region & $\nu_{\rm R}$ & $S_{\nu}(\nu_{\rm R})$ &$F_{\rm soft}$     &   $F_{\rm med}$   &   $F_{\rm hard}$  & $F_{\rm tot}$     & $\rho$       \\
      &      &(arcsec)&   GHz         &     mJy                &10$^{-15}$ cgs     &10$^{-15}$ cgs     &10$^{-15}$ cgs     &10$^{-15}$ cgs     &              \\   
\hline
\noalign{\smallskip}
{\bf 3C\,295}        &     &              &      &        &             &             &             &             &              \\
n1.9                 & hs2 & c(0.5x0.5)   & 4.99 & 2233.4 & 1.90(0.54)  & 2.29(0.43)  & 4.30(1.03)  & 8.49(1.24)  & 0.08(0.01)   \\
s2.5                 & hs2 & c(0.5x0.5)   & 4.99 & 1987.1 & 0.90(0.34)  & 1.55(0.34)  & 1.98(0.81)  & 4.43(0.94)  & 0.05(0.01)   \\
\hline
\noalign{\smallskip}
{\bf 3C\,296}        &     &              &      &        &             &             &             &             &              \\
n3.2                 & k1  & c(1.0x1.0)   & 8.44 & 2.1    & 0.24(0.39)  & 0.87(0.32)  & 1.21(0.93)  & 2.32(1.06)  & 13.28(6.02)  \\
n6.0                 & k1  & c(1.0x1.0)   & 8.44 & 4.7    & 0.18(0.17)  & 0.78(0.24)  & 2.58(0.13)  & 0.97(0.32)  & 2.56(0.85)   \\
\hline
\noalign{\smallskip}
{\bf PKS\,1421-490}  &     &              &      &        &             &             &             &             &              \\
n5.9                 & hsq & c(1.0x1.0)   & 8.64 & 3088.8 & 7.66(0.83)  & 16.10(1.10) & 57.70(4.07) & 81.46(4.30) & 0.36(0.02)   \\
\hline
\noalign{\smallskip}
{\bf 3C\,303}        &     &              &      &        &             &             &             &             &              \\
e17.0                & k1  & c(2.0x2.0)   & 1.45 & 5.8    & 5.53(0.16)  & 0.45(0.42)  & 1.05(1.05)  & 1.50(1.14)  & 17.79(13.54) \\
w2.8                 & k1  & c(1.25x1.25) & 1.45 & 24.1   & 0.95(0.67)  & 0.00(0.00)  & 0.00(0.00)  & 0.95(0.67)  & 2.98(2.11)   \\
w5.6                 & k1  & c(1.25x1.25) & 1.45 & 22.0   & 1.14(0.47)  & 0.73(0.36)  & 3.52(2.07)  & 5.39(2.15)  & 17.26(6.78)  \\
w8.8                 & k1  & c(1.25x1.25) & 1.45 & 17.7   & 1.58(0.56)  & 0.95(0.48)  & 0.00(0.00)  & 2.53(0.74)  & 10.54(3.07)  \\
w17.0                & k1  & c(2.0x2.0)   & 1.45 & 745.0  & 4.25(0.87)  & 5.90(1.19)  & 16.60(3.82) & 26.75(4.09) & 2.53(0.38)   \\
\hline
\noalign{\smallskip}
{\bf 1508+572}       &     &              &      &        &             &             &             &             &              \\
w2.6                 & kqc & c(1.2x1.2)   & 1.43 & 1.3    & 1.36(0.21)  & 1.62(0.27)  & 2.87(0.75)  & 5.85(0.83)  & 331.72(45.86)\\
\hline
\noalign{\smallskip}
{\bf PKS\,1510-089}  &     &              &      &        &             &             &             &             &              \\
s2.4                 & kqc & r(1.7x1.0)   & 4.86 & 11.5   & 5.10(1.47)  & 10.20(2.12) & 19.40(7.85) & 34.70(8.26) & 67.99(15.18) \\
s5.0                 & kqc & r(2.0x1.7)   & 4.86 & 17.0   & 3.29(1.07)  & 6.33(1.61)  & 11.60(5.16) & 21.23(5.51) & 28.11(6.85)  \\
\hline
\noalign{\smallskip}
{\bf 3C\,321}        &     &              &      &        &             &             &             &             &              \\
n147.0               & hs2 & c(2.0x2.0)   & 1.51 & 74.0   & 0.27(0.14)  & 0.12(0.12)  & 0.89(0.52)  & 1.28(0.55)  & 1.21(0.50)   \\
s137.0               & hs2 & c(3.0x3.0)   & 1.51 & 489.6  & 0.42(0.17)  & 0.33(0.19)  & 1.88(0.86)  & 2.64(0.90)  & 0.38(0.12)   \\
\hline
\noalign{\smallskip}
{\bf 3C\,327}        &     &              &      &        &             &             &             &             &              \\
e99.0                & hs2 & c(5.0x5.0)   & 8.47 & 136.3  & 0.58(0.24)  & 0.77(0.30)  & 7.72(1.86)  & 9.07(1.90)  & 0.81(0.17)   \\
\hline
\noalign{\smallskip}
{\bf 4C\,+00.58}    &     &              &      &        &             &             &             &             &              \\
e3.0                & k1  & c(1.2x1.2)   & 4.86 & 43.8   & 1.21(0.23)  & 1.28(0.25)  & 3.44(0.68)  & 5.93(0.76)  & 3.10(0.38)   \\
\hline
\noalign{\smallskip}
{\bf 3C\,330}        &     &              &      &        &             &             &             &             &              \\
e30.0                & hs2 & c(2.5x2.5)   & 8.45 & 703.7  & 0.29(0.13)  & 0.81(0.24)  & 1.08(0.55)  & 2.17(0.62)  & 0.04(0.01)   \\
w32.0                & hs2 & c(2.5x2.5)   & 8.45 & 189.0  & 0.09(0.06)  & 0.27(0.13)  & 0.00(0.00)  & 0.35(0.15)  & 0.02(0.01)   \\
\hline
\noalign{\smallskip}
{\bf NGC\,6251}      &     &              &      &        &             &             &             &             &              \\
w6.0                 & k1  & c(1.48x1.48) & 1.49 & 6.5    & 1.26(0.46)  & 1.13(0.39)  & 2.93(1.32)  & 5.32(1.45)  & 60.04(15.59) \\
w9.0                 & k1  & c(1.48x1.48) & 1.49 & 3.4    & 1.18(0.32)  & 0.79(0.32)  & 2.88(1.11)  & 4.85(1.20)  & 104.92(24.45)\\
w15.5                & k1  & c(1.48x1.48) & 1.49 & 6.5    & 1.98(0.37)  & 1.74(0.36)  & 2.74(0.88)  & 6.46(1.02)  & 75.42(11.23) \\
w20.0                & k1  & c(1.48x1.48) & 1.49 & 9.3    & 0.85(0.25)  & 0.93(0.27)  & 3.15(1.01)  & 4.93(1.07)  & 38.35(8.00)  \\
w23.0                & k1  & c(1.48x1.48) & 1.49 & 10.9   & 0.66(0.24)  & 0.78(0.24)  & 1.38(0.71)  & 2.82(0.78)  & 19.09(5.01)  \\
\hline
\noalign{\smallskip}
{\bf 1642+690}       &     &              &      &        &             &             &             &             &              \\
s2.9                 & kqc & c(0.9x0.9)   & 4.86 & 48.2   & 2.00(0.83)  & 2.84(1.30)  & 16.60(6.42) & 21.44(6.60) & 9.51(2.84)   \\
\hline
\noalign{\smallskip}
{\bf 3C\,345}        &     &              &      &        &             &             &             &             &              \\
n2.8                 & kqc & c(1.0x1.0)   & 4.86 & 241.8  & 5.80(1.47)  & 0.31(1.60)  & 18.70(6.66) & 24.81(7.01) & 2.15(0.60)   \\
\hline
\noalign{\smallskip}
{\bf 3C\,346}        &     &              &      &        &             &             &             &             &              \\
e2.0                 & k1  & c(0.8x0.8)   & 1.53 & 538.4  & 1.39(0.35)  & 1.07(0.33)  & 3.37(1.19)  & 5.83(1.28)  & 0.77(0.16)   \\
\hline
\noalign{\smallskip}
{\bf 3C\,349}        &     &              &      &        &             &             &             &             &              \\
s38.0                & hs2 & c(2.0x2.0)   & 1.5 & 112.4   & 0.42(0.42)  & 0.70(0.50)  & 1.53(1.53)  & 2.65(1.66)  & 1.61(0.99)   \\
s43.0                & hs2 & c(2.0x2.0)   & 1.5 & 911.1   & 0.32(0.32)  & 1.63(0.81)  & 1.01(1.01)  & 2.96(1.34)  & 0.22(0.10)   \\
\hline
\noalign{\smallskip}
{\bf 3C\,351}        &     &              &      &        &             &             &             &             &              \\
n25.0                & hsq & c(2.0x2.0)   & 1.42 & 548.2  & 5.84(0.55)  & 7.23(0.68)  & 12.50(1.82) & 25.57(2.02) & 3.39(0.26)   \\
n27.0                & hsq & c(2.0x2.0)   & 1.42 & 1245.4 & 4.97(0.50)  & 6.13(0.62)  & 8.45(1.59)  & 19.55(1.78) & 1.15(0.10)   \\
\hline
\noalign{\smallskip}
{\bf 3C\,353}        &     &              &      &        &             &             &             &             &              \\
e21.0                & k2  & c(1.2x1.2)   & 8.44 & 2.0    & 0.04(0.06)  & 0.21(0.11)  & 0.81(0.57)  & 1.06(0.58)  & 6.66(3.53)   \\
e23.0                & k2  & c(1.2x1.2)   & 8.44 & 5.5    & 0.24(0.12)  & 0.95(0.22)  & 1.29(0.53)  & 2.48(0.59)  & 5.97(1.34)   \\
e70.0                & k2  & c(1.5x1.5)   & 8.44 & 7.9    & 0.00(0.00)  & 0.00(0.00)  & 1.57(0.73)  & 1.57(0.73)  & 2.35(1.10)   \\
e73.0                & k2  & c(1.5x1.5)   & 8.44 & 16.9   & 0.17(0.10)  & 0.10(0.07)  & 0.30(0.34)  & 0.57(0.36)  & 0.49(0.27)   \\
e88.0                & k2  & c(1.5x1.5)   & 8.44 & 15.3   & 0.20(0.11)  & 0.67(0.20)  & 1.51(0.65)  & 2.38(0.69)  & 2.01(0.55)   \\
w47.0                & k2  & c(2.0x2.0)   & 8.44 & 17.1   & 0.27(0.12)  & 0.61(0.19)  & 2.48(0.77)  & 3.37(0.80)  & 2.51(0.57)   \\
\hline
\noalign{\smallskip}
{\bf 4C\,+62.29}     &     &              &      &        &             &             &             &             &              \\
s1.4                 & kql & c(0.5x0.5)   & 4.86 & 14.4   & 6.43(1.16)  & 9.26(1.51)  & 20.6(4.38)  & 36.29(4.78) & 54.37(6.96)  \\
s2.4                 & kql & c(0.5x0.5)   & 4.86 & 36.1   & 2.09(0.66)  & 3.42(0.92)  & 4.22(2.11)  & 9.73(2.39)  & 5.86(1.39)   \\
\hline
\noalign{\smallskip}
{\bf 1800+440}       &     &              &      &        &             &             &             &             &              \\
w3.2                 & kqc & c(1.3x1.3)   & 1.51 & 39.4   & 5.67(1.67)  & 9.29(1.93)  & 14.70(5.32) & 29.66(5.90) & 52.09(10.11) \\
\noalign{\smallskip}
\hline
\end{tabular}\\
\label{tab:app5}
Note: Col. (2) reports the class component accordingly with the definition of Section~\ref{sec:sample}. In Col. (3), where the region size for the 
flux measurements is shown, the letters `c', `r', and `e' indicate a circular, a rectangular and an elliptical region, respectively. Col. (10) reports the value of the
ratios as defined in Section~\ref{sec:ratios}.  
\end{table}

\begin{table}
\caption{}
\begin{tabular}{|lccccccccc|}
\hline
Component & type & region & $\nu_{\rm R}$ & $S_{\nu}(\nu_{\rm R})$ &$F_{\rm soft}$     &   $F_{\rm med}$   &   $F_{\rm hard}$  & $F_{\rm tot}$     & $\rho$       \\
      &      &(arcsec)&   GHz         &     mJy                &10$^{-15}$ cgs     &10$^{-15}$ cgs     &10$^{-15}$ cgs     &10$^{-15}$ cgs     &              \\   
\hline
\noalign{\smallskip}
{\bf 3C\,371}        &     &              &      &        &             &             &             &             &              \\
w1.7                 & kbl & c(0.7x0.7)   & 4.86 & 41.5   & 25.20(1.65) & 29.80(2.10) & 34.70(5.87) & 89.70(6.45) & 48.52(3.28)  \\
w3.1                 & kbl & c(1.0x1.0)   & 4.86 & 98.8   & 16.30(1.14) & 16.80(1.29) & 24.20(3.08) & 57.30(3.53) & 12.99(0.76)  \\
\hline
\noalign{\smallskip}
{\bf 3C\,380}        &     &              &      &        &             &             &             &             &              \\
n1.2                 & kqc & c(0.6x0.6)   & 1.66 & 2821.6 & 36.80(6.03) & 52.60(7.90) & 84.00(23.00)&173.40(25.06)& 4.08(0.55)   \\
\hline
\noalign{\smallskip}
{\bf 3C\,390.3}      &     &              &      &        &             &             &             &             &              \\
n103.0               & k2  & c(2.5x2.5)   & 1.45 & 209.8  & 4.55(0.54)  & 7.94(0.88)  & 14.80(2.65) & 27.29(2.84) & 9.42(0.95)   \\
s87.0                & hs2 & c(5.0x5.0)   & 1.45 & 2786.9 & 1.42(0.34)  & 2.86(0.57)  & 5.49(2.14)  & 9.77(2.24)  & 0.25(0.06)   \\
\hline
\noalign{\smallskip}
{\bf 1849+670}       &     &              &      &        &             &             &             &             &              \\
n5.0                 & kqc & c(1.81.8)    & 1.40 & 4.6    & 1.67(0.75)  & 1.52(0.89)  & 0.00(0.00)  & 3.19(1.16)  & 58.88(21.07) \\
\hline
\noalign{\smallskip}
{\bf 4C\,73.18}      &     &              &      &        &             &             &             &             &              \\
s2.6                 & kqc & c(1.0x1.0)   & 1.43 & 30.1   & 4.79(1.53)  & 8.53(2.32)  & 0.73(4.45)  & 14.05(5.25) & 39.60(12.97) \\
\hline
\noalign{\smallskip}
{\bf 3C\,403}        &     &              &      &        &             &             &             &             &              \\
e28.0                & k2  & c(2.0x2.0)   & 8.47 & 44.0   & 1.34(0.30)  & 2.58(0.43)  & 4.73(1.20)  & 8.65(1.31)  & 2.78(0.38)   \\
e48.0                & hs2 & c(1.23x1.23) & 8.47 & 13.2   & 0.40(0.16)  & 0.60(0.23)  & 0.84(0.60)  & 1.83(0.67)  & 2.08(0.66)   \\
e51.0                & hs2 & c(1.23x1.23) & 8.47 & 31.0   & 1.07(0.28)  & 0.72(0.23)  & 1.39(0.62)  & 3.18(0.72)  & 1.65(0.33)   \\
\hline
\noalign{\smallskip}
{\bf Cygnus A}       &     &              &      &        &             &             &             &             &              \\
e53.2                & hs2 & c(1.23x1.23) & 4.53 & 3078.0 & 0.00(0.00)  & 0.87(0.54)  & 3.37(1.91)  & 4.24(1.98)  & 0.033(0.01)  \\
e57.7                & hs2 & c(2.5x2.5)   & 4.53 & 59741.4& 10.8(0.95)  & 42.70(2.21) & 105.00(6.92)& 158.50(7.33)& 0.076(0.003) \\
w63.3                & hs2 & c(1.48x1.48) & 4.53 & 9109.6 & 1.28(0.30)  & 5.43(0.90)  & 14.90(2.96) & 21.61(3.12) & 0.067(0.009) \\
w67.0                & hs2 & c(2.5x2.5)   & 4.53 & 45903.4& 8.37(0.82)  & 27.50(1.76) & 76.70(5.79) & 112.57(6.11)& 0.071(0.003) \\
\hline
\noalign{\smallskip}
{\bf 2007+777}       &     &              &      &        &             &             &             &             &              \\
w5.0                 & kbl & c(1.2x1.2)   & 1.49 & 2.5    & 0.30(0.30)  & 0.74(0.33)  & 3.86(1.39)  & 4.90(1.46)  & 137.67(40.46)\\
w8.5                 & kbl & c(1.2x1.2)   & 1.49 & 3.0    & 1.04(0.32)  & 1.11(0.38)  & 5.25(1.42)  & 7.40(1.50)  & 180.90(34.86)\\
w11.6                & kbl & c(1.2x1.2)   & 1.49 & 1.9    & 0.33(0.20)  & 0.71(0.29)  & 1.92(0.98)  & 2.96(1.03)  & 114.95(38.02)\\
\hline
\noalign{\smallskip}
{\bf PKS\,2101-490}  &     &              &      &        &             &             &             &             &              \\
e5.0                 & kqc & c(1.2x1.2)   & 8.64 & 5.6    & 0.39(0.23)  & 0.67(0.26)  & 2.33(0.97)  & 3.39(1.03)  & 7.25(2.16)   \\
e8.0                 & kqc & c(1.2x1.2)   & 8.64 & 6.3    & 0.97(0.33)  & 0.99(0.28)  & 1.84(0.75)  & 3.80(0.87)  & 7.37(1.64)   \\
e12.0                & hsq & c(2.0x2.0)   & 8.64 & 46.3   & 1.26(0.39)  & 1.80(0.41)  & 1.86(0.92)  & 4.92(1.07)  & 1.30(0.28)   \\
\hline
\noalign{\smallskip}
{\bf PKS\,2153-69}   &     &              &      &        &             &             &             &             &              \\
s26.0                & hs2 & c(2.5x2.5)   & 8.64 & 338.2  & 2.58(0.73)  & 3.02(0.88)  & 1.34(1.34)  & 6.94(1.76)  & 0.26(0.06)   \\
n10.3                & k2  & c(1.0x1.0)   & 8.64 & 8.3    & 0.93(0.42)  & 2.13(0.75)  & 1.27(0.90)  & 4.33(1.25)  & 6.39(1.80)   \\
n47.0                & hs2 & c(2.5x2.5)   & 8.64 & 89.3   & 1.04(0.47)  & 2.10(0.76)  & 2.32(1.64)  & 5.46(1.87)  & 0.74(0.25)   \\
\hline
\noalign{\smallskip}
{\bf 2155-152}       &     &              &      &        &             &             &             &             &              \\
s3.5                 & kqc & c(1.2x1.2)   & 1.40 & 92.1   & 1.16(0.82)  & 1.46(1.04)  & 3.30(2.84)  & 5.92(3.13)  & 4.81(2.47)   \\
\hline
\noalign{\smallskip}
{\bf PKS\,2201+044}  &     &              &      &        &             &             &             &             &              \\
n2.2                 & kqc & c(0.9x0.9)   & 8.46 & 15.1   & 11.90(0.94) & 14.20(1.15) & 20.60(2.76) & 46.70(3.13) & 39.44(2.53)  \\
\hline
\noalign{\smallskip}
{\bf 2209+080}       &     &              &      &        &             &             &             &             &              \\
s4.4                 & kqc & c(1.5x1.5)   & 4.86 & 64.1   & 1.53(0.55)  & 1.70(0.65)  & 4.00(1.83)  & 7.23(2.02)  & 2.53(0.67)   \\
\hline
\noalign{\smallskip}
{\bf 2216-038}       &     &              &      &        &             &             &             &             &              \\
s8.0                 & kqc & c(2.0x2.0)   & 1.40 & 105.9  & 0.79(0.57)  & 2.39(0.99)  & 7.51(3.82)  & 10.69(3.99) & 7.49(2.72)   \\
s12.4                & kqc & c(2.0x2.0)   & 1.40 & 39.7   & 0.00(0.00)  & 1.50(0.75)  & 9.67(4.32)  & 11.17(4.38) & 20.3(7.91)   \\
\hline
\noalign{\smallskip}
{\bf 3C\,445}        &     &              &      &        &             &             &             &             &              \\
s275.0               & hs2 & c(9.0x9.0)   & 1.43 & 564.7  & 4.69(0.60)  & 5.36(0.70)  & 10.30(2.26) & 20.35(2.44) & 2.70(0.31)   \\
\hline
\noalign{\smallskip}
{\bf 3C\,452}        &     &              &      &        &             &             &             &             &              \\
w130.0               & hs2 & c(3.0x3.0)   & 8.47 & 54.0   & 0.33(0.13)  & 0.53(0.16)  & 0.16(0.35)  & 1.02(0.41)  & 0.29(0.10)   \\
\hline
\noalign{\smallskip}
{\bf 3C\,454.3}      &     &              &      &        &             &             &             &             &              \\
n2.3                 & kqc & c(0.8x0.8)   & 8.46 & 19.0   & 5.95(1.25)  & 7.72(1.55)  & 10.30(3.98) & 23.97(4.45) & 16.93(2.91)  \\
w5.4                 & kqc & c(1.2x1.2)   & 8.46 & 138.3  & 6.94(1.18)  & 10.60(1.43) & 28.60(4.86) & 46.14(5.20) & 4.28(0.46)   \\
\hline
\noalign{\smallskip}
{\bf 3C\,465}        &     &              &      &        &             &             &             &             &              \\
n5.5                 & k1  & c(1.5x1.5)   & 8.47 & 5.2    & 1.76(0.44)  & 0.75(0.31)  & 2.07(0.85)  & 4.58(1.01)  & 11.79(2.48)  \\
n14.6                & k1  & c(1.5x1.5)   & 8.47 & 2.2    & 0.17(0.16)  & 0.20(0.18)  & 0.24(0.48)  & 0.61(0.53)  & 3.62(2.98)   \\
\noalign{\smallskip}
\hline
\end{tabular}\\
\label{tab:app6}
Note: Col. (2) reports the class component accordingly with the definition of Section~\ref{sec:sample}. In Col. (3), where the region size for the 
flux measurements is shown, the letters `c', `r', and `e' indicate a circular, a rectangular and an elliptical region, respectively. Col. (10) reports the value of the
ratios as defined in Section~\ref{sec:ratios}.  
\end{table}

\begin{thebibliography}{}
\bibitem[Aldcroft et al. 2003]{aldcroft03} Aldcroft, T.L., Siemiginowska, A., Elvis, M., Mathur, S., Nicastro, F., \& Murray, S.S. 2003 ApJ, 597 751                            
\bibitem[Barthel 1989]{barthel89} Barthel, P. D. 1989 ApJ, 336, 606
\bibitem[Bergamini et al. 1967]{bergamini67} Bergamini, R., Londrillo, P., Setti, G. 1967, NCimB, 52, 495
\bibitem[Birkinshaw et al. 2002]{birkinshaw02} Birkinshaw, M., Worrall, D.M., \& Hardcastle, M.J. 2002 MNRAS, 335, 142                                                          
\bibitem[Bondi et al. 2004]{bondi04} Bondi, M., Brunetti, G., Comastri, A., \& Setti, G. 2004, MNRAS, 354, L43                                                                  
\bibitem[Bridle 1986]{bridle86} Bridle, A. H. 1986, Can. J. Phys., 64, 353
\bibitem[Brunetti et al. 2002]{brunetti02} Brunetti, G., Bondi, M., Comastri, A., \& Setti, G. 2002, A\&A, 381, 795                                                             
\bibitem[Celotti et al. 2001]{celotti01} Celotti, A., Ghisellini, G., \& Chiaberge, M.\ 2001, MNRAS, 321, L1 
\bibitem[Chartas et al. 2002]{chartas02} Chartas, G., Gupta, V., Garmire, G., Jones, C., Falco, E.E., Shapiro,I.I., \& Tavecchio, F. 2002, ApJ, 565, 96                         
\bibitem[Cheung 2004]{cheung04} Cheung, C.C. 2004, ApJL, 600, L23                                                                                                               
\bibitem[Cheung et al. 2006]{cheung06} Cheung, C.C., Stawarz, L., \& Siemiginowska, A. 2006 ApJ, 650, 679                                                                       
\bibitem[Chiaberge et al. 2003]{chiaberge03} Chiaberge, M., Gilli, R., Macchetto, F.D., Sparks, W.B., \& Capetti, A. 2003 ApJ, 582, 645                                         
\bibitem[Comastri et al. 2003]{comastri03} Comastri, A., Brunetti, G., Dallacasa, D., Bondi, M., Pedani, M., \& Setti, G. 2003, MNRAS, 340, L52                                 
\bibitem[Crawford \& Fabian 2003]{crawford03} Crawford, C.S., \& Fabian, A.C. 2003, MNRAS, 339, 1163                                                                            %
\bibitem[Dermer 1995]{dermer95} Dermer, C. D. 1995 ApJ, 446, L63
\bibitem[Donahue et al. 2003]{donahue03} Donahue, M., Daly, R.A., \& Horner, D.J. 2003, ApJ, 584, 643                                                                          
\bibitem[Dunkley et al. 2009]{dunkley09} Dunkley, J., et al. 2009 ApJS, 180, 306
\bibitem[Evans et al. 2005]{evans05} Evans, D.A., Hardcastle, M.J., Croston, J.H., Worrall, D.M., \& Birkinshaw, M. 2005, MNRAS, 359, 363                                       
\bibitem[Fabian et al. 2003a]{fabian03a} Fabian, A.C., Celotti, A., \& Johnstone, R.M. 2003a MNRAS, 338, L7                                                                     
\bibitem[Fabian et al. 2003b]{fabian03b} Fabian, A.C., Sanders, J.S., Crawford, C.S., \& Ettori, S. 2003b MNRAS, 341, 729                                                        
\bibitem[Fanaroff \& Riley 1974]{fanaroff74} Fanaroff, B.~L. \& Riley J.~M. 1974, MNRAS, 167, P31
\bibitem[Finoguenov et al. 2008]{finoguenov08} Finoguenov, A., Ruszkowski, M., Jones, C., BrŸggen, M., Vikhlinin, A., Mandel, E.
\bibitem[Gelbord et al. 2005]{gelbord05} Gelbord, J.M. et al. 2005, ApJ, 632, L75                                                                                               
\bibitem[Hardcastle et al. 2001a]{hardcastle01a} Hardcastle, M.J., Birkinshaw, M., \& Worrall, D.M. 2001 MNRAS, 323, L17                                                        
\bibitem[Hardcatle et al. 2001b]{hardcastle01b} Hardcastle, M.J., Birkinshaw, M., \& Worrall, D.M. 2001 MNRAS, 326, 1499                                                        
\bibitem[Hardcastle et al. 2002a]{hardcastle02a} Hardcastle, M.J., Worrall, D.M., Birkinshaw, M., Laing, R.A., \& Bridle, A.H. 2002a MNRAS, 334, 182                            
\bibitem[Hardcastle et al. 2002b]{hardcastle02b} Hardcastle, M.J., Birkinshaw, M., Cameron, R.A., Harris, D.E., Looney, L.W., \& Worrall, D.M. 2002b, ApJ, 581, 948             %
\bibitem[Hardcastle e al. 2004]{hardcastle04} Hardcastle, M.~J., Harris, D.~E., Worrall, D.~M., Birkinshaw, M. 2004, ApJ, 612, 729
\bibitem[Hardcastle et al. 2005a]{hardcastle05a} Hardcastle, M.J., Worrall, D.M., Birkinshaw, M, Laing, R.A., \& Bridle, A.H. 2005a MNRAS, 358, 843                             
\bibitem[Hardcastle et al. 2005b]{hardcastle05b} Hardcastle, M.J., Sakelliou, I., \& Worrall, D.M. 2005b, MNRAS, 359, 1007                                                      
\bibitem[Hardcastle et al. 2007]{hardcastle07} Hardcastle, M.J., Croston, J.H., \& Kraft, R.P. 2007 ApJ, 669, 893                                                               %
\bibitem[Harris et al. 2000]{harris00} Harris, D.E., et al. 2000 ApJ, 530, L81                                                                                                  
\bibitem[Harris \& Krawczynski 2002]{harris02} Harris, D.~E., Krawczynski, H. 2002, ApJ, 565, 244
\bibitem[Harris et al. 2002]{harris02a} Harris, D.E., Krawczynski, H., \& Taylor, G.B. 2002a ApJ, 578, 740                                                                        
\bibitem[Harris et al. 2004]{harris04} Harris, D.E., Mossman, A.E., \& Walker, R.C. 2004a ApJ, 615, 161                                                                           
\bibitem[Harris \& Krawczynski 2006]{harris06} Harris, D.~E., Krawczynski, H. 2006, \araa, 44, 463
\bibitem[Harris et al. 2010]{harris10} Harris, D.~E., Massaro, F., Cheung, C.~C. 2010, AIPC, 1248, 355
\bibitem[Harris et al. 2011]{harris11} Harris, D.~E., Massaro, F., Cheung, C.~C. 2011 in prep.
\bibitem[Hine \& Scheuer 1980]{hine80} Hine, R. G. \& Scheuer, P. A. G. 1980 MNRAS, 193, 285
\bibitem[Hodges-Kluck et al. 2010]{hodgeskluck04} Hodges-Kluck, E.J., Reynolds, C.S., Miller, M.C., Cheung, C.C. 2010 ApJL, 717, L37                                            
\bibitem[Hogan et al. 2011]{hogan11} Hogan, B.S., Lister, M.L., Kharb, P., Marshall, H.L., Cooper, N.J. 2011 ApJ, 730, 92                                                       %
\bibitem[Hogg 2000]{hogg00} Hogg, D. W. 	arXiv:astro-ph/9905116v4
\bibitem[Hogg et al. 2002]{hogg02} Hogg, D. W., Baldry, I. K., Blanton M. R. \& Eisenstein D. J. arXiv:astro-ph/0210394v1
\bibitem[Hough \& Readhead 1989]{hough89} Hough, D.~H. \& Readhead, A.~C.~S. 1989, AJ, 98, 1208
\bibitem[Kendall \& Stuart 1979]{kendall79} Kendall, M., \& Stuart, A., 1979, ``The Advanced Theory of Statistics,'' Mac Millan, New York 
\bibitem[Kalberla et al. 2005]{kalberla05} Kalberla, P.M.W., Burton, W.~B., Hartmann, D., et al. 2005, A\&A, 440, 775
\bibitem[Kataoka et al. 2003a]{kataoka03a} Kataoka, J., Edwards, P., Georganopoulos, M., Takahara, F., \& Wagner, S. 2003a, A\&A, 399, 91                                        
\bibitem[Kataoka et al. 2003b]{kataoka03b} Kataoka, J., Leahy, J.P., Edwards, P.G., Kino, M., Takahara, F., Serino, Y., Kawai, N., \& Martel, A.R. 2003b A\&A, 410, 833         
\bibitem[Kataoka \& Stawarz 2005]{kataoka05} Kataoka, J. \& Stawarz, \L. 2005, ApJ, 622, 797
\bibitem[Kataoka et al. 2008]{kataoka08} Kataoka, J. et al. 2008, ApJ, 685, 839                                                                                                
\bibitem[Kraft et al. 2000]{kraft00} Kraft, R.P. et al. 2000 ApJ, 531, L9                                                                                                       
\bibitem[Kraft et al. 2005]{kraft05} Kraft, R.P., Hardcastle, M.J., Worrall, D.M., \& Murray, S.S. 2005, ApJ, 622, 149                                                          
\bibitem[Kraft et al. 2007]{kraft07} Kraft, R.P., Birkinshaw, M., Hardcastle, M.J., Evans, D.A., Croston, J.H., Worrall, D.M., \& Murray, S.S. 2007 ApJ, 659, 1008              
\bibitem[Jester et al. 2006]{jester06} Jester, S., Harris, D.~E., Marshall, H.~L., Meisenheimer, K. 2006, ApJ, 648, 900
\bibitem[Jorstad \& Marscher 2004]{jorstad04} Jorstad, S., \& Marscher, A. 2004 ApJ, 614, 615                                                                                   
\bibitem[Jorstad \& Marscher 2006]{jorstad06} Jorstad, S.G. \& Marscher, A.P. 2006, Astr. Nach., 327, 227                                                                       %
\bibitem[Landt et al. 2004]{landt04} Landt, H., Padovani, P., Perlman, E. S., Giommi, P.  2004 MNRAS, 351, 83
\bibitem[Landt et al. 2006]{landt06} Landt, H., Perlman, E. S., Padovani, P. 2006 ApJ, 637, 183
\bibitem[Leahy et al. 1997]{leahy97} Leahy, J.~P. et al. 1997, MNRAS, 291, 20
\bibitem[Ly et al. 2005]{ly05} Ly, C., De Young, D., \& Bechtold, J. 2005, ApJ, 618, 609                                                                                        
\bibitem[Mackay 1971]{mackay71} Mackay, C.~D. 1971, MNRAS, 154, 209
\bibitem[Marshall et al. 2001]{marshall01} Marshall, H. et al. 2001 ApJ, 549, L167                                                                                              
\bibitem[Marshall et al. 2002]{marshall02} Marshall, H.L., Miller, B.P., Davis, D.S., Perlman, E.S., Wise, M., Canizares, C.R., Harris, D.E., \& Biretta, J.A. 2002, ApJ, 564, 683  
\bibitem[Marshall et al. 2005]{marshall05} Marshall, H.L. et al. 2005, ApJS, 156, 13                                                           %
\bibitem[Marshall et al. 2011]{marshall11} Marshall, H.L. 2011 ApJS, 193, 15
\bibitem[Massaro et al. 2009a]{massaro09a} Massaro, E., et al. 2009a, A\&A, 495, 691
\bibitem[Massaro et al. 2009b]{massaro09b} Massaro, F., Chiaberge, M., Grandi, P., et al. 2009b, ApJ, 692, L123
\bibitem[Massaro et al. 2009c]{massaro09c} Massaro, F., Harris, D.~E., Chiaberge M. et al. 2009c, ApJ, 696, 980
\bibitem[Massaro et al. 2010a]{massaro10a} Massaro, F., Cheung, C.~C., Harris, D.~E. 2010a, AIPC, 1248, 475 
\bibitem[Massaro et al. 2010b]{massaro10b} Massaro, F., et al. 2010b, ApJ, 714, 589
\bibitem[Massaro et al. 2011a]{massaro11a} Massaro, F., Cheung, C.~C., Harris, D.~E. 2011a, IAUS, 275, 160
\bibitem[Massaro et al. 2011b]{massaro11b} Massaro, F., Siemiginowska, A., et al. 2011b, in prep.
\bibitem[Massaro et al. 2011c]{massaro11c} Massaro, F., Cheung, C.~C., Harris, D.~E. 2011c, in prep.
\bibitem[Meisenheimer et al. 1989]{meisenheimer89} Meisenheimer, K., Roser, H.-J., Hiltner, P. R., Yates, M. G., Longair, M. S., Chini, R., Perley, R. A. 1989 A\&A, 219, 63
\bibitem[Miller et al. 2006]{miller06} Miller, B.P., et al. 2006, ApJ, 652, 163                                                                                                 
\bibitem[Miller \& Brandt 2009]{miller09} Miller, B.P., \& Brandt, W.N. 2009, ApJ, 695, 755                                                                                     
\bibitem[Orr \& Browne 1982]{orr82} Orr, M.~J.~L. \& Browne, I.~W.~A. 1982, MNRAS, 200, 1067
\bibitem[Park et al. 2006]{park06} Park, T., Kashyap, V. L.; Siemiginowska, A., van Dyk, D. A., Zezas, A.; Heinke, C. W., Bradford J. 2006 ApJ, 652, 610
\bibitem[Perlman et al. 2010a]{perlman10a} Perlman, E.S., Georganopoulos, M., May, E.M., Kazanas, D. 2010a ApJ, 708, 1                                                          
\bibitem[Perlman et al. 2010b]{perlman10b} Perlman, E.S., Padgett, C.A., Georganopoulos, M., et al. 2010b, ApJ, 708, 171                                                        
\bibitem[Pesce et al. 2001]{pesce01} Pesce, J.E., Sambruna, R.M., Tavecchio, F., Maraschi, L., Cheung, C.C., Urry, C.M. \& Scarpa, R. 2001 ApJL, 556, L79                       
\bibitem[Sambruna et al. 2002]{sambruna02} Sambruna, R.M., Maraschi, L., Tavecchio, F., Urry, C.M., Cheung, C.C., Chartas, G., Scarpa, R., \& Gambill, J.K. 2002 ApJ, 571, 206  %
\bibitem[Sambruna et al. 2004]{sambruna04} Sambruna, R.M., Gambill, J.K., Maraschi, L., Tavecchio, F., Cerutti, R., Cheung, C.C., Urry, C.M., \& Chartas, G. 2004 ApJ, 608, 698 %
\bibitem[Sambruna et al. 2007]{sambruna07} Sambruna, R.M., Donato, D., Tavecchio, F., Maraschi, L., Cheung, C.C., \& Urry, C.M. 2007 ApJ, 670, 74                               
\bibitem[Sambruna et al. 2008]{sambruna08} Sambruna, R.M., Donato, D., Cheung, C.C., Tavecchio, F., Maraschi, L. 2008, ApJ, 684, 862                                            
\bibitem[Scarpa \& Urry 2002]{scarpa02} Scarpa, R. \& Urry, M. C. 2002 NewAR, 46, 405
\bibitem[Schwartz et al. 2000]{schwartz00} Schwartz, D.~A., et al. 2000, ApJ, 540, L69
\bibitem[Schwartz et al. 2006]{schwartz06} Schwartz, D.A., Marshall, H.L., Lovell, J.E.J., et al. 2006, ApJL, 641, L107                                                         
\bibitem[Siemiginowska et al. 2002]{siemiginowska02} Siemiginowska, A., Bechtold, J., Aldcroft, T.L., Elvis, M., Harris, D. E., \& Dobrzycki, A. 2002, ApJ, 570, 543            
\bibitem[Siemiginowska et al. 2003a]{siemiginowska03a} Siemiginowska, A. et al. 2003 ApJ, 595, 643                                      
\bibitem[Siemiginowska et al. 2003b]{siemiginowska03b} Siemiginowska, A., Smith, R. K., Aldcroft, Thomas L., Schwartz, D. A., Paerels, F., Petric, A. O. 2003 ApJ, 598L, 15 
\bibitem[Sun et al. 2005]{sun05} Sun, M., Jerius, D., \& Jones, C. 2005 ApJ, 633, 165                                                                                           
\bibitem[Tavecchio et al. 2000]{tavecchio00} Tavecchio, F., Maraschi, L., Sambruna, R.~M., Urry, C.~M. 2000, ApJ, 544, L23
\bibitem[Tavecchio et al. 2007]{tavecchio07} Tavecchio, F., Maraschi, L., Wolter, A., Cheung, C.C., Sambruna, R.M., \& Urry, C.M. 2007, ApJ, 662, 900                           
\bibitem[Taylor 2005]{taylor2005} Taylor, M.~B. 2005, ASP Conf.~Ser., 347, 29 
\bibitem[Urry \& Padovani 1995]{urry95} Urry, C.~M., \& Padovani, P. 1995, PASP, 107, 803
\bibitem[Wilson et al. 2000]{wilson00} Wilson, A.S., Young, A.J., \& Shopbell, P.L. 2000 ApJL, 544, L27                                                                         
\bibitem[Wilson et al. 2001]{wilson01} Wilson, A.S., Young, A.J., \& Shopbell, P.L. 2001 ApJ, 547, 740                                                                          
\bibitem[Worrall et al. 2001]{worrall01} Worrall, D.M., Birkinshaw, M., \& Hardcastle, M.J. 2001 MNRAS, 326, L7                                                                 
\bibitem[Worrall et al. 2003]{worrall03} Worrall, D.M., Birkinshaw, M., \& Hardcastle, M.J. 2003 MNRAS, 343, 73                                                                 
\bibitem[Worrall et al. 2005]{worrall05} Worrall, D.M., \& Birkinshaw, M. 2005, MNRAS, 360, 926                                                                                  
\bibitem[Worrall 2009]{worrall09} Worrall, D.~M. 2009, A\&ARv, 17, 1
\bibitem[Wright 2006]{wright06} Wright E.~L. 2006, PASP, 118, 1711
\end{thebibliography}
\end{document}